\documentclass[apj,numberedappendix]{emulateapj}
\usepackage{graphicx}
\usepackage{csquotes}

\usepackage{apjfonts}

\usepackage{graphicx}
\usepackage{color}
\usepackage{amsbsy}
\usepackage{mathrsfs}
\usepackage{mathpazo,bm}
\usepackage{amsmath}
\usepackage{xfrac}
\usepackage{multirow}

\newlength{\hfwidth}
\newlength{\hfwidthsingle}
\newlength{\figspace}
\newcommand{\beq}{\begin{equation}}
\newcommand{\eeq}{\end{equation}}
\newcommand{\beqn}{\begin{eqnarray}}
\newcommand{\eeqn}{\end{eqnarray}}
\newcommand{\pderiv}[2]{\frac{\partial{#1}}{\partial{#2}}}

\newcommand{\ptderiv}[1]{\frac{\partial{#1}}{\partial{t}}}

\newcommand{\Rey}{\mathrm{Re}}

\newcommand{\Ro}{\mathrm{Ro}}
\newcommand{\ttimes}[1]{10^{#1}}

\newcommand{\vt}[1]{\boldsymbol{\mathrm{#1}}} 
\renewcommand{\v}[1]{{\boldsymbol{#1}}}

\newcommand{\ksi}{\xi}

\newcommand{\del}{\v{\nabla}}
\newcommand{\grad}{\del}
\newcommand{\Div}{\del\cdot}
\newcommand{\curl}{\del\times}

\newcommand{\hatr}{\hat{\v{r}}}
\newcommand{\hatphi}{\hat{\v{\phi}}}

\newcommand{\va}{v_{_{\rm A}}}

\newcommand{\cv}{c_{_{V}}}
\newcommand{\cp}{c_p}

\newcommand{\OmegaPV}{\varOmega_{_{\rm PV}}}
\newcommand{\OmegaGPV}{\varOmega_{_{\rm GPV}}}
\newcommand{\betaPV}{\beta_{_{\rm PV}}}
\newcommand{\mS}{\mathcal S}
\newcommand{\mG}{\mathcal G}
\newcommand{\mA}{\mathcal A}

\newcommand{\Eq}[1]{Eq.~(\ref{#1})}

\newcommand{\eq}[1]{\Eq{#1}}
\newcommand{\eqp}[1]{(Eq.~\ref{#1})}

\newcommand{\Fig}[1]{Fig.~\ref{#1}}
\newcommand{\fig}[1]{\Fig{#1}}

\renewcommand{\table}[1]{Table~\ref{#1}}
\newcommand{\sect}[1]{Sect.~\ref{#1}}

\shorttitle{{Hydrodynamical instabilities in protoplanetary disks}}
\shortauthors{Lyra \& Umurhan}

\slugcomment{Draft version}

\begin{document}

\title{The initial conditions for planet formation :\\ 
Turbulence driven by hydrodynamical instabilities in disks around young stars.}

\author{Wladimir Lyra\altaffilmark{1,2} and Orkan M. Umurhan\altaffilmark{3,4,$\bigstar$}}

\altaffiltext{1}{California State University, Northridge. Department
  of Physics and Astronomy 18111 Nordhoff St, Northridge, CA 91330, wlyra@csun.edu.}
\altaffiltext{2}{Jet Propulsion Laboratory, California Institute of Technology, 4800
Oak Grove Drive, Pasadena, CA, 91109, wlyra@jpl.nasa.gov.}
\altaffiltext{3}{NASA Ames Research Center, Space Sciences Division, Planetary Sciences Branch, Moffatt Field, CA 94035}
\altaffiltext{4}{SETI, Carl Sagan Center, 190 Bernardo Way, Mountain View, CA 94043}
\altaffiltext{$\bigstar$}{Corresponding Author: orkan.m.umurhan@nasa.gov}

\date{Received ; Accepted}

\begin{abstract}
This review examines recent theoretical developments in our understanding of turbulence in cold, non-magnetically active, planetesimal forming regions of protoplanetary disks which we refer to throughout as ``Ohmic zones". We give a brief background introduction to the subject of disk turbulence followed by a terse pedagogical review of the phenomenology of hydrodynamic turbulence.  The equations governing the dynamics of cold astrophysical disks are given and basic flow states are described. We discuss the Solberg-H{\o}iland conditions required for stability, and the three recently identified turbulence generating mechanisms possibly active in protoplanetary disk Ohmic zones, namely, (i) the Vertical Shear Instability, (ii) The Convective Overstability and (iii) the Zombie Vortex Instability.  We summarize the properties of these processes, identify their limitations and discuss where and under what conditions these processes are active in protoplanetary disk Ohmic zones.  
\end{abstract}

\section{Introduction}
\label{sect:introduction}

Planet formation is simultaneously one of the oldest and one of the newest concerns of human inquiry. ``How did the Earth come to be?'' is a question that almost invariably appears in the cosmogonies of the ancients. They not always had a clear idea of what ``Earth'' meant, but this is a question that, in one form or another, virtually every society in recorded history has at some point asked itself. Particularly interesting are the ideas of Leucippus (480-420? B.C.E.) who, according to testimonial, is to have said \citep{Diels_Kranz_1961}
\begin{quote}
{\emph{The worlds come into being as follows: many bodies of all sorts and shapes move from 
the infinite into a great void; they come together there and produce a single whirl, in 
which, colliding with one another and revolving in all manner of ways, they begin 
to separate like to like.}}
\par
\medskip
\noindent
Diogenes Laertius IX, 31
\footnote{Scholars of the classical period note that nothing but third person accounts survive of Leucippus' words.}
\end{quote}

\noindent This vision strikes surprisingly modern, and not without foundation within the modern theory of planet formation. Substitute ``many bodies of all sorts and shapes'' by {\it gas and dust}, then ``single whirl'' by {\it protoplanetary disk} and finally ``revolving in all manner of ways'' by {\it turbulence}, and it could have figured in the introduction of a paper in the latest issue of a major astronomy journal. This attests not to clairvoyance of the ancient Greeks, but to the antiquity of the question. Given the huge sample space, some of the educated guesses of the time are bound to contain some truth.

By the 18th century, Newtonian gravity and the orbits of the planets were understood in enough detail to realize that the low inclinations of the orbits implied that the easiest way to attain that configuration was if the planets have formed in a disk that orbited the proto-Sun (Kant 1755). Because Jupiter and Saturn are gas giant planets, this disk must have been a disk of gas. Early mathematical considerations by Laplace (1796) applied Newton's theory of universal gravitation and laws of motion to a slowly rotating spherical cloud, implying that it should collapse under its own weight. Due to conservation of angular momentum, the gas settles into a flat disk orbiting the condensing proto-Sun in the center. In this {\it solar nebula}, planets are taking shape.\par
 C. F. von Weizs{\"a}cker extended these fundamental notions and pointed out that eddies in the forming solar nebula ought to increase with distance from the Sun prefiguring the role these may have in the formation of planetesimals out of dust \citep{Gamow_Hynek_1945}.  Rather presciently,  Weizs{\"a}cker also reasoned that many
other star systems should harbor similar kinds of nebulae like our own Sun did after its birth.

Modern searches have now revealed such disks around young stars \citep{ElsasserStaude78,Rucinski85,Aumann85,SargentBeckwith87,Strom+89, Beckwith+90, O'dellWen94, McCaughreanO'Dell96,Ricci+08}, now called {\it protoplanetary disks} or, in symmetry with the solar nebula, {\it extrasolar nebulae} or {\it exonebulae}.  {In general, protoplanetary disks 
range in radius from 10s to 100s of AU \citep[e.g.][]{Ansdell+18}, in density from $10^{13}$ to $10^{15}$ cm$^{-3}$ {\bf (for mean molecular weight $\mu=2.5$)}, in mass from $10^{-3}$ to $10^{-1} M_\odot$, and from 1000\,K near the star to 10\,K in the outskirts of the disk.} These disks form while the star is being formed, as a consequence of the cloud's gravitational collapse, and here we already find one of the first formidable problems of star and planet formation: as interstellar clouds are huge in size, even the slightest rotation means far too much angular momentum \citep{Mestel65a,Mestel65b}.  {Indeed, if a 
proto-star was to accommodate all the angular momentum in a ring at 1AU containing about 10\% of its mass, it 
would achieve break-up velocity, a problem that also exists in similar form for the gas giant planets in the Solar System \citep{TakataStevenson96,Bryan+18,Batygin18}. In order to accrete, the gas must somehow find a way to transfer its angular momentum and if a gas parcel does this in nearly circular orbits, it must simultaneously find a way to lose energy in the process. }

Because gravity is a radial force, and an axisymmetric disk has no pressure forces in the azimuthal direction, the only process that produces change in angular momentum is viscous stresses. Angular momentum transport is mediated by viscosity; without it, gas will simply orbit the star, not changing its radial position, and star formation will not complete. Although molecular viscosity is much too small to account for the observed mass accretion rate, it was recognized \citep{ShakuraSunyaev73,Lynden-BellPringle74} that the stochastic behavior of turbulence would lead to diffusion of momentum, similarly to viscosity. Moreover, while molecular viscosity acts in the small scales of the flow, turbulence acts all the way up to the integral scale, generating much more powerful stresses. The question of angular momentum transport is thus a question of how the disk becomes turbulent. 

The 1970s and 1980s saw several possible candidates to generate turbulence and angular momentum transport in disks, such as convection \citep{Cameron78,LinPapaloizou80} or nonlinear instability \citep{Shakura+78}. Yet, it was only with the re-discovery of the magnetorotational instability \cite[MRI;][]{BalbusHawley91} that the question seemed settled. The MRI entails a combination of a weak (subthermal) magnetic field and the shear present in the Keplerian rotation the gas, that de-stabilizes the flow. The instability is powerful, seeming to explain the required accretion rates.

The MRI also has a positive effect on planet formation. Starting with micron-sized dust grains, coagulation models \citep{Brauer+07} predict growth to centimeter size by electromagnetic hit-and-stick mechanisms (mostly van der Walls forces). However, growth beyond this size is halted, for two reasons. First, collisions between pebbles lead to destruction rather than growth \citep{Benz00}. Second, because of the balance between pressure, rotation and gravity, the gas orbits the star slightly slower than an independent body at the same distance would. Consequently, pebbles tend to outpace the gas. The resulting headwind drains their angular momentum, leading them into spiral trajectories towards the star, in timescales as short as a hundred years at 1AU \citep{Weidenschilling77a}. A distinct possibility to solve these problems is gravitational instability of the layer of solids \citep{Safronov72,Lyttleton72,GoldreichWard73,YoudinShu02}. When the dust aggregates had grown to centimeter size, the gas drag is reduced and the solids are pushed to the midplane of the disk due to the stellar gravity. Although such bodies do not have enough mass to attract each other individually, the sedimentation increases the solids-to-gas ratio by orders of magnitude when compared to the interstellar value of $10^{-2}$. It was then hypothesized \citep{Safronov72} that due to the high densities of this midplane layer, the solids could collectively achieve critical number density and undergo direct gravitational collapse. Such a scenario has the advantage of occurring on very rapid timescales, thus avoiding the radial drift barrier.

This picture was nonetheless shown to be simplistic, in the view that even low levels of turbulence in the disk preclude the midplane layer of solids from achieving densities high enough to trigger the gravitational instability \citep{Weidenschilling80}. Even in the absence of self-sustained turbulence such as the one generated by the MRI, the solids themselves can generate turbulence due to the backreaction of the drag force onto the gas. Such turbulence can be brought about by Kelvin-Helmholtz instabilities due to the vertical shear the dense layer of solids induces on the gas \citep{Weidenschilling80,WeidenschillingCuzzi93,Sekiya98,Johansen+06}, or by streaming instabilities induced by the radial migration of solids particles \citep{YoudinGoodman05,Paardekooper06,YoudinJohansen07,JohansenYoudin07}. In the turbulent motion, the solids are stirred up by the gas, forming a vertically extended layer where the stellar gravity is balanced by turbulent diffusion \citep{Dubrulle+95,GaraudLin04}. But if turbulence precludes direct gravitational collapse through sedimentation, it was also shown that it allows for it in an indirect way. As solid particles concentrate in high pressure regions \citep{HaghighipourBoss03}, the solids-to-gas ratio can be enhanced in the transient turbulent gas pressure maxima, potentially reaching values high enough to trigger gravitational collapse. Numerical calculations by \cite{Johansen+07} show that this is indeed the case, with the particles trapped in the pressure maxima generated by the MRI collapsing into dwarf planets when the gravitational interaction between particles is considered. While the MRI is not strictly necessary, the streaming instability does not seem to operate effectively for solar and subsolar metallicities, requiring more solids than initially present in the Solar Nebula. 

 {However, the conditions for the magnetorotational instability are not met in a huge portion of the disk \citep{BlaesBalbus94}. Being a mechanism that depends on magnetization, it also depends on ionization; as a result, if the ionization fraction is such that resistive effects become important, the MRI may shut down, defining a MRI-dead zone \citep{BlaesBalbus94,Gammie96}. The partially ionized gas can be seen as a three-fluid system, composed of the ion fluid, the electron fluid, and the neutral fluid. In ideal MHD, where density and ionization fraction are sufficiently high, ions and electrons are tied to the magnetic field and they drag the neutrals along with them; deviations from ideal conditions occur when these fluids start to decouple. In the ambipolar diffusion limit the magnetic field is tied to the electrons and ions, and drifts with them through the neutrals \citep{BlaesBalbus94,MacLow+95}. In the Ohmic limit collisions with neutrals are efficient in dragging the ions and electrons, and the magnetic field is not frozen in any fluid. In between these limits lies the domain of Hall MHD, where the electrons are coupled to the magnetic field, but collisions with neutrals decouple the ions \citep{Wardle99,BalbusTerquem01,SalmeronWardle05,PandeyWardle06,Wardle07,PandeyWardle08,WardleSalmeron12}. At constant magnetic field, this is a progression as density increases, that first the neutrals decouple from magnetic effects (ambipolar diffusion), then the ions (Hall MHD), and finally the electrons (Ohmic resistivity). 

We focus on the Ohmic zone, deep into the resistive regime and any magnetic effect is irrelevant. We could have used ``hydrodynamical'' zone,  yet we keep the name Ohmic for juxtaposition to the other non-ideal MHD effects. It is in this non-magnetic regime that the hydrodynamical instabilities we review exist. 

The physics of the Hall-dominated and ambipolar-dominated zones is not covered in this review, but we point out their importance in providing boundary conditions for the Ohmic zone. The upper atmosphere of the disk will be dominated by ambipolar diffusion, with the emission of a magnetocentrifugal wind \citep{BaiStone11, Gressel+15}. As such, it will provide a ``lid'' to the instabilities here discussed. Until their interaction with the ambipolar zone is defined, we cannot assume that these instabilities are felt in the disk atmosphere; if they are not, they should not be seen in infrared observations. The Hall-dominated zone has a dynamical instability, the Hall shear instability, if the angular momentum vector and the magnetic field are aligned \citep{Kunz08,KunzLesur13,Lesur+14,Bai15}. The reader should keep in mind that our knowledge of the physics is these zones is still taking shape, and it is unclear how the 
hydrodynamical instabilities we review behave in the ambipolar and Hall zones.}

The  {Ohmic} zone is extended. In the inner disk, ionization is provided by thermal collisions. Down to $\approx$900K, thermal velocities have enough energy to ionize the alkali metals.  {This temperature corresponds to a distance of 0.1\,AU, depending on the underlying disk model, so only inwards of it will the disk be sufficiently ionized for the MRI. Concurrently, outward of $\approx$30\,AU (again depending on the underlying disk model), even though the temperatures are low, the column density is low enough that stellar X-rays can ionize the disk throughout.} Between these limits, a large dead zone exists. This, incidentally, is squarely where we expect planets to form, at least in the Solar System. If we need turbulence to generate accretion and form planets, we have to look beyond the magnetorotational instability. If star formation necessitates turbulent transport through the accretion disk, and if planet formation needs turbulent pressure maxima to trigger the streaming instability in solar metallicity, we need to look beyond the MRI. 

The past few years have seen a number of processes being proposed for purely hydrodynamical instabilities in disks:  {two linear instabilities}, the {\it Vertical Shear Instability} \citep{Nelson+13,Stoll_Kley_2014} and the {\it Convective Overstability} \citep{KlahrHubbard14,Lyra14}, and  {a non-linear instability} the {\it Zombie Vortex Instability} \citep{Marcus+15,Marcus+16}. These processes were linked to different regimes of opacities by \cite{Malygin+17}. It is the 
purpose of this review to frame these disparate processes into a coherent picture of hydrodynamical processes in Ohmic dead zones of accretion disks. This review is organized as follows. In the next section we briefly review the physics of accretion disks. On \sect{sect:turbulence} we give an overview of turbulence, introducing the hydrodynamical instabilities on \sect{sect:instabilities}. A synthesis, the main goal of this review, is given on \sect{sect:synthesis}. Following it, \sect{sect:future} presents our opinion for the directions of the field in the next few years, finally concluding on \sect{sect:conclusions}. 

\begin{table*}
\caption[]{Symbols used in this work.}
\label{table:symbols}
\begin{center}
\begin{tabular}{lll c lll}\hline
Symbol & Definition & Description & & Symbol & Definition & Description\\\hline
$R$ & & cylindrical radial coordinate&&                                                            $N_z$ & \eq{eq:bruntZ} & vertical Brunt-V\"ais\"al\"a frequency \\                   
$\phi$ & & azimuth &&                                                                              $N_R$ & \eq{eq:bruntR} & radial Brunt-V\"ais\"al\"a frequency \\                        
$z$ && vertical coordinate&&                                                                       $\v{F}$ & & body force \\                                                                                     
$r$ && spherical radial coordinate&&                                                               $\ell$ & $\approx 2\pi/k$ & vortex eddy length scale \\                                                   
$\lambda$ && wavelength&&                                                                          $u_\ell$ & & typical eddy speed \\                                                                        
$k$ &$=2\pi/\lambda$& wavenumber&&                                                                 $t_\ell$ & $\approx \ell/u_\ell$ &  eddy overturn time \\                                                 
$m$ & & azimuthal wavenumber&&                                                                     $\ell_0$ & & energy injection scale \\                                                                    
$t$&  & time &&                                                                                    $\ell_{\rm diss}$ & & dissipation length scale \\                                                                                   
$\rho$ & & density&&                                                                               $\varepsilon$ & erg\,s$^{-1}$\,g$^{-1}$& energy dissipation rate per unit mass\\                                                    
$\v{u}$ & & velocity&&                                                                             ${\rm E}_k$ & & energy per mass in eddies between $k$ and $k+\delta k$\\                                                            
$T$& & temperature&&                                                                               ${\cal E}$ &$={\rm E}_k/k$ & energy per mass density\\                                                                              
$\gamma$ & & adiabatic index&&                                                                     $\v{\varomega}$&$=\curl{\v{u}}$&vorticity\\                                                               
$\cp$ & & specific heat at constant pressure &&                                                    ${\cal Z}$ &$=\varomega^2$& enstrophy \\                                                                  
$\cv$ &$=\cp/\gamma$& specific heat at constant volume &&                                          $\varPsi$ & $u = \curl{\varPsi\hat{\v{z}}}$ & streamfunction\\                                            
$c_s$ &$=\left[T \ \cp (\gamma-1)\right]^{1/2}$& sound speed&&                                     $\OmegaPV$ & $=\rho^{-1}\v{\varomega}\cdot \del s  $ & potential vorticity \\                             
$p$&$=\cv (\gamma-1) \rho T $ & pressure &&                                                        $\betaPV$ &$\equiv \pderiv{\OmegaPV}{\rm q}$  & potential vorticity gradient \\                                                     
$s$&=$\cv\ln(p/\rho^{\gamma})$ & specific entropy&&                                                $l$ & & latitude\\                                                                                                                          
$\varPhi$&$=-GM_\star/r$ & gravitational potential&&                                               $\OmegaGPV$ & \eq{eq:general_PV} & generalized potential vorticity \\                                                               
$G$& & gravitational constant&&                                                                    $\overline{u_i' u_j'}$& & spatial correlation of component velocity fluctuations\\                                                     
$M_\star$& & stellar mass&&                                                                        $\sigma_{\rm SB}$ & & Stefan-Boltzmann constant\\                                                         
$\mu$& & mean molecular weight&&                                                                   $\alpha$ & = $\overline{u_r' u_\phi'}\big/c_s^2 $& measure of turbulent intensity\\                                                 
${\rm R}$& & gas constant&&                                                                        $\nu_t$ & $ = \alpha c_s H$ & effective turbulent viscosity\\                                                                       
$\v{{\rm T}}$ && viscous stress tensor &&                                                          $L$ & $=R^2\varOmega$ &  angular momentum\\                                                                                         
$\varsigma$& & dynamic shear molecular viscosity&&                                                 $K$ & &  spring constant\\                                                                                
$\zeta$& & dynamic bulk molecular viscosity&&                                                      $\rm q$ && generalized coordinate\\                                                                                                  
$\nu$&$=\varsigma/\rho$ & kinematic viscosity &&                                                   $\v{\ksi}$ && Lagrangian displacement\\                                                                                             
$\Rey$&$=\mathcal{U}\mathcal{L}/\nu$ & Reynolds number&&                                           $\v{B}$ && magnetic field\\                                                                                                             
$\mathcal{U}$& & representative velocity&&                                                         $\v{\va}$ & $=\v{B}/\sqrt{4\pi\rho}$ & Alfv\'en velocity\\                                                                          
$\mathcal{L}$& & representative length&&                                                           $\beta$ &$= 2c_s^2/\va^2$ & plasma beta \\                                                                                              
$\mathcal{Q}$& & heat source&&                                                                     $\eta$ && resistivity\\                                                                                                             
$\varOmega$ &$=\sqrt{GM_\star/R^3}$ & Keplerian angular frequency &&                               $\v{J}$&&current\\                                                                                                                   
$H$ &$= c_s/\varOmega$ & disk scale height &&                                                      $c$ && speed of light\\                                                                                   
$h$ &$=H/R$ & disk aspect ratio &&                                                                 ${\rm Re_M}$ &$=\mathcal{U}\mathcal{L}/\eta$ & Magnetic Reynolds number\\                   
$\varSigma$ & $\propto \rho H$ & column density  &&                                                $\varLambda$ &$=\va^2/\varOmega\eta$ & Elssaser number\\                                    
$q$ & $=-d\ln \varOmega/d\ln R $ & shear parameter &&                                              $Z$&&charge multiplicity\\                                                                                
$q_\rho$ & $=-d\ln\rho/d\ln R $ & radial density gradient &&                                       $\gamma_i$ & $\v{f} \equiv \gamma_i \rho_i \rho (\v{u}_i-\v{u})$ & ion-neutral drag coefficient\\                                           
$q_T$ & $=-d\ln T/d\ln R $ & radial temperature gradient &&                                        $\sigma_{\rm coll}$ & & collisional cross section\\                                                                                       
$q_\Sigma$ & $=-d\ln \varSigma/d\ln r $ & column density gradient  &&                              $n$ & & number density\\
$\omega$ && complex eigenfrequency&&                                                               $x$ & $\equiv n_e/n$ & ionization fraction\\                                                                  
$\sigma$ & $={\rm Im}(\bar\omega)$ & growth rate &&                                                $\kappa_R$ & & Rossland mean opacity\\                                                                        
${\rm Ma}$ &$={\cal U}/c_s$& Mach number &&                                                        $\ell_r$ & $=(\kappa_r\rho)^{-1}$ & photon mean free path\\                                                   
$\kappa_{\rm ep}$ & $ = \varOmega\sqrt{2(2-q)}$ & epicyclic frequency &&                           $\tau_r$ & \eq{eq:relaxationtimes} & thermal relaxation time\\                                                
$\mA_k$ & $=c_s \ \partial_k \ln \rho$ &normalized density gradient&&                                                             $a_{\rm rad}$ & $=4\sigma_{SB}/c$ & Radiation constant\\                                                      
$\mG_k$ & $=\gamma^{-1}c_s\ \partial_k \ln p$ &normalized pressure gradient&&                                                      $\chi$ && Coefficient of thermal diffusion\\                                                                  
$\mS_k$ & $=\mG_k-\mA_k$  & normalized entropy gradient&&                 ${\rm Ro}$ && Rossby number\\\hline                                                                                 
\end{tabular}
\end{center}
\end{table*}

\section{Astrophysical disks: equations and steady states} 
\label{sect:disks}

Disks are objects in steady state. They are gaseous objects, so we need to solve the equations of hydrodynamics in a central potential. The equations to solve are 

\beqn
\pderiv{\rho}{t} &=& - \left(\v{u}\cdot\del\right)\rho - \rho \Div{\v{u}},\label{eq:continuity}\\
\rho\pderiv{\v{u}}{t} &=& - \rho\left(\v{u}\cdot\del\right)\v{u} - \grad{p} 
- \rho\grad\varPhi + \grad \cdot {\v{{\rm T}}} + \v{F}, \label{eq:navier-stokes}\\
p&=& \rho c_s^2/\gamma\label{eq:eos}\\
\varPhi  &=& -\frac{GM_\star}{r},
\eeqn
\noindent where $\rho$ is density, $\v{u}$ is the velocity, $p$ is the pressure (assumed to be that of an ideal gas law), $\varPhi$ is the gravitational potential, $c_s$ is the sound speed in which $c_s^2 \equiv \gamma {\rm R} T/\mu$ where $T$ is the gas temperature, $\mu$ the gas mean molecular weight, R is universal gas constant, and $\gamma$ is the adiabatic index given by the ratio of specific heats at constant pressure and volume respectively. Additional body forces are represented by the vector $\v{F}$.  Finally, $G$ is the gravitational constant, $M_\star$ is the stellar mass, and $r$ the astrocentric distance. The mathematical symbols used in this work are listed in \table{table:symbols}. 
\par
The symmetric viscous stress tensor is $\v{{\rm T}}$, and in Cartesian coordinates it is expressed according to the Einstein index convention form
\beq
\v{{\rm T}} \leftrightarrow 
{\rm T}_{ik} \equiv \varsigma\left[\left(\pderiv{u_i}{x_k} + 
\pderiv{u_k}{x_i}\right)
-\frac{2}{3}\delta_{ik}\pderiv{u_m}{x_m}\right] + \zeta \delta_{ik}\pderiv{u_m}{x_m},
\eeq 
and recalling that according to this convention, repeated indices are summed over from 1 to 3. Thus, 
\beq
\grad \cdot {\v{{\rm T}}} \leftrightarrow 
\pderiv{{\rm T}_{ik}}{x_k} =
\pderiv{{\rm T}_{i1}}{x_1} + \pderiv{{\rm T}_{i2}}{x_2} + \pderiv{{\rm T}_{i3}}{x_3}.
\eeq
The quantities $\varsigma$ and $\zeta$ are the dynamic shear and bulk {\emph {molecular}} viscosities whose values are rooted in the mean free paths of the molecules of the gas (e.g., of molecular hydrogen).
 {From a topological point of view, there are two kinds of material motions, either purely compressive or purely incompressible.  Each type of motion will have different properties with respect to the transfer or momentum at the molecular level wherein compressible motions will have a corresponding bulk viscosity while incompressible motions will have a corresponding shear viscosity.  Since most motions of interest for disks have a primarily incompressible character it is usual practice to ignore the bulk viscosity.}
 As such, then we may
define a viscous diffusion $\nu \equiv \varsigma/\rho$.  If further ${\mathcal U}$ and ${\mathcal L}$ 
respectively represent scales typifying unsteady flow speeds and length scales of the 
fluid system, then one can characterise the relative importance of viscosity through
the definition of the Reynolds number
\beq
\Rey \equiv {\mathcal U}{\mathcal L}\big/\nu.
\label{eq:reynolds}
\eeq
A very large value of Re indicates that viscosity is relatively weak.  For protoplanetary disks,
values of $\Rey$ can be as high as 10$^{14}$ which is based on (i) the values of H$_2$'s viscosity, (ii) the assumption that
local soundspeeds give an upper limit of the fluctuating velocity scales and, (iii) the local disk pressure scale height describes relevant length scales.
In practice then, owing to the small values of these viscosities and the large scales of interest, protoplanetary disk flows are treated as being inviscid and these stress terms are dropped.  However we will eventually come full-circle: owing to the strong linear instabilities found in these flows, we must replace these stress terms with an equivalent {\emph{turbulent}} viscosity model in order to account for the flow of turbulent energy (see discussion in \sect{sect:turbulence}).  

Lastly, these equations must be supplemented with one describing the energy/entropy content of the flow.  Typically, we start with the evolution equation for the entropy,

\beq
\rho T \left(\pderiv{s}{t} + \v{u}\cdot\del s\right) = {\cal Q}.
\label{eq:entropy}
\eeq

\noindent where the specific entropy $s$ is defined based on the specific heat at constant volume

\beq
s \equiv \cv \ln(p/\rho^{\gamma}).
\eeq

The heat source term ${\cal Q}$ will usually have a viscous heating and either a radiative cooling or thermal diffusion term. For  {locally} isothermal disks, one simply sets $\gamma=1$ in \eq{eq:eos} and not solve \eq{eq:entropy}.

Writing \eq{eq:continuity} and \eq{eq:navier-stokes} in cylindrical coordinates ($R$,$\phi$, $z$) we find, without the viscous terms

\beqn
\pderiv{\rho}{t} &=& - u_R \pderiv{\rho}{R} - \frac{u_\phi}{R}\pderiv{\rho}{\phi} - u_z\pderiv{\rho}{z} \nonumber\\
&&- \rho\left( \pderiv{u_R}{R} + \frac{u_R}{R} + \frac{1}{R}\pderiv{u_\phi}{\phi} + \pderiv{u_z}{z}\right)\label{eq:continuity-cyl}\\
\pderiv{u_R}{t} &=& - u_R \pderiv{u_R}{R} - \frac{u_\phi}{R}\pderiv{u_R}{\phi} - u_z\pderiv{u_R}{z} + \frac{u_\phi^2}{R} - \frac{1}{\rho}\pderiv{p}{R} - \frac{GM}{r^3}R\nonumber\\
\label{eq:radial-momentum-cyl}\\
\pderiv{u_\phi}{t} &=& - u_R \pderiv{u_\phi}{R} - \frac{u_\phi}{R}\pderiv{u_\phi}{\phi} - u_z\pderiv{u_\phi}{z} - \frac{u_\phi u_R}{R} - \frac{1}{\rho R}\pderiv{p}{\phi} \label{eq:azimuthal-momentum-cyl}\\
\pderiv{u_z}{t} &=& - u_R \pderiv{u_z}{R} - \frac{u_\phi}{R}\pderiv{u_z}{\phi} - u_z\pderiv{u_z}{z} - \frac{1}{\rho}\pderiv{p}{z} - \frac{GM}{r^3}z \label{eq:vertical-momentum-cyl}
\eeqn

We look for steady state solutions, so $\partial_t = 0$.  We also assume azimuthal symmetry, setting $\partial_\phi = 0$. Consider also that the disk is in vertical hydrostatic equilibrium, so $u_z=0$. Plus, for the moment we assume that the centrifugal balance is maintained, so $u_r=0$. With these constrains, the continuity and angular momentum equations admit only the trivial solution, and we are left with quite little 

\beqn
\frac{GM}{r^3} R & = & - \frac{1}{\rho}\pderiv{p}{R} + \varOmega^2 R, \label{eq:centrifugal-balance}\\
\frac{GM}{r^3} z &=& - \frac{1}{\rho}\pderiv{p}{z}.  \label{eq:hydrostatic-equilibrium}
\eeqn

\noindent The first equation above is what is left of the radial momentum equation, and the second one from the vertical momentum equation. These equations highlight why disks are flat. It is the influence of the centrifugal force: while  {pressure counterbalancing} gravity makes things spherical, the centrifugal force makes things cylindrical.  {The relative strength of the pressure and the centrifugal terms defines if the object will appear more on the spherical side, in hydrostatic equilibrium, or more on the flat side, in centrifugal steady state. A pressure-supported object is spherical, whereas a centrifugally-supported object is flat. \eq{eq:hydrostatic-equilibrium} gives the condition of vertical hydrostatic equilibrium, which we turn to now. }

\begin{figure}
  \begin{center}
    \resizebox{\columnwidth}{!}{\includegraphics{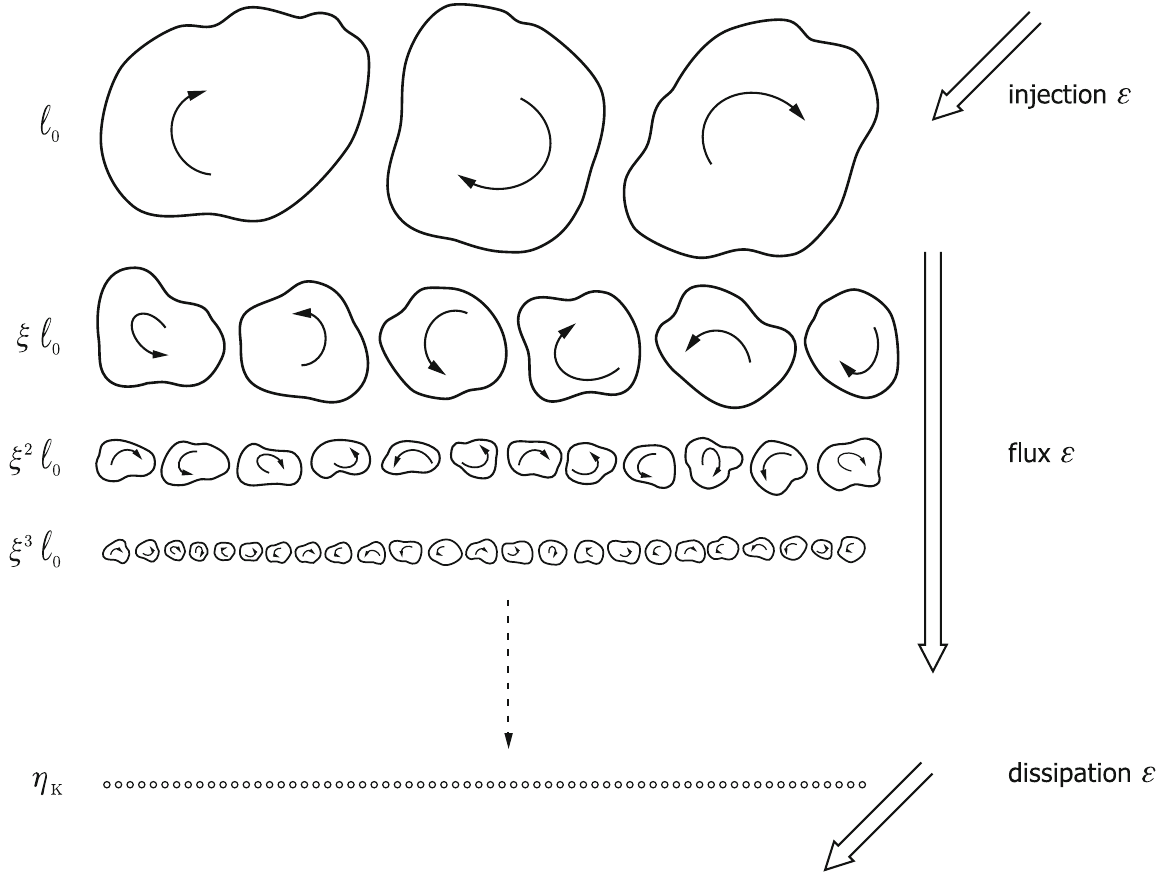}}
\end{center}
\caption[]{A cartoon image of turbulent cascade as cast into verse by L.F. Richardson (1881-1953)
``{\emph{Big whirls have little whirls/That feed on their velocity,/And little whirls have lesser whirls/And so on to viscosity.}}" Each daughter generation is scaled by its parent size by $\xi$.
$\eta_K$ is the same as the dissipation scale $\ell_{\rm diss}$, sometimes also known as the Kolmogorov microscale.  Throughout this inertial range, there is no build up of power -- only a flux of $\varepsilon$ shuttling energy from larger to smaller scales via nonlinear interactions. [Figure from \citet{Regev_Umurhan_Yecko_2016}, by permission.]} 
\label{fig:richardson_cascade}
\end{figure}

  \begin{figure}
  \begin{center}
    \resizebox{\columnwidth}{!}{\includegraphics{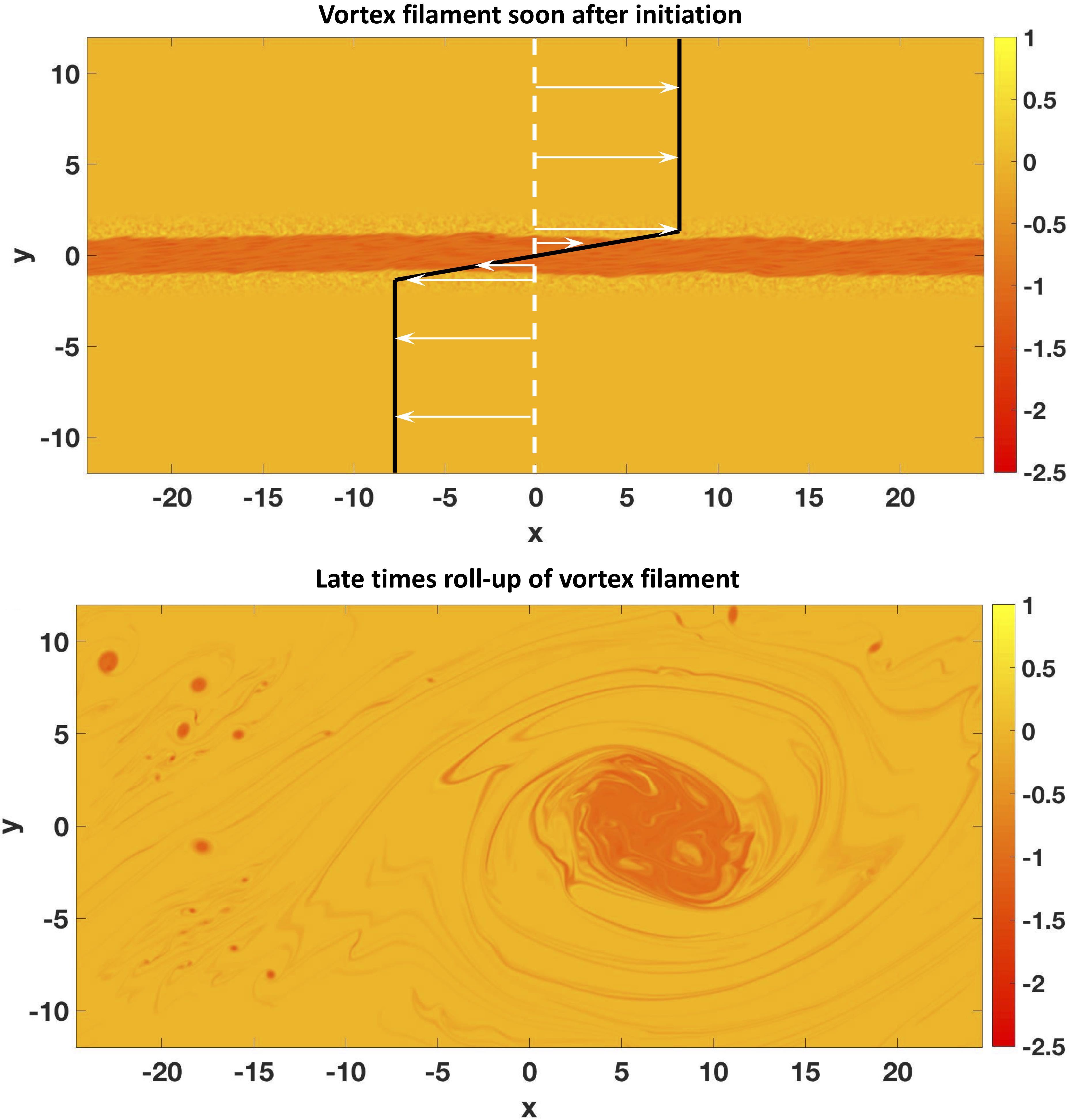}}
\end{center}
\caption[]{Roll-up of 2D vortex strip.  Top panel shows vortex strip (shear layer) soon after initiation with vorticity shown in side intensity panel. Overlain is the sense of the mean velocity field $\v{u}(y)$.  Bottom panel shows the result of the roll-up of this shear layer. The final resulting structure is larger in scale compared to the initial shear layer indicating an inverse cascade of energy.  However, the roll-up has also generated small scale vortex filaments -- indicating the generation of vorticity on scales much smaller than the original strip.  
[Figure adapted from \citet{BiancofioreUmurhan18}, by permission.]} 
\label{fig:shear_layer}
\end{figure}

\subsection{Vertical hydrostatic equilibrium}
\label{sect:vert_equilibrium}

If we assume an isothermal equation of state, or simply that the sound speed does not depend on $z$, so we can write $c_s = c_s(R)$, we can solve for the vertical structure

\beq
\pderiv{\ln \rho}{z} = -\frac{GM}{c_s^2} \frac{z}{r^3} 
\eeq

\noindent which is integrated to yield

\beq
\rho (R,z) = \rho (R) {\rm exp} \left[ \frac{GM}{c_s^2} \left(\frac{1}{\sqrt{R^2+z^2}} - \frac{1}{R}\right)\right].
\label{eq:density}
\eeq

We can write $\sqrt{R^2+z^2} = R\sqrt{1+(z/R)^2}$, and expand in Taylor series to first order to find 

\beq
\rho(R,z) = \rho(R)  {\rm exp} \left( -\frac{GM}{2c_s^2 R^3} z^2\right).
\eeq

The quantity $c_s^2 R^3 / GM$ has dimension of length (squared) and thus defines a scale height. $GM/R^3$ is the square of the Keplerian angular frequency 

\beq
\varOmega_K \equiv \sqrt{\frac{GM}{R^3}}
\eeq 

So 

\beq
H \equiv \frac{c_s}{\varOmega_K},
\label{eq:Scale_Height_Relationship}
\eeq

\noindent and we can write the density stratification compactly 

\beq
\rho(R,z) = \rho(R)  e^{-\frac{z^2}{2H^2}}.
\label{eq:density-stratification}
\eeq

Note that $c_s=\varOmega_K H$ and $u_K=\varOmega_K R$, so 

\beq
\frac{H}{R} = \frac{c_s}{u_K} \equiv \frac{1}{\rm Ma}
\eeq

At the position of Jupiter, $T \approx 180$K and $u_K \approx$ 10
km/s, so $c_s \approx 500$ m/s and thus $H/R \approx 0.05$, rendering the
disk thin.  This quantity, $h\equiv H/R$, is usually called the disk aspect ratio. Another feature to notice is that given $\varOmega \sim R^{-3/2}$, and if temperature dependence is described by $T\sim R^{-q_T}$, where $q_T$ is an index (see below), then for values of $q_T<3/2$ we see that the scale height $H$ increases with distance. For $q_T=1$  {it follows that the aspect ratio $h=H/R$ is constant.}
 
\subsection{Radial Centrifugal balance}

Having solved for the disk vertical structure, we turn to the radial equation, \eq{eq:centrifugal-balance}. We can write the pressure gradient as 

\beq
\frac{1}{\rho} \pderiv{p}{R} = c_s^2 \left[ \pderiv{\ln\rho}{R}  + \pderiv{\ln c_s^2}{R} \right]
\eeq

\noindent and given the hydrostatic solution for density, \eq{eq:density}

\beq
\frac{1}{\rho} \pderiv{p}{R} = -\frac{c_s^2}{R}(q_\rho+q_T) + \frac{GM}{R^2} - \frac{GM}{r^3}R - \frac{GM}{R^3}z^2 \frac{q_T}{2R} 
\eeq

\noindent where we also assumed radial power laws for density and temperature 

\beqn
\rho(R) &\propto& R^{-q_\rho}, \label{eq:labelqrho}\\
T(R) &\propto& R^{-q_T} \label{eq:labelqT}.
\eeqn

Substituting the pressure gradient on \eq{eq:centrifugal-balance} and solving for $\varOmega$

\beq
\varOmega^2 = \frac{GM}{R^3} - \frac{c_s^2}{R^2}(q_\rho+q_T) - \frac{GM}{R^3}z^2\frac{q_T}{2R}.
\eeq

\noindent Finally, substituting the definition of $\varOmega_K$, we find

\beq
\varOmega = \varOmega_K \left\{ 1 - \frac{1}{2}\left(\frac{H}{R}\right)^2\left[q_\rho+q_T + \frac{q_T}{2}\left(\frac{z}{H}\right)^2\right]\right\}
\label{eq:vertical-shear}
\eeq

Notice that if $q_\rho=q_T=0$, i.e., no pressure gradient, the angular frequency returns to Keplerian. Also, because $H/R$ is small, the deviations from Keplerian are also small. 

Before we address how this steady-state configuration transitions via linear and nonlinear instabilities into full-blown protoplanetary disk turbulence, we must give a brief overview of how turbulence is confronted in this astrophysical context.

 {
\subsection{Ertel's theorem and Rossby Waves}
\label{sect:PV}

A theorem and quantity of singular importance emerges from an analysis of the equations of motion. If we  assume both no heat gain or losses in the flow (${\cal Q} = 0$, which means that the gas specific entropy $s$ is materially conserved), and we neglect gas viscosity by setting ${\v{{\rm T}}} = 0$, then a manipulation of the resulting equations of motion reveals a conserved quantity

\beq
\left(\pderiv{}{t} + \v{u} \cdot \del\right) \frac{\v{\varomega} \cdot \del s}{\rho} = 0,
\label{eq:Ertels_Theorem}
\eeq

\noindent where $\v{\varomega} \equiv \curl{\v{u}}$ is the vorticity. The conserved quantity is called {\it potential vorticity}

\beq
\OmegaPV \equiv \frac{\v{\varomega}\cdot \del s}{\rho},
\eeq

\noindent and \eq{eq:Ertels_Theorem} is known as Ertel's theorem; a succinct derivation
may be found in \cite{Regev_Umurhan_Yecko_2016}.  The dynamical importance of the conservation of potential vorticity (often known as ``PV") and its use as a tool to interpret flows -- especially 
those that are strongly rotating and/or ones with one spatial dimension severely limited like in
planetary atmospheres and disks --
cannot be overstated \citep{Pedlosky82}.  \par
While a thorough survey of its consequences is not possible here
\citep[for a good introductory view see][]{Vallis06}, it is important to discuss a dynamical phenomenon
that emerges as a consequence of perturbations of \eq{eq:Ertels_Theorem}, namely, the ubiquity of the {\it Rossby wave}.  To best illustrate the Rossby wave, we consider a very shallow constant density fluid layer on a rotating sphere with constant radially directed gravity.  If the rotation vector of the spinning sphere of radius $R_0$ is $\varOmega {\v{\hat z}}$, where ${\v{\hat z}}$ is the spin axis and the constant $\varOmega_0 > 0$,
and if the fluid layer depth is $H$ (and constant for our purpose), then the entropy gradient is a delta function with radial unit vector ${\v{\hat r}}$.  Since for a given observer at latitude $l$, the
inner product ${\v{\hat r}}\cdot {\v{\hat z}} = \cos l$, it follows that $\OmegaPV$ is non-zero
at the location of the entropy gradient which is taken to be the surface of the sphere for the sake of simplicity.  The PV then has the form $2\varOmega \cos l$.  It can be seen that global
PV increases with increasing latitude. Introduction of perturbations to an otherwise static rotating atmosphere will result in a wave response known as the Rossby wave.  To see this consider a small latitudinal zone centered at some given latitude $l_0$.  If we expand $\OmegaPV$
\beq
=2\varOmega_0 + \betaPV(y-y_0),
\eeq
where
\beq
2\varOmega_0 = 2\varOmega \cos l_0 \quad
\betaPV \equiv \left.\frac{1}{R}\pderiv{\OmegaPV}{l}\right|_{l = l_0},
\eeq
in which $y$ has units of length,
then we can perform
an approximate linear perturbation analysis of \eq{eq:Ertels_Theorem}
assuming strictly two-dimensional incompressible flow whose details can be found in \citet{Pedlosky82}.  Perturbations of the form 
$\exp{\left(i\omega t + ik_x x + ik_y y\right)}$ in which $\omega$ is the frequency response, $x$ is
the longitudinal distance and where $k_x$ and $k_y$ are the corresponding wavevector of the disturbance
results in a Rossby wave frequency response given by
\beq
\omega_{_{\rm RW}} = \frac{\betaPV \, k_x}{k_x^2 + k_y^2}.
\label{eq:RW_frequency}
\eeq
We might imagine a circumplanetary level-set of constant PV being perturbed with a wavelength $k_x$ and $k_y$.  The frequency response indicates that the longitudinal wave pattern propagates in the negative longitudinal (westerly) direction.  This is a general feature of systems supporting Rossby waves: in a right-handed coordinate system, positive latitudinal gradients in PV results in a negative pattern speed of the resulting wave.
\par
Rossby waves are also prevalent in protoplanetary disks and this has been extensively written about \citep{Lovelace+99,Li+00,Umurhan10}.  There are notable differences, the main one being: while accretion disks exhibit gradients in their mean PV simply owing to the mean Keplerian flow, Rossby waves are mostly supported 
where there are anomalous changes in the mean Keplerian flow, 
like locations where the disk material supports pressure extrema 
\citep{Lovelace+99,Li+00,Meheut+10,Lin14} or where there are sharp edges like near disk gaps 
\citep{Koller+03,deValBorro+07,Lyra+09b,Yellin-Bergovoy+17}, or potentially in places where the disk goes from being Ohmic to significantly magnetized \citep{VarniereTagger06,LyraMacLow12}.  However, the Rossby wave frequency response estimated based
on our simplified rotating sphere model, \eq{eq:RW_frequency}, may be also a useful guide
for disks as well \citep{Sheehan+99}.  At a local radial position $R_0$ in a Keplerian disk with rotation 
$\varOmega = \varOmega_0 (R_0/R)^{3/2}$, in which the disk material moves only in the azimuthal-radial directions and the fluid is idealized to be of constant density of finite disk thickness, 
then we find that 
\beq
\OmegaPV = \frac{1}{R} \pderiv{\left(\varOmega R\right)}{R} = \frac{\varOmega}{2},
\eeq
and following \citet{Sheehan+99}, adopting the same line of thinking we can estimate the $\betaPV$-parameter to be
\beq
\betaPV \sim  \left.\pderiv{\OmegaPV}{R}\right|_{R=R_0} = -\frac{3}{4}\frac{\varOmega}{R}.
\label{eq:disk_beta}
\eeq
}

\section{Turbulence: a terse overview.}
\label{sect:turbulence}

Fluid turbulence remains one of last unsolved problems of classical physics \citep{Davidson_2004}
and despite over 100 years of research by thousands of scientists and engineers, very little can be said.  Many who work in the field, when asked ``{\it What is turbulence?}'', respond with ``{\it I do not really know, but I know it when I see it}.'' Fluid flows that undergo turbulent transition can be said to display widespread structural disorder and chaos.  They are inherently unpredictable and highly nonlinear.  Unstable fluid structures in the grips of turbulence can completely fill up the totality of the fluid volume containing the turbulence (like what happens in the bounded regions containing unstable jet flows or wakes) or display sublimity like spatiotemporal intermittency, i.e., regions and pockets in space and/or time where the flow exhibits laminar behavior surrounded by regions of high intensity vortex tube twisting and attendant chaos.\par
There are several possible routes to turbulence in any fluid setting.  In the context   of protoplanetary disk Ohmic zones, we focus on the picture of {\emph{supercritical transition}} wherein the system of interest, generally in some steady state, will experience a linear instability if physical conditions in the fluid are met -- we refer to this as the {\emph{primary instability}}. If the physical criteria controlling the onset are only weakly surpassed by the physical state of the fluid system, then often times the flow will reconfigure itself into some new nonlinear state -- often time-dependent with an identifiable set of periods -- but nevertheless orderly structured and predictable.   If the physical conditions greatly surpass the criterion for onset, then even this nonlinear state may experience an instability and when this happens we refer to it as the {\emph{secondary instability}}.  This transition might result in another, even more complicated, nonlinear state to emerge -- or -- this transition could lead to a catastrophic cascade and a complete breakdown of any ordered flow structure. \par
 Of many consequences of such chaotic situations, an important one is that fluid vortical structures emerging on every scale will vigorously exchange angular momentum with one another.  Another feature to keep in mind is that every time a transition occurs in the flow, the system generally loses some symmetry it once possessed before the instability.  A fully turbulent state has lost all semblance of symmetries that are permissible in the governing equations of motion.
 Ironically, these lost symmetries return when the flow is viewed from a statistical and/or spatiotemporally averaged point of view. 
\par
For example, in dissipative MRI turbulence the primary instability leads to the emergence of a radial component to the basic Keplerian state which periodically reverses sign as a function of distance away from the disk midplane.  Because of the emergence of a strong vertical shear in this radial velocity component, the new flow state undergoes a secondary instability via the excitation of parasitic instabilities \citep{Goodman_Xu_1994} which is an element of the wider class of shear instabilities.  This result is a dynamical cascade into widespread and very strong turbulence that effectively transports angular momentum.

\par
From this general perspective and for our purposes here we assume that (i) there is a continuous energy source for the turbulence, (ii) it enters the system at some integral scale of the system $\ell_0$ and, (iii) it steadily injects energy on these scales at a given rate, $\varepsilon$ -- a quantity typically expressed in units of energy per unit time per unit mass (erg\,s$^{-1}$\,g$^{-1}$, or  cm$^2$/s$^3$).\par

From the vantage point of astrophysical flows like protoplanetary disks, the presence or absence of turbulence has several profound effects.  Of importance to us (and as intimated above) is the ability of turbulent structures to transport angular momentum.  In regions centered on the midplane at 5AU in a disk surrounding a 1 solar mass star, the Reynolds number  $\Rey={\mathcal U}{\mathcal L}/\nu$ with velocities ${\mathcal U}$ of the order of the sound speed and length ${\mathcal L}$ of the order of the local scale height $H$ is roughly $10^{13-14}$, based on molecular hydrogen's viscous diffusivity $\nu \sim 10^3$cm$^2$\,s$^{-1}$. The corresponding viscous dissipation scale \citep{Regev_Umurhan_Yecko_2016} is $\ell_{\rm diss} \sim {\rm Re}^{-3/4} H \approx 0.1-1$km. This means that the character of the unsteady flow between the scales that drive the turbulence from the generation scales (an order unity fraction of $H$) down to the dissipation scale $\ell_{\rm diss}$ must somehow be characterized. Currently, only a crude mixing-length theory characterization exists for disks (see \sect{sect:alpha_disk}).\par

\subsection{Kinetic energy spectra}

 \begin{figure}
  \begin{center}
    \resizebox{\columnwidth}{!}{\includegraphics{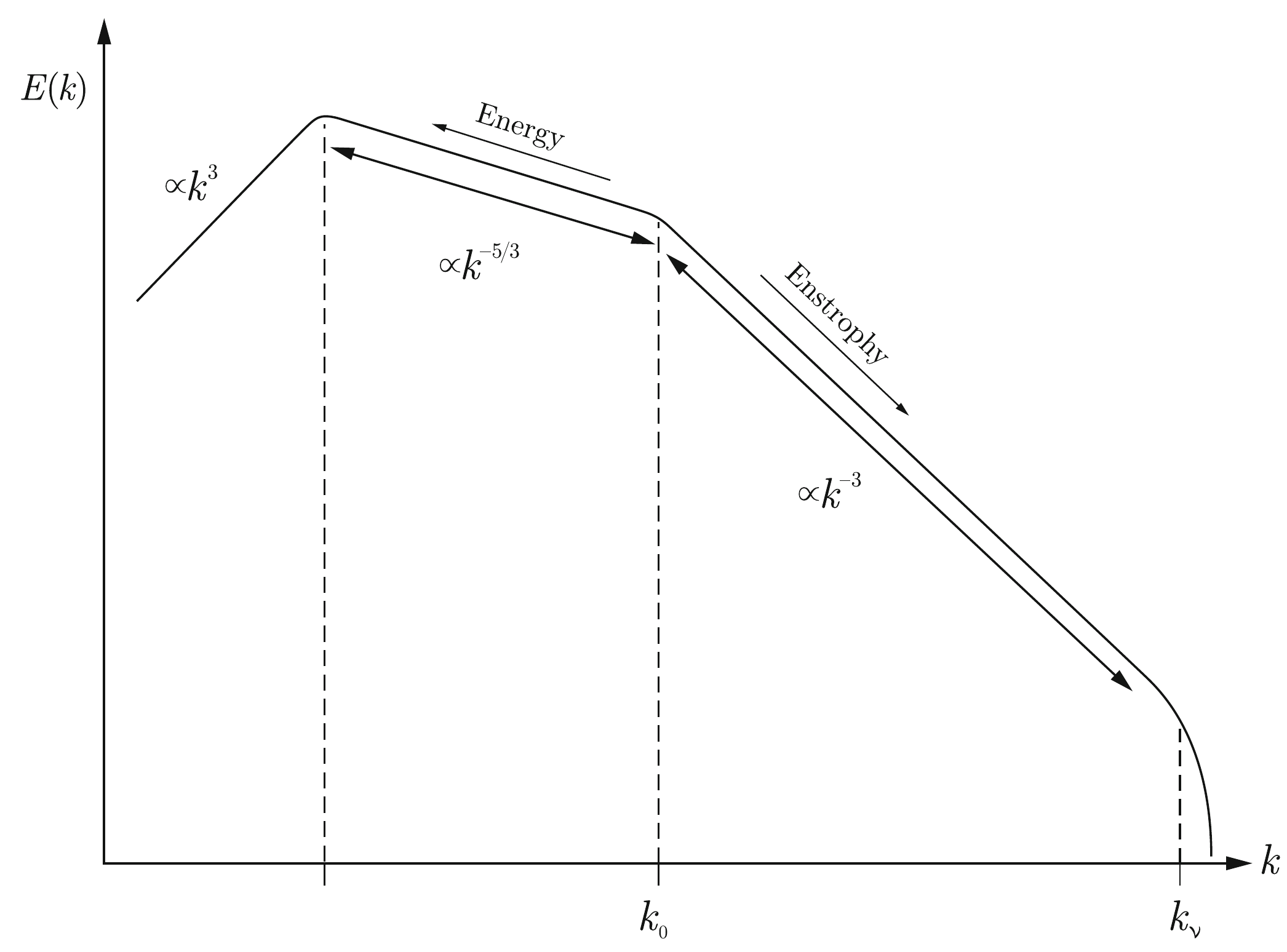}}
\end{center}
\caption[]{The double cascade spectrum of the Kraichnan-Batchelor theory of 2D steady driven turbulence.  The injection scale is $k_0 = 2\pi/\ell_0$.  The two cascade regimes are indicated. The inclusion of linear dissipation acting only at the largest scales accounts for the $k^3$ character of the energy spectrum in that regime.  Simulations of 
\citet{Boffetta_Musacchio_2010} largely bear out the character of this theory.
[Figure from \citet{Regev_Umurhan_Yecko_2016}, by permission.]} 
\label{fig:double_cascade}
\end{figure}

Flows which exhibit turbulence can be identified (at least in part) by their statistical properties.  The flow is presumed to have a mean state plus a spatiotemporally fluctuating field, respectively given by $\v{\bar{u}}$ and $\v{u'}$. In most instances, one examines the kinetic energy contained in fluctuating quantities by constructing its {\emph{energy spectrum}}.  This is most readily seen if we assume $\v{u'}$ to be periodic in each spatial dimension.  Provided the dimensions containing the fluid, ${\mathcal L}$, is sufficiently larger than the injection scales, then
we can identify all components of the fluctuating field exhibiting spatial structure with absolute value of the wavenumber $\left|\v{k}\right| = k  \equiv 2\pi/\ell$, where $\ell$ is the spatial scale of the fluctuating quantity.  Let us define ${\rm E}_k$ as being the total specific energy in the flow contained in fluid structures whose wavenumber lies between $k$ and $k+\delta k$. The specific
energy is understood to be the energy per unit mass (cm$^2$/s$^2$) -- for the purposes of discussion we assume the flow has constant density and we will consider the corresponding specific kinetic energy.
The {\emph{power}} in this specific energy is 
formally defined through the relationship expressing the total specific energy contained in the fluctuating flow field whose wavenumbers lie between $k$ and $k+\delta k$, i.e.,
\beq
{\rm E}_{k} = {\cal E}(k) \delta k \equiv \int_k^{k+\delta k} {\cal E}d k, \qquad
\lim_{\delta k \rightarrow 0} {\cal E}(k) \rightarrow \pderiv{{\rm E}_{k}}{k}.
\eeq
 ${\cal E}(k)$ has units of cm$^3$/s$^2$. Evidently, the system integrated total specific energy, E, is the integral of ${\cal E}$ over all wavenumbers, i.e.,
 \beq
 {\rm E} = \int_0^\infty {\cal E} dk.
 \eeq
   When ${\rm E}_{k}$ behaves as a power law (at least locally in k), a quick estimate for ${\cal E}$ is $\approx {\rm E}_{k}/k$.\par
 ${\cal E}(k)$ is a diagnostic quantity reflecting the state of the fluid at any given instant.  If the turbulence is in a statistically steady state -- where injected power is ultimately drained from the system through either by molecular viscosity on the smallest scales, or by Ekman pumping and/or radiative losses to space on the largest scales -- then ${\cal E}(k)$ will be steady over some suitable time average.
  Because fluid systems like the Navier-Stokes equations are highly nonlinear,  energy contained in one narrow band of wavenumbers can be traded to other fluid structures of differing wavenumber through the process of nonlinear interactions.  Sometimes the energy exchange is local in wavenumber space, but other times the exchange can occur between
  structures of vastly differing length scales.

\subsection{3D isotropic turbulence: Kolmogorov-Obukhov theory and Richardson cascade}
\label{sect:3D_Kolmogorov}

What little that can be said about freely evolving turbulence of the 3D incompressible Navier-Stokes equations relate to its observed statistical properties expressed as some signature quality of ${\cal E}(k)$.
The Kolmogorov-Obukhov theory of 3D isotropic turbulence offers an explanation 
predicated on certain assumptions about the flow, namely that it is (i) homogenous and isotropic,
(ii) self-similar and, (iii) the dissipation rate is constant.  A detailed discussion about this theory can be found in \citet{Davidson_2004}.  We offer a crude sketch to convey the essence of the theory as describing statistically steady state turbulence.  The notion that the fully developed state is homogenous and isotropic is to be understood from this time-averaged statistical point of view.  \par
In this light we consider what happens to structures whose size $\ell$ is much smaller than
the injection scale $\ell_0$ but much larger than the dissipation scales $\ell_{\rm diss}$.  We call this size range the {\emph{inertial regime}}.  Thus,
the first two of these assumptions give rise to the so-called Richardson cascade (see \fig{fig:richardson_cascade}) governing what happens in the inertial regime, in which a given 3D twisting vortex of some size $\ell$, with rotation speed $u_\ell$, undergoes a nonlinear bifurcation that generates smaller-sized (scaled on $\ell$ by $0<\xi<1$) daughter 3D vortex structures, also see discussion also found in \citet{Frisch_1995}.  We can think of the vortical structure on scale $\ell$ as "living" only for one rotation time before undergoing "destruction", i.e., $t_\ell =\ell/u_{\ell}$.  In this steady state picture, the rate of energy entering and exiting every length scale must be the same, i.e., $\varepsilon$ is independent of $\ell$.  Thus, we say that this energy rate is given by its energy divided by this rotation time.  In terms of the specific energy this means to say,
\beq
\varepsilon = \frac{u_\ell^2/2}{t_{\ell}} \sim \frac{u_\ell^3}{\ell}, \ \ \longrightarrow \  u_\ell \sim \left(\varepsilon \ell\right)^{-1/3} \sim \varepsilon^{1/3}k^{-1/3}.
\eeq
It immediately follows that the energy contained on the length scale $\ell$
\beq
{\rm E}_k = u_\ell^2/2 \sim \varepsilon^{2/3}k^{-2/3}.
\eeq
This is sometimes known as the Kolmogorov two-thirds rule.  The amount of power in the 
small wavelength bin containing $k$ is therefore expressed by ${\cal E} \approx {\rm E}_k/k 
\sim \varepsilon^{2/3}k^{-5/3}$. This is the famous "universal $k^{-5/3}$" shape characterizing the inertial spectrum of freely evolving 3D turbulence.  This property has been assessed in numerous experiments of 3D turbulence including wind tunnel experiments and others \citep{Frisch_1995}.
\par    
Identifying a fluid system as being turbulent generally requires resolving its inertial spectrum.    Reproducing this $k^{-5/3}$ shape is now considered a standard benchmark for for validating most direct numerical simulations (DNS) which purport to model turbulence.
For numerical experiments, in turn, this means capturing {\emph{at least}} 1$-$2 decades of the inertial spectrum, i.e., there ought to be a factor of 10$-$100 scale difference between
the injection and dissipation scales.
We shall return to this matter and how it bears upon simulations relating to protoplanetary disk turbulence in \sect{sect:future}.

\subsection{2D turbulence}
\label{sect:2D_turbulence}

Freely decaying 3D turbulence can be considered a possible end-member state contained in the idealized abstract space of turbulent flows.  By its construction, there are no body forces acting upon the fluid.  Body forces, like gravity, often act upon the fluid by constraining one dimension or enforcing a symmetry on the flow.  One might expect the presence of a body force adds more complexity to the problem and that, in comparison to freely evolving 3D turbulence, even less might be said about the statistical quality of any emergent state.  In the idealized case of 2D turbulence, the opposite is true.  2D turbulence is the case of the Navier-Stokes equations in which one dimension is restricted {\emph{by fiat}} -- henceforth we assume it is the $z$-direction.  Atmospheric flow (constrained by both gravity and rotation) and the motion along bubble surfaces (constrained by surface tension) are close, but not exact, physical instances of idealized 2D turbulence.  \par
Such flows are known to support a {\emph{double cascade}}.  The Kraichnan-Batchelor theory of steady driven 2D turbulence says that flows with power injected on some scale $\ell_0$, which is much less than scale of the system $L$ and much larger than the dissipative scales $\ell_{\rm diss}$, will exhibit a {\emph{direct cascade of enstrophy}} and an {\emph{inverse cascade of energy}}.  With the vorticity 
as defined $\v{\varomega}$, strictly 2D flow results in 
\beq
\v{\varomega} = \varomega_z \hat{z} \equiv \left[\pderiv{u_x}{y} - \pderiv{u_y}{x}\right]\hat{z}.
\eeq

\noindent Enstrophy is defined to be the square of the vorticity, i.e., ${\cal Z} \equiv \v{\varomega}^2$. In the inverse cascade regime ($L\ge\ell\ge\ell_0$) the energy spectra has the familiar form, ${\cal E} \sim k^{-5/3}$. Unlike the 3D case discussed above, energy shifts from smaller scales to larger scales.  Often times this can be identified by eye as the numerical experiment of an unstable 2D shear layer in \fig{fig:shear_layer} shows:  a thin vortex strip rolls up under its own induced flow and eventually turns into a larger tumbling structure containing a jumble of fine scale vortex filaments even smaller than the original filament.  These developed finer scale filamentary structures qualitatively exhibits the forward cascade of enstrophy.  The emergent large scale vortex structure exhibits the inverse cascade of energy.\par
In the forward enstrophy cascade regime
of steady driven 2D turbulence, the wavenumber dependence of the energy spectrum is much steeper, in fact ${\cal E} \sim k^{-3}$.   These properties are summarized in \fig{fig:double_cascade}. The trends predicted in the Kraichnan-Batchelor theory of steady driven turbulence have yet to be disproven \citep{Boffetta_Ecke_2012}: the features of the theory summarized in \fig{fig:double_cascade} are asymptotically approached as recent highly resolved DNS indicate
\citep{Boffetta_Musacchio_2010}.

\par
The direct enstrophy cascade can be rationalized by the following argument.  For 2D incompressible Navier Stokes flow, the equation governing its evolution is greatly simplified to
\beq
\frac{d\varomega_z}{dt} = \nu\nabla^2 \varomega_z,
\label{eq:2D_vorticity_eqn}
\eeq 
in which the incompressibility constraint means that all quantities may be expressed in terms of a single scalar, called the streamfunction $\varPsi$, in which
\[
u_x = -\pderiv{\varPsi}{y}, \ \ u_y = \pderiv{\varPsi}{x}, \longrightarrow \ \ \varomega_z = \nabla^2 \varPsi.
\]
If the dissipation scale is indeed very small, then \eq{eq:2D_vorticity_eqn} states that in the inertial range viscosity is negligible and that, consequently, the vorticity is a materially conserved quantity.  In this case we can follow the same physically motivated logic administered in our discussion of freely evolving 3D turbulence.  By assuming isotropy, homogeneity and self-similarity we state that patches of vorticity can undergo a Richardson type of cascade by preserving the  vorticity with which it was endowed.  Calling this vorticity $\varomega_0$, we find therefore that mother-daughter structures must preserve this quantity all the way down to the dissipation scales, therefore
\beq
\varomega_0 \sim \frac{u_\ell}{\ell} = {\rm constant}, \ \ \longrightarrow
u_\ell = \varomega_0 \ell \sim \varomega_0 k^{-1}. 
\eeq
This means that the energy of the structure is $E_k \sim \varomega_0^2 k^{-2}$, which means that in the forward enstrophy cascade zone ${\cal E}(k) \sim E_k/k \sim \varomega_0^2 k^{-3}$.

\subsection{The quality of turbulence in strongly rotating and stratified flows and Rhines scales}

Turbulence in strongly rotating and stratified flows is an unsettled topic.  A case in point is the Earth's atmosphere which one might consider to be a close approximation to 2D flow.  
\citet{Nastrom_Gage_1985} analyzed wind data based measurements made on thousands of commercial airplane flights,  and they found that the atmosphere indeed exhibits a $k^{-3}$ spectrum on scales ranging from the planetary ($\sim$ 10,000km) through the  synoptic scales ($\sim$ 1000km).  However, in the   mesoscales range, which starts at about 500km and goes down to about a few km, the spectral shape $\sim k^{-5/3}$.  While the large scale energy injection mechanism is known (baroclinic instability $\sim$1000 km scales), a robust physical explanation for the $k^{-5/3}$ spectrum, including what are the energy sources and sinks, remains a subject of much debate with
several candidate mechanisms, like gravity wave breaking, moist convection and others  \citep{Tulloch_Smith_2006,Marino_etal_2014,Sun_etal_2017}.
While the large scale trend may be rationalized in terms of the inverse cascade picture of driven 2D turbulence, the switch over into what ostensibly exhibits 3D statistical character occurs on length scales that are much larger than the vertical scales of the atmosphere ($\sim$ 10km, corresponding to the height of the tropopause).
Similar trends have recently been identified for Jupiter's weather layer based on analysis of imaging data acquired during the Cassini spacecraft's Jupiter gravity assist maneuver \citep{Young_Read_2017}.
\par
 {
The inverse cascade is also arrested on the large {\it{Rhines scales}}.  Rhines (1979) demonstrated that for 2D turbulence on a rotating sphere there exists a scale $\lambda_b$ spanning the north-south direction on which the differences of the planetary vorticity across this length scale equals the overturn time of the upscale cascading turbulent eddies denoted by $\varomega_{\rm eddy} \sim {\cal U}_b/\lambda_b$.  Disturbances in 
the planetary PV give rise to Rossby waves (as described in \sect{sect:PV}).  
If the frequency response of the planetary scale Rossby wave equals that of the turbulent eddy overturn time, then
 instead of these eddies sending their kinetic energy to larger length scales 
 as demanded by the inverse cascade process, these $\lambda_b$-scale turbulent eddies will transfer their
energy into large scale Rossby waves\footnote{ {Note that a steady rotating global atmosphere sitting atop flat topography will 
have variations in its potential vorticity as a function of latitude.  Potential vorticity will
also show variations if there is some amount of significant topography.  Thus,  large scale inverse-cascading turbulent eddies encountering topography will also result in some of their energy
being transferred into Rossby waves.  This latter feature is relevant to planetary atmospheres.}}.
If we assume the turbulent eddies are of longitudinally-latitudinally equi-dimensional, then by setting $k_x = k_y = k_b \equiv 2\pi/\lambda_b$ in \eq{eq:RW_frequency}, we find that setting $\omega_{_{\rm RW}} = \varomega_0$ implies
\beq
\frac{\betaPV}{2 k_b} = {\cal U}_b k_b, \quad \longrightarrow \quad
k_b = \sqrt{{\betaPV}\big/{2 {\cal U}_b}}.
\label{eq:def_of_Rhines_scale}
\eeq
These special scales are observed in simulations and in data of planetary atmospheres.  These large scale waves may themselves nonlinearly develop and/or play a role in generating large scale circulation in the form of global zonal flows through the process of wave-mean flow interactions.  For a deeper and expanded discussion regarding wave-mean flow processes see \citet{Pedlosky82} and \citet{Vallis06}.
The consequences are clear: systems that are nearly 2D and turbulent, that exhibit both an inverse cascade and other
features like strong rotation, can result in the organization of such disorder into global scale organized motions like jets and waves.  If protoplanetary disk dynamics indeed exhibits the qualities of 2D turbulence, then there will arise a similarly motivated Rhines scale \citep{Sheehan+99}.  We reflect on this further below.}

\subsection{The alpha disk model}
\label{sect:alpha_disk}

Despite the great abundance of observations and data available for the Earth's atmosphere, an understanding of the nature of its turbulence is still cloudy.  The situation with regards to accretion disk turbulence remains much worse -- although the developments of the last five years raise hope that inroads toward its characterization is on the near term horizon.
\par
In lieu of this, and for a great many number of years, the astrophysical community has relied on a mixing-length model to represent accretion disk turbulence.  The so-called $\alpha$-disk model \citep{ShakuraSunyaev73,Lynden-BellPringle74} takes a heuristic approach toward handling this situation.
It assumes that
 there exists a dynamically sustained process that generates disordered velocity fluctuations about the mean Keplerian state.  If these disordered fluctuations are indeed turbulent then
 they will be engaged in some kind of nonlinear energy transfer which translates into
 them exhibiting non-zero spatial correlations.
These correlations are assessed by calculating spatial averages of the corresponding Reynolds stresses.  Following the original example, we consider $u_r'$ and $u_\phi'$ respectively as the fluctuating radial and azimuthal velocities about the basic Keplerian (azimuthal) velocity $U_K \sim R^{-1/2}$.  Denoting spatial averages with overbars, the negative of the averaged radial-azimuthal component of the Reynolds stress is equated with an effective linear momentum diffusion through the following identification:
\beq
\nu_t \pderiv{U_K}{R} \equiv -\overline{u_r' u_\phi'}.
\eeq
The left hand side expression of the above is motivated from the definition of viscosity from the molecular dynamics point of view which says that neighboring parcels of gas will exchange momentum in proportion to the gradient of their localized bulk (mean) velocities.  In this case the mean/bulk velocity is taken as the Keplerian one.  But the key ingredient in this characterization is the measured correlation $\overline{u_r' u_\phi'}$.  Thus, the effective turbulent diffusion
($\nu_t$, in units of cm$^2$/s) is expressed in terms of the product of the local sound speed and scale height: $\nu_t \equiv \alpha c_s H$.  Thus $\alpha$ is the measure of the turbulent intensity
and we have
\beq
\alpha \sim \frac{\overline{u_r' u_\phi'}}{c_s^2},
\eeq
which follows from  the definition of $H$ in \eq{eq:Scale_Height_Relationship}.  In this form,
$\alpha$ measures the square of the local Mach number of the fluctuating disk velocities which are (presumably) dominated by the velocities of the injection scales.  
\par
Quoting this value of $\alpha$ is the main characterization used to assess the {\emph{degree of turbulence}} in protoplanetary disks.  It says nothing about the statistics of the actual small scale motions driven downscale by the larger scale driven dynamics -- i.e.,  
to date ${\cal E}(k)$ is still not known.  This has direct consequences for the accumulation of disk dust.\par
Nonetheless one may make estimates.  Assuming the turbulence is steady and the energy cascade is 3D, as in the sense of Kolmogorov-Obukhov theory, then a cascade rate $\varepsilon_\alpha$ based on the measured value of $\alpha$ may be estimated by following
the physical reasoning in \sect{sect:3D_Kolmogorov} to be approximately,
\beq
\varepsilon \approx \varepsilon_\alpha \equiv {\delta v}^3/H_{{\rm oturn}} = \alpha^{3/2} c_s^3/H.
\label{eq:approximate_energy_cascade_rate}
\eeq
where we have identified the square of the fluctuation velocities with 
the Reynolds stress: $\delta u^2 \equiv\left| \overline{u_r' u_\phi'} \right|$. Implicit in this assumption is that the overturn scales $H_{{\rm oturn}}$ are approximately that of a scale height.  The three large-scale eddy producing linear instabilities described in the upcoming sections generate structures whose fundamental length scales are generally 
a fixed fraction of the local $H$.  Interestingly,  
as noted in \sect{sect:vert_equilibrium},
if the mean temperature profile in a protoplanetary disk Ohmic zone
has a value of $q_T<3/2$, then the scale height $H$ increases with distance and, therefore,
the overturn scales similarly enlarge with radial distance from the central star, 
much inline with von Weiszk\"aker's predictions \citep{Gamow_Hynek_1945}.\par
 {If, on the other hand, the protoplanetary disk turbulence exhibits a 2D energy cascade at the largest scales, then energy will cascade until it encounters its corresponding Rhines scale.  3D simulations of buoyant convection in disks performed by \citet{Cabot+96} indeed show that the small scale forcing sends power from the small scales and drives into place 2D large-scale zonal jets.  The length scale of 
the emergent jets from these simulations $L_{jet}$ were shown by \citet{Sheehan+99} to be consistent with the corresponding Rhines scale appropriate to the local disk section.  Specifically, characterizing the turbulent large scale eddies by the previously motivated $\alpha$ prescription, then the Rhines scale vortices will have circulation speeds of ${\cal U}_b \sim \alpha^{1/2} c_s$.  Thus, taking this together with the definition of the Rhines scale found in \eq{eq:def_of_Rhines_scale} and the disk appropriate value of the $\betaPV$-parameter found in \eq{eq:disk_beta}, then we find that the corresponding 
predicted value of $k_b$ to be
\beq
\lambda_b \sim \frac{\pi}{k_b} = \pi\sqrt{\frac{8 \alpha^{1/2}}{3} HR}
=\pi\sqrt{\frac{8 \alpha^{1/2}}{3 h}} H.
\label{eq:disk_Rhines}
\eeq
The convective turbulence simulated in \citet{Cabot+96} predicted values of $\alpha$ in the range $10^{-2}$ to $10^{-4}$.  Assuming disk aspect ratios $h\sim 0.05 - 0.1$,  plugging in these values into \eq{eq:disk_Rhines} show that $\lambda_b$ is predicted to be an order one fraction
of the local pressure scale height $H$, which is in rough agreement with the measured values of the
jet widths $L_{jet}$ observed to emerge, presumably as the end-state of the inverse cascade of \citet{Cabot+96}.  Finally we can estimate the energy cascade rate toward the Rhines scale following
the same line of reasoning leading to \eq{eq:approximate_energy_cascade_rate}
\beq
\epsilon \approx \epsilon_\alpha = \frac{\alpha^{5/4}}{\pi} \sqrt{\frac{3h}{8}} \frac{c_s^3}{H}.
\eeq

}
\begin{figure*}
  \begin{center}
    \resizebox{.7\textwidth}{!}{\includegraphics{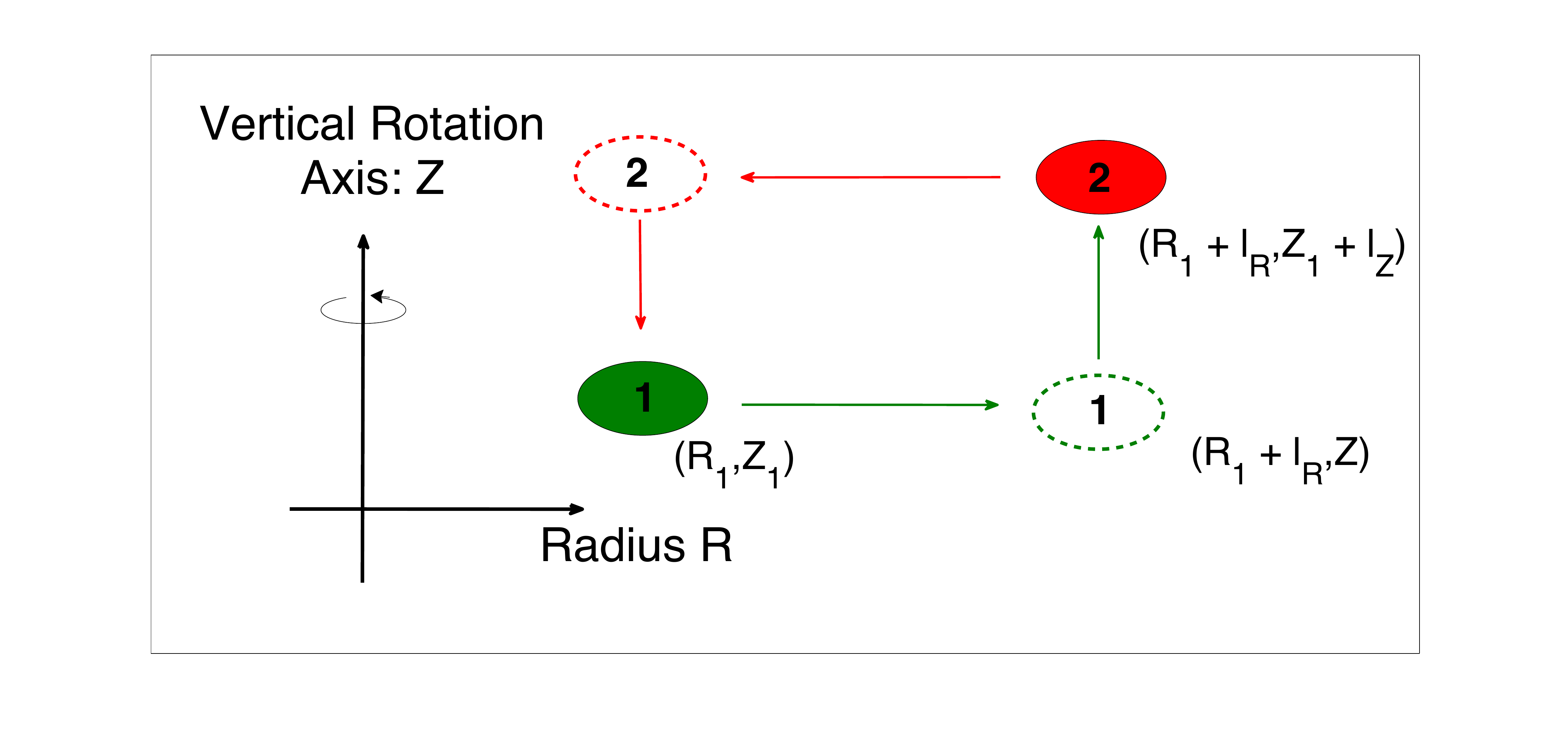}}
  \end{center}
  \caption[]{ {Rayleigh instability. Consider two annuli 1 and 2, with the first at
  radial-vertical position $R_1, Z_1$ and the second at $R_2,Z_2$.  The energy argument, as originally found in \cite{Kippenhahn_etal_2012} and discussed recently in \cite{Lin_Youdin_2015}, goes as follows:
  one interchanges the positions of the two circular annuli by preserving their individual angular momentum.  If the total energy of the new system is less upon interchange then setup is
  regarded as unstable.  In the diagram, the fluid ring 1 moves to radial position $R_1 + \lambda_R = R_2$
  and then onto vertical position $Z_1 + \lambda_z = Z_2$ while ring 2 executes the opposite motion.
  A circular final state requires energy dissipation either in the form of heat or radiative losses -- a feature that is implicit in this energy argument.  In the case of purely barotropic Keplerian flow,
  which is stable according to this analysis,
  the interchange in the vertical direction is superfluous as the rotational speeds are the same. 
  The vertical interchange matters when analyzing the conditions for the VSI as the vertical
  shear in the Keplerian flow figures prominently in driving the instability.
  Figure adapted from \cite{Umurhan+13}.}
    }
\label{fig:rayleigh}
\end{figure*}

\section{The Prime Mover: Linear instability}
\label{sect:instabilities}

In this section we review the Rayleigh criterion (\sect{sect:Rayleigh_criterion}), the magneto-rotational instability (\sect{sect:MRI}), and introduce the hydrodynamical instabilities (\sect{sect:thermal}). 

\subsection{Rayleigh criterion}
\label{sect:Rayleigh_criterion}

The Rayleigh criterion is paramount to understanding what instabilities can be present in the Ohmic zone. Starting from the momentum equations, \eq{eq:radial-momentum-cyl} and \eq{eq:azimuthal-momentum-cyl}, without the pressure gradient, and restricting ourselves to the midplane 

\beqn
\pderiv{u_R}{t} + u_R\pderiv{u_R}{R} + u_\phi \pderiv{u_R}{\phi} - \frac{u_\phi^2}{R} &=& \varOmega^2 R \\
\pderiv{u_\phi}{t} + u_R\pderiv{u_\phi}{R} + u_\phi \pderiv{u_\phi}{\phi} + \frac{u_\phi u_R}{R} &=& 0.
\eeqn

We can decompose the flow into base (time-independent) and perturbation in 

\beq
\v{u}(\v{x},t) = \bar{\v{u}}(\v{x}) + \v{u}^\prime(\v{x},t),
\eeq

\noindent with the base flow $\bar{u}_R = 0$ and $\bar{u}_\phi = \varOmega R \gg u_\phi^\prime$, the perturbation equations become 

\beqn
\pderiv{u_R^\prime}{t} - 2\varOmega u_\phi &=& 0 \\
\pderiv{u_\phi^\prime}{t} + (2-q)\varOmega u_R &=& 0 
\eeqn

where 

\beq
q\equiv -\pderiv{\ln\varOmega}{\ln R}
\eeq

Now consider the perturbation as a Fourier mode $\psi^\prime = \hat\psi_0 e^{-i(\omega t -\v{k}\cdot\v{x})}$, so that 

\beqn
-i\omega \hat u_R - 2\varOmega \hat u_\phi &=& 0 \\
-i\omega \hat u_\phi  + (2-q)\varOmega u_R &=& 0 
\eeqn

\noindent which leads to the solution for the complex eigenfrequency

\beq
\omega^2 = 2\varOmega^2 (2-q)
\eeq

This frequency is called epicyclic frequency, 

\beq
\kappa_{\rm ep} \equiv \varOmega\sqrt{2 (2-q)}
\eeq

\noindent if $q<2$ then the flow is stable. Since the Keplerian flow has $q=1.5$, then undisturbed Keplerian flow is unconditionally stable. Attempts at turning the Keplerian flow unstable have to focus on how
the Rayleigh criterion is violated. 

It is an equivalent statement to the Rayleigh criterion formulated above that angular momentum must increase outward. Given the angular momentum $L=R^2\varOmega$, then its derivative is

\beq
\frac{dL}{dr} = \varOmega R (2-q)
\eeq

\noindent and thus the condition $q<2$ is equivalent to $dL/dR>0$, i.e., angular momentum should increase with distance.  {{The classical way to understand this criterion is by appealing to an annular ring interchange argument as summarized in the cartoon Fig. \ref{fig:rayleigh}.  
Consider two annuli one at an inner radius and disk height, $R_1,Z_1$ respectively,
and the second at an outer radial and disk height position, $R_2 = R_1 + \lambda_R, Z_2 = Z_1 + \lambda_Z$ respectively.  We posit the following dynamical action: we interchange the positions of the two annuli
under the condition that their angular momenta are conserved.  We place the interchanged annuli into circular orbits and we assess the new total energy of the configuration.  If the energy in this final state is lower than the original arrangement then this new configuration is preferable and the system is unstable.  If upon interchange the total energy is higher, then this system is not preferred and the system is stable.  The key feature here is that we posit that the annuli are placed into circular orbits and this implicitly means that energy must be either gained or lost by the system as a whole.  For barotropic Keplerian flows, the vertical interchange has no effect as the angular momentum is constant
on cylinders (i.e., $dL/dZ = 0$) and, thus, the criterion for a lower energy state is equivalent to $dL/dR>0$.  The details of this relatively straightforward analysis may be found in \cite{Tassoul_1978}
or \cite{Kippenhahn_etal_2012}.}}

\subsection{The Solberg-H{\o}iland criteria}
\label{sect:SH}

In the presence of buoyancy, the Rayleigh criterion is substituted for the Solberg-H{\o}iland criteria. These criteria have been derived elsewhere \citep{Tassoul_1978,Abramowicz+84,Ogilvie16}, but we repeat it here for completeness largely in part because these criteria figure prominently in the processes discussed in this review. If we include the adiabatic energy equation, stated in terms of pressure, 

\beqn
\ptderiv{\rho} + \left(\v{u}\cdot\del\right)\rho &=& -\rho\Div{\v{u}},\\
\ptderiv{\v{u}} + \left(\v{u}\cdot\del\right)\v{u} &=& -\frac{1}{\rho}\grad{p} + \v{g},\\
\ptderiv{p} + \left(\v{u}\cdot\del\right)p &=& -\gamma p \Div{\v{u}},
\eeqn

\noindent and linearize the system $u_R=u_R^\prime$, $u_\phi = \varOmega(R,z) R + u_\phi^\prime$, $u_z=u_z^\prime$, $p=p_0(R,z) + p^\prime$, and $\rho=\rho_0(R,z) + \rho^\prime$, then 

\beqn
-i\omega\rho^\prime + u_R^\prime (\partial_R \rho_0 + ik_R\rho_0) + u_z^\prime(\partial_z \rho_0 + ik_z\rho_0) &=& 0, \nonumber\\
&&\\
-i\omega u_R^\prime -2\varOmega u_\phi^\prime + \frac{ik_R}{\rho_0} p^\prime - \frac{\rho^\prime}{\rho_0^2} \partial_R p_0 &=& 0, \nonumber\\
&&\\
-i\omega u_\phi^\prime +\varOmega(2-q)u_R^\prime + \partial_z(\varOmega r)  u_z^\prime &=& 0, \nonumber\\
&&\\
-i\omega u_z^\prime +\frac{ik_z}{\rho_0}p^\prime - \frac{\rho^\prime}{\rho_0^2}\partial_z p_0 &=& 0, \nonumber\\
&&\\
-i\omega p^\prime +u_R^\prime (\partial_R p_0 + \rho_0  c_s^2ik_R) + u_z^\prime (\partial_z p_0 + \rho_0 c_s^2 i k_z) &=& 0, \nonumber\\ 
&&
\eeqn

\noindent where we have also substituted $p_0 = \rho_0 c_s^2 / \gamma$.  We define

\beqn
\mA_k&\equiv&c_s \ \partial_k\ln \rho,\\
\mG_k&\equiv&\gamma^{-1} c_s \ \partial_k\ln p,\\
\mS_k&\equiv&\mG_k-\mA_k, 
\eeqn

\noindent as normalizations of the density, pressure, and entropy gradients in units of frequency. Solving the system yields the dispersion relation

\beqn
&&\omega^4 - \omega^2\left[ 2(2-q)\varOmega^2 + \mA_z \mG_z + \mA_R \mG_R +c_s^2 (k_z^2+k_R^2)\right] \nonumber \\
&+&c_s^2 \big\{ 2(2-q)\varOmega^2 k_z^2 - \mG_R\mS_R k_z^2 - \mG_z\mS_z k_R^2 \nonumber\\
&&-  k_Rk_z  \big[2 \varOmega \partial_z(\varOmega r)  - \mG_z\mS_R - \mG_R\mS_z\big]\big\} \nonumber\\
&+&C_1 + i C_2 = 0,
\eeqn

\noindent with

\beqn
C_1 &=& 2(2-q)\mA_z \mG_z \varOmega^2 -2\mA_R \mG_z\varOmega \partial_z(\varOmega r), \\
C_2 &=& c_s  (k_z\mG_R-k_R\mG_z) \left[ 2 \varOmega \partial_z(\varOmega r) +\mA_z \mG_R - \mA_R \mG_z \right].
\eeqn

Keeping only the leading order terms in $c_s$, i.e, the $c_s^2$ terms, we
filter out the acoustic modes. The resulting anelastic dispersion
relation is

\beqn
\omega^2 (k_R^2 + k_z^2) & = & k_z^2 \left(\frac{1}{r^3}\frac{\partial L^2}{\partial r} - \mG_R\mS_R\right)
- k_R ^2 \ \mG_z \mS_z \nonumber\\
&+& k_R k_z \left(-\frac{1}{r^3}\frac{\partial L^2}{\partial z} + \mG_R\mS_z + \mG_z\mS_R\right), \label{eq:dispersion-anelastic}
\eeqn

\noindent where we substituted $L=\varOmega r^2$ for the angular momentum and recognize

\beqn
\kappa_{\rm eq}^2 &\equiv& 2(2-q)\varOmega^2 = \frac{1}{r^3}\frac{\partial L^2}{\partial r},\\
\kappa_z^2 &\equiv&2\varOmega\partial_z(\varOmega r) = \frac{1}{r^3}\frac{\partial L^2}{\partial z},
\eeqn

\noindent as the squares of the epicyclic frequency $\kappa_{\rm eq}$ and a vertical
oscillation frequency $\kappa_z$. Notice that we can cast the dispersion relation \eq{eq:dispersion-anelastic} into the following matrix form, 

\beq
\omega^2  = \frac{\v{k}^T \vt{P} \v{k}}{\v{k}^T\v{k}},
\label{eq:omega2}
\eeq

\noindent where $\v{k}$ is the disturbance wavevector, and the matrix $\vt{P}$ is a 2$\times$2 matrix 

\beq
\vt{P}=\left[\begin{array}{cc}
A & B\\
C & D\\
    \end{array}\right],
    \qquad
A = -\mG_z\mS_z,\ \ \
D = \kappa^2 - \mG_R\mS_R.    
    \label{eq:matrix2}
\eeq

The partial derivatives of pressure ($\mG$) and entropy ($\mS$) combine
into the Brunt-V\"ais\"al\"a frequencies

\beqn
N_R^2 &\equiv& -\mG_R \mS_R = -\frac{1}{\rho c_p}\pderiv{p}{R} \pderiv{s}{R} \label{eq:bruntR},\\
N_z^2 &\equiv& -\mG_z \mS_z = -\frac{1}{\rho c_p}\pderiv{p}{z} \pderiv{s}{z} \label{eq:bruntZ}, 
\eeqn

\noindent so $A=N_z^2$ and $D=\kappa^2 +N_R^2$. As for $B$ and $C$, there is freedom to choose them from the $k_Rk_z$ terms in \eq{eq:dispersion-anelastic}

\beq
B+C = -r^{-3} \partial_z L^2 + \mG_R\mS_z + \mG_z\mS_R.
\eeq

Notice that we can construct $\vt{P}$ as symmetric by setting $B = C = (B+C)/2$. The eigenvectors of symmetric $\vt{P}$ form an orthogonal basis, and $\v{k}$ can be represented from linear combinations of the eigenvectors. \eq{eq:omega2} thus means that $\omega^2$ is an eigenvalue of $\vt{P}$. The characteristic equation ${\rm Det}(\vt{P}-\omega^2\vt{I})=0$ is

\beq
\omega^4 - (A+D)\omega^2 + (AD-B^2)  = 0 
\eeq

\noindent i.e., a biquadratic with $a=1$, $b=- {\rm Tr}(\vt{P})$, and $c= {\rm Det}(\vt{P})$. The eigenvalues are thus

\beq
2 \omega^2  = {\rm Tr}(\vt{P}) \pm \sqrt{{\rm Tr}^2(\vt{P}) - 4\,{\rm Det}(\vt{P})}
\eeq

\noindent they will be positive if $(i)$ ${\rm Tr}(\vt{\vt{P}}) > 0$; and if the square root is always greater than {\rm Tr}(\vt{\vt{P}}), that is, if $(ii)$ ${\rm Det}(\vt{P}) > 0$. Conditions $(i)$ and $(ii)$ are the necessary and sufficient conditions for stability.  The trace condition gives the first Solberg-H{\o}iland criterion
 
\beq
\kappa_{\rm eq}^2 - \frac{1}{\rho c_p}\grad{p} \cdot \grad{s}  > 0
\label{eq:SH1}
\eeq

For the determinant condition, we need to define the term $B=C$.  We split the $k_zk_R$ terms in \eq{eq:dispersion-anelastic} as

\beqn
B &=& -\kappa_z^2 + \mG_R\mS_z, \\
C &=& \mG_z\mS_R. 
\eeqn

That $B=C$ follows from the thermal wind equation which can be proven by the following: in equilibrium we have 

\beq
\varOmega^2 r \hatr = \frac{1}{\rho} \grad{p}  + \grad{\varPhi}, 
\label{eq:thermalwind-orig}
\eeq

\noindent and taking the curl of \eq{eq:thermalwind-orig} produces

\beq
\grad{\left(\varOmega^2 r\right)} \times \hatr   = -\frac{1}{\rho^2}\grad{\rho} \times \grad{p}.
\label{eq:curlthermal}
\eeq

\noindent The term in the RHS is the baroclinic term. From the definition of entropy we write $\rho$ in terms of $p$ and $s$ 

\beq
\grad{\ln\rho}  = \frac{1}{\gamma}\grad\ln p + \frac{1}{\cp}\grad{s},
\eeq

\noindent so the baroclinic term is 
\beqn 
\frac{1}{\rho^2}\grad{\rho} \times \grad{p} &=&\frac{1}{\rho\cp}\grad{s} \times \grad{p}\nonumber\\
&=& \v{\mS} \times \v{\mG}.\label{eq:baroclinic}
\eeqn

\noindent Substituting \eq{eq:baroclinic} into \eq{eq:curlthermal}, we recover for the $\hatphi$ direction

\beq
-\kappa_z^2 + \mG_R\mS_z = \mG_z\mS_R ,
\eeq

\noindent proving that the matrix $\vt{P}$ is symmetric. The determinant condition ${\rm Det}(\vt{P})>0$ becomes 

\beq
-\mG_z\mS_z(\kappa_{\rm eq}^2 - \mG_R\mS_R)  + \mG_z\mS_R (\kappa_z^2 - \mG_R\mS_z) > 0.
\eeq

\noindent Factoring out $\mG_z$ from both terms and simplifying reveals

\beq
-\mG_z \left(\mS_z\kappa_{\rm eq}^2 - \mS_R \kappa_z^2 \right)  > 0. 
\eeq

\noindent Substituting for the gradients leads us to the second condition

\beq
-\frac{\partial p}{\partial z} \left(\frac{\partial s}{\partial z}
  \frac{\partial L^2}{\partial r} - \frac{\partial s}{\partial
    r}\frac{\partial L^2}{\partial z}  \right)  > 0.
\label{eq:SH2}
\eeq

\eq{eq:SH1} and \eq{eq:SH2} are the Solberg-H{\o}iland criteria for stability. If either criterion is violated, then it means that there are sets of disturbances -- characterized by some range wavevectors $\v{k}$ -- for which the system must be linearly unstable.

We call attention to an often overlooked consequence of the Solberg-H{\o}iland criteria. The determinant condition $AD-B^2> 0$ implies that $A$ and $D$ must have the same sign. Since the trace condition $A+D > 0$, then $A$ and $D$ must each be positive, which implies

\beqn
N_z^2 & > & 0, \label{eq:SH01} \\
\kappa_{\rm eq}^2 + N_R^2 & > & 0 \label{eq:SH02},
\eeqn

\noindent have to be satisfied {\it independently} for stability. In \eq{eq:SH1} the dot product notation is a mathematically compact way to write the condition; physically, the $\kappa^2_{\rm ep} + N_R^2$ and $N_z^2$ terms are working on different modes. If either of them are negative, then the trace can still be positive but the second condition \eqp{eq:SH2} is negated.

\eq{eq:SH01} is the usual Schwarzschild criterion assessing 
the buoyant stability of a vertically stratified fluid. Notice that it follows directly from \eq{eq:dispersion-anelastic} by setting $k_z=0$.

\eq{eq:SH02}, which
follows from \eq{eq:dispersion-anelastic} by setting $k_R=0$,
 states that $k_z$ oscillations that are stable to epicyclic motions can be de-stabilized by buoyancy, and vice-versa. Notice that the criterion is derived for adiabatic motion, so it will be modified in the presence of finite thermal diffusion or cooling, in principle allowing for unstable growth even when the adiabatic criterion is obeyed. Indeed, that is the driving force of the convective overstability (\sect{sect:cov}).

\eq{eq:SH2} pertains to the stability of the mixed $k_R k_z$ mode. Violation of \eq{eq:SH2} is the driving mechanism for the vertical shear instability, (\sect{sect:vsi}). \par
\medskip

{\emph{Physical Interpretation}}:  The Solberg-H{\o}iland criteria may be rationalized
in the same way the Rayleigh criterion is understood by appealing to the fluid-ring/parcel interchange analysis motivated in \sect{sect:Rayleigh_criterion}.  In particular, the contribution to the criterion by the buoyancy term is often referred to as the Schwarzschild criterion:
Referring once again to \fig{fig:rayleigh}, we consider the angular momentum conserving interchange of the two fluid rings as depicted.  In their new interchanged circular orbits, we allow each fluid ring to adiabatically expand and reach pressure equilibrium with its new environment.  As a result of the expansion, the fluid rings now exhibit different densities which are entirely dictated by the entropy gradient of the fluid, as indicated in \eq{eq:dispersion-anelastic}. 
This interchanged state is unstable if the total resulting energy -- which is now a sum of both the kinetic energy and the gravitational potential energy --  is less than what it was in the undisturbed configuration. With respect to \fig{fig:rayleigh}, if the gravity vector were pointed down and toward the star (indicated by $\grad{p}$), then the interchange would be destabilizing if the adjusted density of parcel 1, having been translated to parcel 2's original position, is less than parcel 2's original density before the interchange.  Similarly, the interchange would signal potential instability if the adjusted density of parcel 2, having been translated to parcel 1's original position, is greater than parcel 1's original density before the interchange.  Such a resulting density change means that the total gravitational potential energy after the interchange is less than with which it began. We are reminded that the interchange physics ultimately rests on the work done by the system upon the interchanged fluid rings and this work is ultimately measured by the terms $\partial_k{p}\partial_k{s}$ found in \eq{eq:dispersion-anelastic}.  Thus, while an adverse entropy gradient is destabilizing according to this energy principle and is the essence of the Schwarzschild criterion, instability of the disk fluid requires that the total energy of the interchanged fluid elements to be less than in the undisturbed state.  As such,  the
Solberg-H{\o}iland criteria, embodied in \eq{eq:SH1} and \eq{eq:SH2}, dictate the properties of the disk gas required to 
 insure that the resulting energy of every possible swapped fluid-ring state 
 is higher than with which it began -- hence, predicting a sufficient condition for stability.

 {\subsection{A magnetic interlude}

Before introducing the hydrodynamical instabilities, it is instructive to recall the basic physics of the MRI and how it violates the Rayleigh criterion.}

\subsubsection{Magnetorotational instability}
\label{sect:MRI}

Consider that the gas parcels pertaining to the rings A and B in a Rayleigh-stable situation (upper panel of \fig{fig:rayleigh}) are connected by a tether with a restoring force $-Kx$, where $K$ is a spring constant. The equations of motion are 

\beqn
\ddot{x} - 2\varOmega y + Kx  &=& 0\\
\ddot{y} + (2-q)\varOmega x + Ky &=& 0
\eeqn

Consider the Lagrangian displacement \[{\rm q}(t) = \bar{\rm q} + \ksi(t),\] with $\bar{\rm q}=0$ and $\ksi={\rm q}_0e^{i\omega t}$, where ${\rm q}$ is a generalized coordinate. The derivatives are 

\beq
\dot{\rm q} = i\omega \ksi \quad {\rm and} \quad \ddot{\rm q} = -\omega^2 \ksi
\eeq

\noindent and the equations of motion are 

\beqn
-\ksi_x \omega^2 - 2\varOmega \ksi_y i\omega + K\ksi_x  &=& 0,\\
-\ksi_y \omega^2 + (2-q)\varOmega \ksi_x i\omega + K\ksi_y  &=& 0.
\eeqn

This system is solved to yield the dispersion relation 

\beq
\omega^4 - \left(2K + \kappa^2\right)\omega^2 + K\left(K-2q\varOmega^2\right)=0
\eeq

\noindent which is a biquadratic equation. The solution is 

\beq
\omega^2 = \frac{\left(2K + \kappa^2\right)^2 \pm \sqrt{4K\left(K-2q\varOmega^2\right)}}{2}.
\eeq

If 

\beq
K - 2q\varOmega^2 < 0 
\label{eq:MRI}
\eeq 

\noindent then $\omega^2$ will have an imaginary component, and thus we have at least one growing root. \eq{eq:MRI} is the condition for instability. We see that if $K = 0$ (no tether), the condition for stability is simply $q>0$, i.e., the angular velocity decreasing outward. This is satisfied in Keplerian disks. 

The situation with the tether is identical to how magnetic fields operate. 
 {Let us include the Lorentz force in \eq{eq:navier-stokes}, i.e.,

\beqn
\v{F} \rightarrow \v{F}_{\rm Lorentz} &=& \frac{1}{4\pi} \left(\curl{\v{B}}\right)\times\v{B}  \nonumber \\
&=&  -\grad{\left(\frac{B^2}{8\pi}\right)} + \frac{\left(\v{B}\cdot\del\right)\v{B}}{4\pi}
\label{eq:lorentz}
\eeqn

\noindent where $\v{B}$ is the magnetic field. The second line above decomposed the Lorentz force into 
magnetic pressure and magnetic tension, respectively. Consider now the magnetic tension with a 
uniform base magnetic field $B_0\v{\hat{e}}$, where $\v{\hat{e}}$ is an arbitrary unit vector, upon which 
we superimpose a Lagrangian field displacement $\v{B}^\prime = B_0 ik\v{\ksi}$. The resulting acceleration is 

\beq
\frac{B_0 \partial \v{B}^\prime}{4\pi\rho} = -\frac{k^2 B_0^2\v{\ksi}}{4\pi\rho} = -(\v{k}\cdot\v{\va})^2 \ \v{\ksi}, 
\label{eq:tension}
\eeq

\noindent where we used $\partial = -ik$ and substituted $\v{\va}=\v{B}/\sqrt{4\pi\rho}$ for the Alfv\'en velocity.} We see from \eq{eq:tension} that the magnetic tension on the perturbation depends on the negative of the displacement, behaving exactly like a spring, with effective spring constant 

\beq
K \equiv (\v{k}\cdot\v{\va})^2.
\eeq

Therefore, the instability condition is 

\beq
(\v{k}\cdot\v{\va})^2 - 2q\varOmega^2 < 0 
\label{eq:mri-condition}
\eeq

 {Because the magnetic field appears multiplied by the wavevector, even weak fields generate significant tension at small enough scales. Indeed the MRI is a weak field instability, and should be generally present when the gas is magnetized.}

\subsubsection{ {Suppression of the MRI}}
\label{sect:deadstuff}

While full ionization is the state of the hot disks around black holes, 
protoplanetary disks are cold and poorly ionized, as noted in the introduction. 
 {In this case, neutrals compose the bulk of the flow, and ions and electrons drift through them. 
The magnetic flux is mostly frozen on the electron fluid, being these the most mobile charge 
carrier. In the case of a weakly ionized gas, the most probable collision a electron would suffer is with a neutral. These processes enter the induction equation as the electromotive force and Ohmic diffusion, respectively

\beq
\ptderiv{\v{B}} = \curl{ \left( \v{u}_e\times\v{B} - \eta \curl{\v{B}} \right)} ,
\eeq

\noindent where $\v{u}_e$ is the velocity of the electron fluid and $\eta$ the Ohmic resistivity. The ratio of the two terms is the magnetic Reynolds number

\beq
\Rey_M = \frac{\mathcal{U}\mathcal{L}}{\eta},  
\eeq

\noindent defined on dimensional grounds. Notice the symmetry with the usual (hydro) Reynolds number, 
with viscosity replaced by resistivity. At any length scale $\mathcal{L}$, magnetic effects will be suppressed 
if this quantity falls below unity. The velocity associated with magnetic fields is the Alfv\'en velocity $\va$, 
and the length of interest is the MRI scale, given by \eq{eq:mri-condition},  $k_{_{\rm MRI}}^{-1} \approx \va\varOmega^{-1}$, so the condition for the operation of the MRI in the presence of resistivity is

\beq
\varLambda \equiv \Rey_M (\va,k_{_{\rm MRI}}) = \frac{\va^2}{\varOmega\eta} > 1
\eeq

\noindent where we defined the {\it Elssaser number}, $\varLambda$, the magnetic Reynolds number 
associated with the MRI. 

Of course, Ohmic resistivity, the drag between electrons and neutrals, is not the only impediment to magnetic coupling. As also stated in the introduction, the drift between electrons and ions when ions are coupled to the neutrals (Hall effect) and ions and neutrals when ions are coupled to the electrons (ambipolar diffusion) are the two other non-ideal effects. We can decompose the electron velocity $\v{u}_e$ in terms of the ion and neutral velocities $\v{u}_i$ and $\v{u}$ as \citep{WardleKonigl93,BalbusTerquem01}

\beq
\v{u}_e = \v{u} + (\v{u}_e - \v{u}_i) + (\v{u}_i - \v{u}) 
\label{eq:electronvel}
\eeq

\noindent the first term is the reference frame velocity (following the neutrals). The second is the Hall effect and the third is ambipolar diffusion. If we consider a singly ionized species, the current $\v{J} = \sum_j n_j Z_j e  \v{u}_j$, where $e$ is the elementary charge and $n_j$ and $Z_je$ are respectively the number density and charge of species $j$, becomes

\beq
\v{J} = n_i e (\v{u}_i - \v{u}_e)
\eeq

As for the ambipolar term, in collision equilibrium the drift between ions and neutrals is set by, in the ion equation of motion, equating the Lorentz force with the neutral-ion collisional drag \citep{BalbusTerquem01,BaiStone17}, so that

\beq
\frac{\v{J} \times \v{B} }{c} = \gamma_{i} \rho_i \rho (\v{u}_i -\v{u})
\eeq

\noindent and \eq{eq:electronvel} can be cast as

\beq
\v{u}_e = \v{u} - \frac{\v{J}}{n_i e} + \frac{\v{J} \times \v{B}}{c \gamma_i\rho_i\rho }, 
\label{eq:electronvel}
\eeq

\noindent where $c$ is the speed of light. The induction equation then becomes

\beq
\ptderiv{\v{B}} = \curl{\left[   \v{u} \times \v{B} - \frac{4\pi \eta \v{J}}{c}  - \frac{\v{J}\times\v{B}}{n_e e}   + \frac{\left(\v{J} \times \v{B}\right)\times\v{B}}{c\gamma_i\rho\rho_i }  \right]}
\label{eq:induction-1}
\eeq

We can cast the Hall and ambipolar terms in terms of an effective resistivity and a current, similarly to the Ohmic term

\beqn
\ptderiv{\v{B}} &=& \curl{\bigg\{   \v{u} \times \v{B} \ - } \nonumber \\
&&\frac{4\pi }{c}   \left[ \eta_O \v{J}  + \eta_H \frac{\v{J}\times\v{B}}{B}   + \eta_A \frac{\v{B} \times \left(\v{J} \times \v{B}\right)}{B^2}  \right] \bigg\} 
\label{eq:induction-2}
\eeqn

\noindent where we now switched $\eta \rightarrow \eta_O$ for symmetry. Comparing \eq{eq:induction-1} and \eq{eq:induction-2}, the Hall and ambipolar resistivities are

\beqn
\eta_H &=& \frac{Bc}{4\pi n_e e}  \\
\eta_A &=& \frac{B^2 }{4\pi \gamma_i \rho \rho_i}
\eeqn

We quote the Ohmic resistivity for completeness \citep{KrallTrivelpiece73,BalbusTerquem01,Wardle07,Lesur+14}

\beq
\eta_O = \frac{m_e c^2}{4\pi e^2}  \frac{n}{n_e} <\sigma_{\rm coll} u>_e  \ \approx \ 234 \ x^{-1} T^{1/2} \ {\rm cm}^2 \,{\rm s}^{-1}
\eeq

\noindent where $m_e$ is the electron mass, $x$ is the ionization fraction, $\sigma_{\rm coll}$ is the 
cross section of electron-neutral collisions, and the angled brackets represent averaging over the 
Maxwellian velocity distribution of the electrons. The resistivities allow for the definition of Hall and ambipolar Elsasser numbers, in symmetry with the Ohmic Elsasser number.

\beq
\varLambda_H \equiv \frac{\va^2}{\varOmega\eta_H} = \frac{e n_e B}{c\rho \varOmega}
\eeq

\beq
\varLambda_A \equiv \frac{\va^2}{\varOmega\eta_A} = \frac{\gamma_i \rho_i}{\varOmega}
\eeq

In general, the MRI will be present if all Elsassers numbers are larger than unity. The dominant resistive effect is that of lowest Elsasser number. For the Minimum Mass Solar Nebula model \citep[MMSN,][]{Hayashi81} and uniform ionization fraction we can write

\beqn
\varLambda_O &=& 3.2 \times 10^{-6} \ x_{13} \ \beta_{4}^{-1} \ \rho_{9}^{-1} \ T_{300}^{-1/2}  \  R_{\rm AU}^{5/4}   \\
\varLambda_H &=& 3.2 \times 10^{-5} \ x_{13} \ \beta_{4}^{-1/2} \ \left(\sfrac{\mu}{2}\right)^{-1} \ R_{\rm AU}^{-1/8}\\
\varLambda_A &=& 1.4 \times 10^{-2} \  x_{13} \ \rho_{9}\ R_{\rm AU}^{-5/4}.
\eeqn

\noindent where $x_{13}$ means ionization fraction in units of 10$^{-13}$, $\beta_{4}$ the plasma beta parameter ($\beta \equiv 2cs^2/\va^2$) in units of 10$^4$, $\rho_9$ the gas volume density 
in units of 10$^{-9}$ g/cm$^3$, $T_{\rm 300}$ the temperature in units of 300\,K, and $R_{\rm AU}$ the distance in AU. Under the approximations made, at 1AU in the MMSN, Ohmic resistivity will be the dominant impedance, but the problem is very model-dependent, and different combinations will yield different Elsasser numbers at different radii. The resistivities do not consider either the presence of grains, which will modify them \citep[][but see \citet{Simon_etal_2018} for a different opinion]{TurnerDrake09}. 

We assume that the disk will have a region that is 
Ohmic-dominated, and concentrate from now on our analysis on this region. 

\subsubsection{An interlude within an interlude: Pure Hall MHD and Generalized Potential Vorticity}

Under conditions in which both the Ohmic and Ambipolar coefficients are greatly dominated by the Hall
term, i.e. when $\eta_O \ll \eta_H, \ {\rm and} \ \eta_A \ll \eta_H$, we might consider the induction equation (\eq{eq:induction-2}) in the identical limit $\eta_O = \eta_A = 0$ and find that it reduces
to

\beq
\ptderiv{\v{B}} = \curl{\left(   \v{u} \times \v{B}  -  \frac{m_e}{x e}\frac{\v{J}\times\v{B}}{\rho} \right)} ,
\label{eq:induction-pure-Hall}
\eeq
where we have restored the terms defining $\eta_H$. If we assume the following, (i) $x$ is a constant,
(ii) ${\cal Q} = 0$ which means that the gas entropy is materially conserved, and (iii) we neglect gas viscosity by setting ${\v{{\rm T}}} = 0$, then Holm (1987) showed that 
combining Eqs. (\ref{eq:continuity}-\ref{eq:navier-stokes}) and Eq.(\ref{eq:entropy}) with Eq. (\ref{eq:induction-pure-Hall})
reveals that the Hall MHD equation set possesses a magnetic generalization of Ertel's theorem.  In particular, there exists a magnetic generalization of potential vorticity, denoted by $\OmegaGPV$, which is materially conserved by the flow.  That is to say the scalar
\beq
\OmegaGPV \equiv \frac{1}{\rho}\left(\v{\varomega} 
+ \frac{x e}{4\pi m_e} \v{B}\right)\cdot \del s,
\label{eq:general_PV}
\eeq
obeys the conservation equation
\beq
\left(\ptderiv{} + {\v{u}\cdot \del} \right) \OmegaGPV = 0.
\label{eq:conservation_of_general_PV}
\eeq
In the limit where the ionization fraction goes to zero, \eq{eq:general_PV}
indicates that $\OmegaGPV$ limits to the classical conservation of fluid vorticity,
$\OmegaPV \equiv \rho^{-1}\v{\varomega}\cdot \del s$ as discussed in \sect{sect:PV}. The magnetic generalization of the potential vorticity quantity has been used to better understand the emergence and stability of flow structures in confined plasmas.  Similar explorations of the role this quantity plays in the emergence of zonal flows in accretion disks subject to Hall MHD have yet to be fully explored.
}
\subsection{Hydrodynamical instabilities}
\label{sect:thermal}

Since the ionization level required to couple the gas to the ambient field is not
always met \citep{BlaesBalbus94}, it leads to zones that are ``dead''
to the MRI \citep{Gammie96,TurnerDrake09}. So, the
quest for hydrodynamical sources of turbulence continues, if only to
provide accretion through this dead zone. In this section, we describe 
the three hydrodynamical instabilities discovered in the past few years and that may 
play a role in disk dynamics in the Ohmic zone. In the next subsection we present 
the Vertical Shear Instability (VSI), followed by the Convective Overstability (COV), concluding with the 
Zombie Vortex Instability (ZVI).
\par
Key to all of the following instabilities is the question of the thermal cooling
times of the primary perturbations that lead to the widespread instability.  For disks
this is controlled by radiative processes.  Radiation driven thermal relaxation
will have two natural limits depending upon the photon mean free path.
We define $\kappa_R$ to be an effective {\emph{material}} Rosseland opacity (cm$^2/$g), in other words, a frequency integrated absorption cross section that accounts for the cumulative absorption 
by both gas and dust per gram.  Letting $\rho$ be the
mass density of the local disk material, then the relaxation times will depend upon
how greatly $\ell_{rad} \equiv (\kappa_R \rho)^{-1}$ exceeds the length scale
of interest, $\ell$, which is the length scale of a fluid disturbance.  When
$\ell_{rad} \gg \ell$ then the thermal relaxation is optically thin while when  
$\ell_{rad} \ll \ell$ then the thermal relaxation is optically thick.  The thermal
relaxation times $\tau_r$ are then given by the following approximate form,
based on the exact formulation found in \citet{Lin_Youdin_2015} based on 
the derived expression found in \citet{Spiegel_1957},
\beq
\tau_r
= 
\left\{
\begin{array}{cr}
\displaystyle
\frac{3\kappa_R \rho^2 \cv \ell^2}{16 \sigma_{\rm SB} T^3}; & \ell_{rad}\ll \ell; \\
&\\
\displaystyle \frac{\cv}{16 \kappa_R \sigma_{\rm SB} T^3}; & \ell_{rad}\gg \ell; 
\end{array}
\right.
\label{eq:relaxationtimes}
\eeq
where $\sigma_{\rm SB}$ is the Stefan-Boltzmann constant.
Obviously, the $\ell^2$ dependence in the optically thick regime reflects the diffusive
nature of the optically thick limit.\par
These two limits are captured in the two extreme thermal relaxation models.  The optically
thin limit is often referred to as Newton's Law of Cooling
in which the source term ${\cal Q}$ in \eq{eq:entropy} is given by
\beq
{\cal Q} = \frac{1}{\tau_r}\big(T_d-T\big),
\eeq
where $T_d$ is some spatially specified forcing temperature representing
how the disk gas is forced by the radiation field from the star, where $T$ in
$\tau_r$ gets replaced with $T_d$.  In the optically thick limit, the heat source/loss
term in \eq{eq:entropy} is expressed as a radiative diffusion expression commonly known
as the Eddington Approximation,
\beq
{\cal Q} = \nabla \left(\frac{a_{\rm rad}c}{3\kappa_R}\nabla T^4\right),
\eeq
where $\sigma_{\rm SB} \equiv a_{\rm rad}c/4,$ and $a_{\rm rad}$ is the radiation density constant. 

\subsubsection{Vertical Shear Instability}
\label{sect:vsi}

\begin{figure}
  \begin{center}
    \resizebox{\columnwidth}{!}{\includegraphics{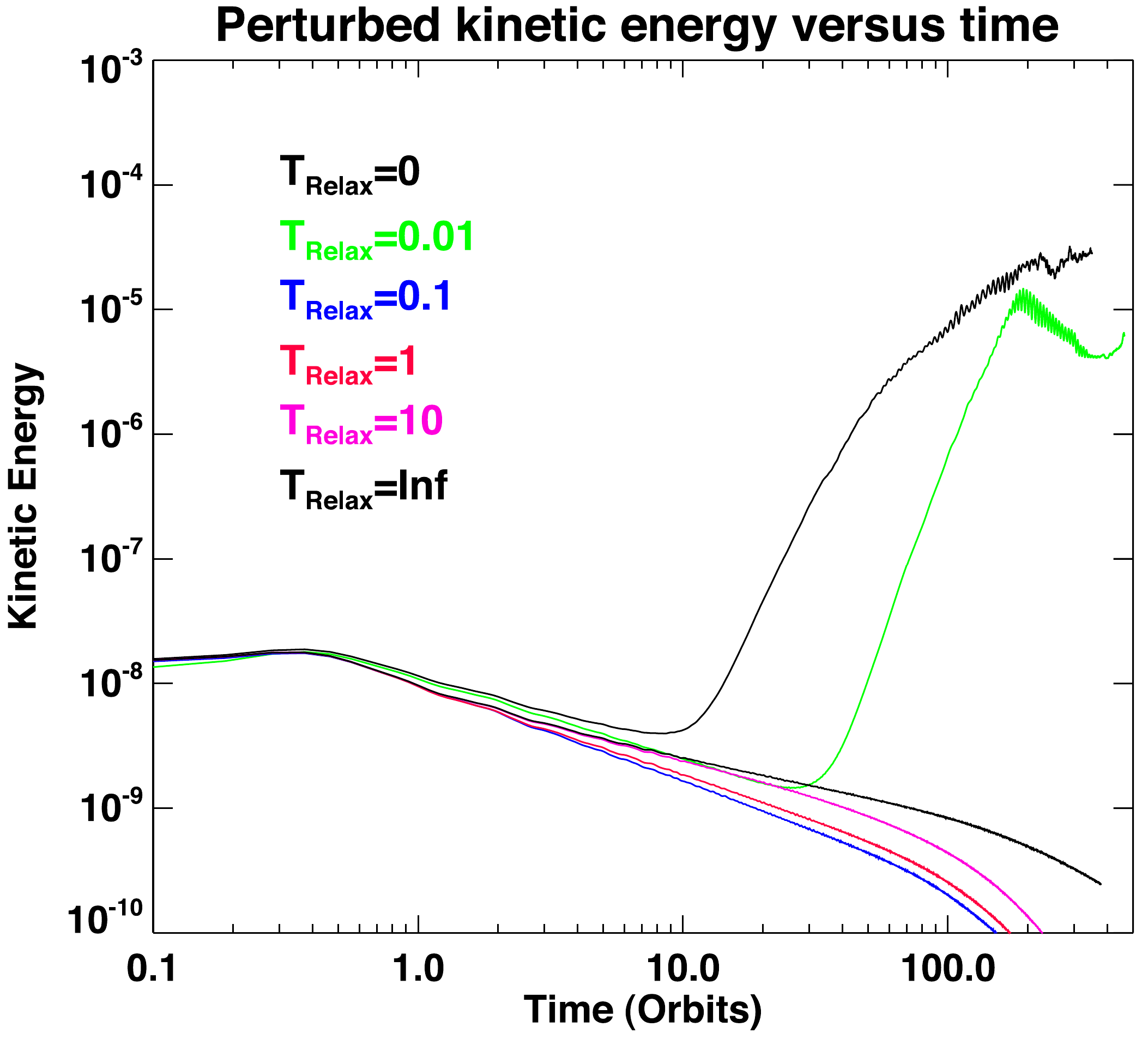}}
\end{center}
\caption[]{Time evolution of the sum of the (normalized) perturbed
  radial and meridional kinetic energies in disks where the
  temperature was initially constant on cylinders, as a function of
  the thermal relaxation time (optically thin cooling law), in units of orbital period, $2\pi\varOmega^{-1}$. Note that
  only the cooling times of 0 and 0.01 cases show growth. Reproduced
  from \citet{Nelson+13}.}
\label{fig:vsi-kinetic}
\end{figure} 

\begin{figure*}
  \begin{center}
    \resizebox{\textwidth}{!}{\includegraphics{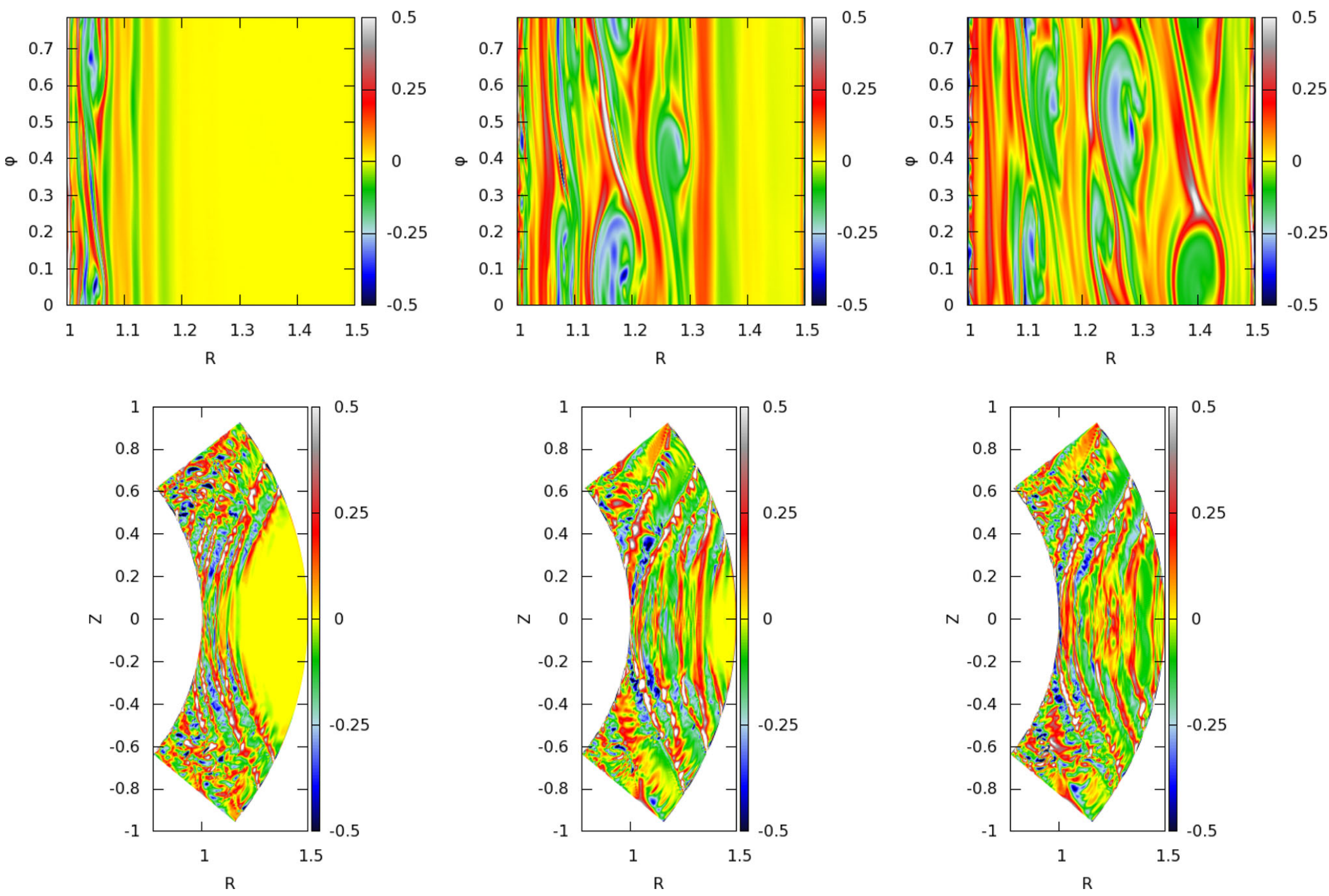}}
\end{center}
\caption[]{Midplane and meridional plane vorticity profiles of the VSI in models utilizing the simple Newton's Law of Cooling formalism.   Snapshots shown:
  after 201, 317 and 401 orbital periods.  In these models $h\equiv H/R = 0.2$ and 
  cooling time $\tau =
  0.5$orb. 
  Figures are reproduced from \citet{Richard_etal_2016}.} 
\label{fig:vsi-richard}
\end{figure*} 

\begin{figure*}
  \begin{center}
    \resizebox{.48\textwidth}{!}{\includegraphics{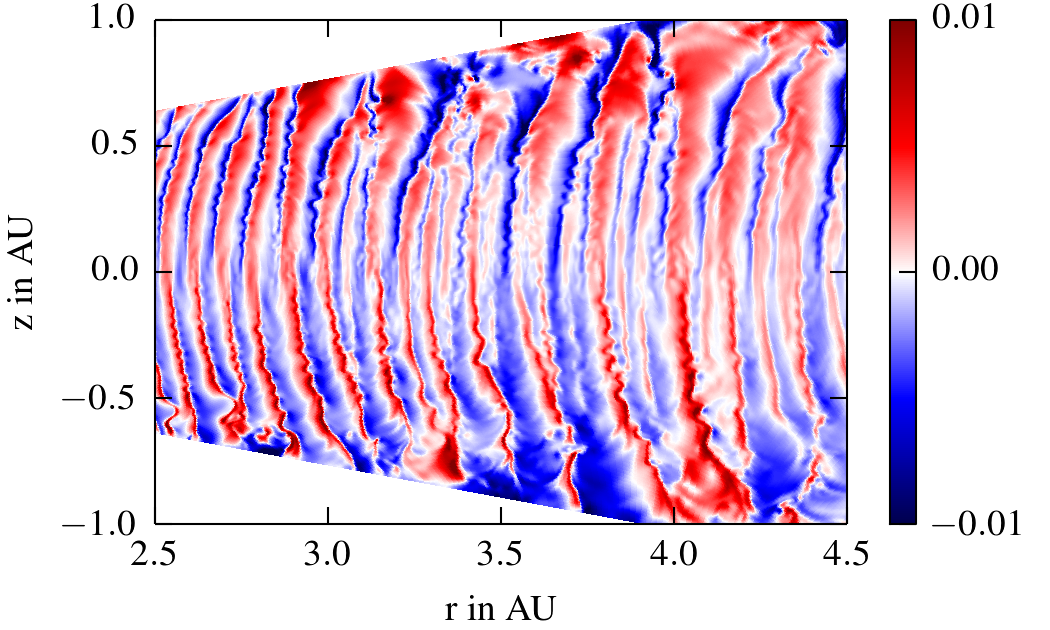}}
    \resizebox{.48\textwidth}{!}{\includegraphics{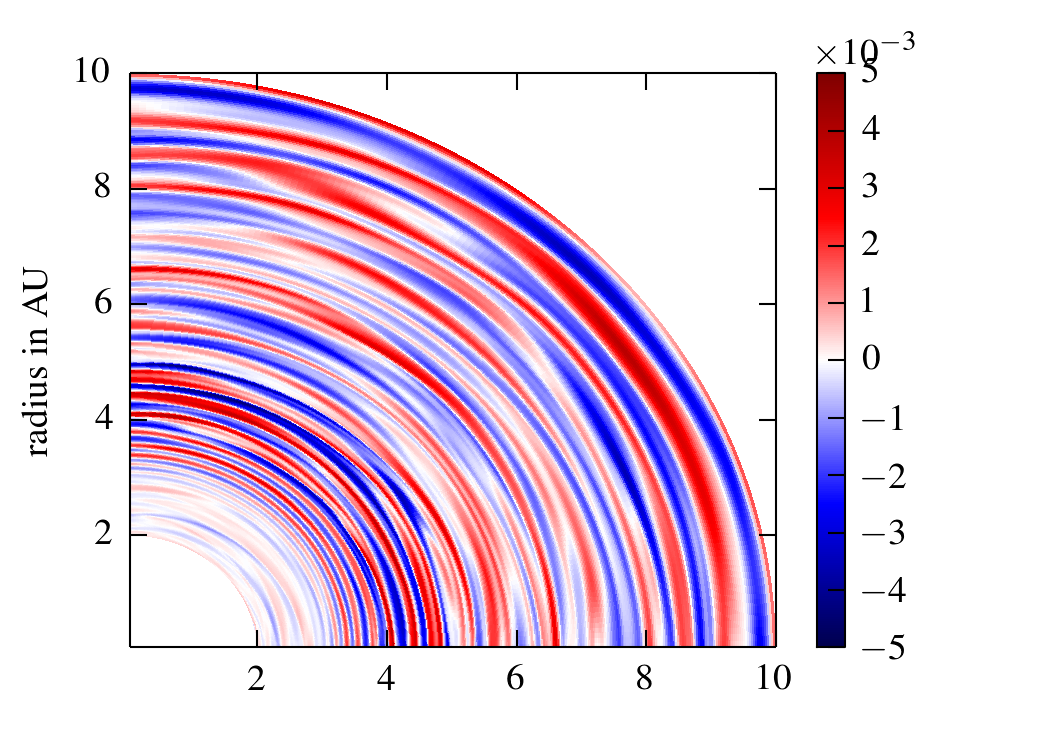}}
\end{center}
\caption[]{Simulations of the VSI with a realistic radiative transfer model \citep{Stoll_Kley_2014}.  Left: Velocity in the meridional direction, $u_\theta$, in
  units of local Keplerian velocity for an isothermal simulation
  without viscosity.  The vertical velocities follow a similar pattern. Right: the vertical velocity in the midplane 
  of the disk for a 3D model after 4000 orbits.
  The nearly axisymmetric property of the saturated instability, even in the ostensibly statistically steady state, is clearly
  visible in these and in simulations of \citet{Nelson_etal_2013}. Figure reproduced from \citet{Stoll_Kley_2014}.} 
\label{fig:vsi-structure}
\end{figure*} 

The vertical shear instability, as the name suggests, draws on the
free energy of vertical shear in disks. According to the
Taylor-Proudman theorem, rotating barotropic flows under the action of
a central conservative force must exhibit constant rotation on cylinders. 
By barotropic we mean to say that the fluid pressure is only a function
of density, $p=p(\rho)$, as is the case in adiabatic flows in which the 
fluid temperature may be written as a function of density.
Under strong stellar
irradiation, like in many parts of protoplanetary disks, the temperature field is
mainly a function of the (cylindrical) radius from the central star, exhibiting very little variation in the disk-vertical extent containing most of its inertia.
Therefore, because temperature and density
isolines do not coincide, the flow configuration is said to be baroclinic which
means
$\nabla p \times \nabla \rho \neq 0$.
Under such baroclinic circumstances, a small amount of vertical shear can be
sustained. 

As evident in \eq{eq:vertical-shear}, the shear is 
proportional to the radial temperature gradient, in a way that the angular
speed and momentum is lowest at the midplane 
and increases in the disk normal directions.
\footnote{This is because the temperature decreases with radial distance.  If the radial temperature profiles were to increase, then the situation
would be reversed with angular speeds and momenta decreasing with increases away from the mid-plane in the disk normal direction.}
\par

So long as some physical agent maintains this baroclinic configuration
(like strong thermal forcing with gravity), then there may exist
a condition in which an angular momenta conserving 
interchange of two annular gas rings of
infinitesimal cross-section can result in a lower 
total energy state \citep{Tassoul_1978,Kippenhahn_etal_2012}.  
This is the physical rationalisation of the Rayleigh
stability of barotropic Keplerian flows, e.g., as discussed
in \sect{sect:Rayleigh_criterion}:  as the vertical uniformity of the mean
flow state means only evaluating purely radial interchanges. As it happens to be the case for Keplerian flows with $\varOmega \sim R^{-3/2}$, this act results in a higher energy configuration and, thus, is not preferable state and the system will want to return to its
original configuration.  This is one way toward understanding both 
the Solberg-H{\o}iland criteria as well as
the physics of axisymmetric disk inertial oscillations (which play a further role  below).\par 
When thermal and viscous diffusion are an important feature of the physics, then
the flow is now baroclinic and the stability properties of rotating flows are no longer controlled by the Solberg-H{\o}iland criteria.
Because the mean-flow now exhibits some vertical variation,
a physical interchange of the type just described could lead to
a configuration with a lower total energy, see once again Fig. \ref{fig:rayleigh}.  
{\emph{With the restriction of
an everywhere isentropic and incompressible gas}} subject to 
wavelike perturbations of the form $\exp(ik_R R + ik_z z)$ in which $k_R$ and $k_z$ are respectively wavenumbers in the radial and disk-vertical directions, a modified
stability criterion, obtained originally by 
\citet{Goldreich_Schubert_1967} and \citet{Fricke_1968},
in application to the radiative zones of differentially rotating stars, 
says that instability is possible when
\beq
\pderiv{L^2}{R} - \frac{k_R}{k_z} \pderiv{L^2}{z} < 0 ,
\label{eq:vsi}
\eeq

\noindent where $L$ is the angular momentum. 
Because for baroclinic disks the rotation rate is a function of
both $R$ and $z$, 
unstable modes are guaranteed to exist since wavevectors with ratios
$k_R/k_z$ that satisfy equation \eq{eq:vsi} can always be found. For a thin
quasi-Keplerian disk, according to \eq{eq:vsi} instability occurs for
modes satisfying $k_R/k_z > h^{-1}$, i.e $\lambda_R < h \lambda_z$,
meaning that unstable modes will have radial wavelengths that are 
much shorter than vertical ones. The instability can be understood
through the mathematical lens of the second of the Solberg-H{\o}iland criteria, \eq{eq:SH02}.  Furthermore, the instability can be physically
seen as arising directly as a consequence of the ring-parcel interchange
diagnostic described in \sect{sect:SH}.  While a purely barotropic Keplerian flow is Rayleigh stable, the interchange procedure can result in a lower energy state after both a radial and a vertical translation of the fluid rings.

\par
Under conditions in which the thermal relaxation times are extremely
short ($\tau_r \varOmega \ll 1$), the above criterion determines the onset of instability.  The possibility
that this process is possible for protoplanetary disks was first suggested 
in the analyses of \citet{Urpin_Brandenburg_1998} and \citet{Urpin_2003}, followed
by the study of \citet{Arlt_Urpin_2004} who gave the first estimate of the growth
rate to be
\beq
\sigma \sim |q| h \varOmega_0.
\eeq
The first unambiguous identification of the instability in a realistic
disk model came in the study of 
\cite{Nelson+13}\footnote{\cite{ArltUrpin04} also performed nonlinear simulations but the disk
model they adopted assumed a Boussinesq equation of state.  The instability, though technically present, is difficult to identify.  Also, using the Boussinesq equation of state 
predicts that onset to instability occurs at zero frequency, 
failing to capture the actual oscillatory
nature of the VSI at onset.}. 
In this work it was found that oscillating inertial
modes become unstable (overstable) and the fastest growing {\emph{body modes}}
have a corresponding radial wavelength $\lambda_R \approx \pi h H_0$.  Body modes
are inertial modes whose power is concentrated around the midplane of the disk and, therefore, constitutes responses affecting the bulk inertia of the disk mass.  A linear analyses in the infinitely fast relaxation time limit (i.e., $\tau_r \rightarrow  0$) found in 
\cite{BarkerLatter15} confirm the growth rate trends reported earlier while raising concerns about the appearance of surface modes
\footnote{\citet{Nelson_etal_2013} and \citet{Barker_Latter_2015} identified two classes of modes.  In addition 
body modes, a second class
of so-called {\emph{surface modes}} were also identified.  These are structures both attached and concentrated on the vertical boundaries of simulations, and they have fast growth rates. However, owing to the very low densities at these locations they hardly affect the emerging dynamics in the bulk of the disk since they contain very little energy and, as such, are considered artefacts of imposing artificial boundary conditions \citep{Umurhan+16VSI}.}, 
while the analysis in \cite{Umurhan+16VSI} verified that wavemode with the fastest growth rates
should occur for finite values of the radial disturbance wavelength, i.e.,
\beq
\lambda_R({\rm max}) = \pi |q| h H_0, \qquad
\sigma({\rm max}) = \sqrt{m} |q| h \varOmega_0/4,
\eeq
where $m$ order the sequence of vertical nodes in the responses. All disturbances with odd values of $m$ are called {\emph{corrugation modes}}, since the disturbance vertical velocities are symmetric about the midplane.   Responses with even powers of $m$ are termed the 
{\emph{breathing modes}} in which the vertical velocities are symmetric with respect to the midplane.   The $m=1$ and $m=2$ mode are the respective fundamental responses -- these contain the majority of the energy in developing VSI.  All normal modes with $m>2$ (the overtones) increasingly exhibit power concentrated further away from the disk midplane.  An immediate consequence is that a majority of the disk motions are projected onto the two fundamentals with very little energy associated with higher overtones. Nonetheless, the linear theory demonstrates that modes show increased growth rates
with increasing node number $m$ and this, in turn, raises some concerns.  Nonlinear numerical simulations show that only the two fundamental modes get expressed as the process grows.  It is possible that the faster growing high vertical node modes, with increased vertical structure with distance from the disk midplane -- saturate at low amplitudes are are suppressed by actual and numerical viscosity.  For a further discussion on this mathematical matter see Barker \& Latter (2015) and
Umurhan et al. (2016b).
\par

The nonlinear simulations of the VSI reported in \citet{Nelson_etal_2013} showed several features of the instability as it develops into a nonlinear saturated state.  They show that early stages of growth is dominated by the fundamental breathing mode but eventually growth switches over to the fundamental corrugation mode.  As such, these two fundamental modes probably constitute the primary injection scales of the resulting turbulence.  When the system reaches a quasi-steady state, the radial scales of the resulting structures are much larger than the radial scales of the fastest growing modes, the latter being approximately $H$ or slightly larger, and being highly suggestive of some amount of inverse energy cascade happening on these scales -- but this has yet to be quantified.  The simulations reported in both \citet{Nelson_etal_2013} and \citet{Richard_etal_2016} utilize the simple thermal "Newton's Law of Cooling" model in which the gas is locally forced to a prescribed radial temperature profile on a cooling time $\tau_c$, usually measured in units of local orbit times (orb$=2\pi/\varOmega_0$).  \citet{Nelson_etal_2013}, who run simulations with $h=0.05$, find that the VSI is most robust for $\tau_c \rightarrow 0$ and will operate with decreasing vigor up to a maximum cooling time $\tau_c = 0.1$orb.  \citet{Richard_etal_2016} show that the
instability survives for up to larger values of $\tau_c$ for correspondingly larger
value of $h$, e.g., see \fig{fig:vsi-structure}. 
\par
When non-isothermal effects are introduced (i.e., $\tau_r>0$), buoyancy stabilizes
otherwise unstable oscillations, according to the Solberg-H{\o}iland
criterion. For this reason, it is the exactly-isothermal response ($\tau = 0$) that is most unstable
one, and the instability ceases to exist when even moderate levels of
adiabaticity are introduced, as shown in \fig{fig:vsi-kinetic}. 
\citet{Lin_Youdin_2015} quantify the cooling time requirement  {for vertically isothermal stratification as }
\beq
\varOmega_k\tau_c < \frac{h|q|}{\left(\gamma-1\right)}
\eeq
where $\gamma$ is the ratio of specific heats.  The growth rates rapidly
diminish and shutoff entirely once cooling times get much longer than this longer cooling times because the
unstable wavelength is shifted deep into the viscous range. 
 {Furthermore, we observe that this condition is valid for the locally isothermal background state. For vertical adiabatic structures, cooling requirements are less stringent.}
\par
Noting that disks are likely optically thick on the length scales on which the VSI operates, \citet{Nelson_etal_2013} equated the critical cooling time ($\tau_c\sim = 0.1$orb)  to a corresponding radiative (diffusive) relaxation time predicting that the VSI ought to be operative for effective disk opacities $\kappa < \kappa_c \approx 1$cm$^2$/gm.
Subsequent realistic simulations handling the thermal response of the gas using a radiative diffusion model have been performed by Stoll and collaborators
\citep{Stoll_Kley_2014,Stoll_Kley_2016, Stoll_Kley_Picogna_2017,Stoll_Picogna_Kley_2017}.
In particular, the important study of \citet{Stoll_Kley_2014} demonstrated that the VSI
remains robust in such realistic models and, furthermore, they show that its turbulent intensity is of medium strength (see further below), arguing also that it likely plays a significant role in the planet formation process as parts of disks in which dust growth occurs likely support the conditions in which this process operates (see also 
\citealt{Flock_etal_2017}).\par

The 3D modeling of the VSI \citep{Nelson_etal_2013,Stoll_Kley_2014,Richard_etal_2016,Flock_etal_2017} 
shows that it develops into long-lived unsteady activity.  Is it turbulence?  At this stage it is not clear.  Large scale unsteady vortices are typical of the final emergent structures.  Despite the successes of these simulations, an inertial range has yet to be resolved since in most simulations the injection scales are resolved up to 10-20 grid points. The need to flesh out the inertial range of this process forms the basis of future directions on the subject.
Nonetheless, the global high-resolution modeling, in particular of
\citet{Richard_etal_2016,Flock_etal_2017}
have quantified the resultant turbulent speeds 
as being a few percent of the local sound speed -- meaning to say that the
midplane ($\alpha\approx 4\times 10^{-5}$-- $10^{-3}$), increasing to about
20\% ($\alpha\approx 10^{-2}$-- $10^{-1}$) in the corona (i.e., $|z| > 4 H_0$).  
Note that the rise
in the apparent values of $\alpha$ far from the midplane is attributable to the corresponding decrease of the mean densities there.\par
Lastly, there are apparently two co-acting secondary transition routes into unsteady activity - one attributable to the roll-up of vortices due to vertical motions of the disk and to azimuthal motions in the
plane of the disk: \citet{LatterPapaloizou18} show in a Boussinesq model that the saturation of the VSI into azimuthally oriented vortices occurs via
parasitic Kelvin-Helmholtz modes \citep{Goodman_Xu_1994}, while \citet{Richard_etal_2016} postulate that a second secondary transition occurs via the Rossby wave instability \citep{Lovelace+99,Li+10}, leads to continual generation of vertically oriented unsteady vortices. 
 {Indeed, \cite{MangerKlahr18} have recently found that long lived vortices can be seeded via RWI in a VSI-unstable disk, provided that the azimuthal domain is large enough ($\geq$ 180$^\circ$).
And finally, with respect to the secondary transition by the Rossby wave instability, it should be noted that the manner in which the basic zonal flows emerge -- upon which this secondary instability operates -- has yet to be fully understood but it is possible that an inverse-cascade like the Rhines mechanism might be partly responsible.}

\subsubsection{Convective Overstability}
\label{sect:cov}

\begin{figure}
  \begin{center}
    \resizebox{.48\textwidth}{!}{\includegraphics{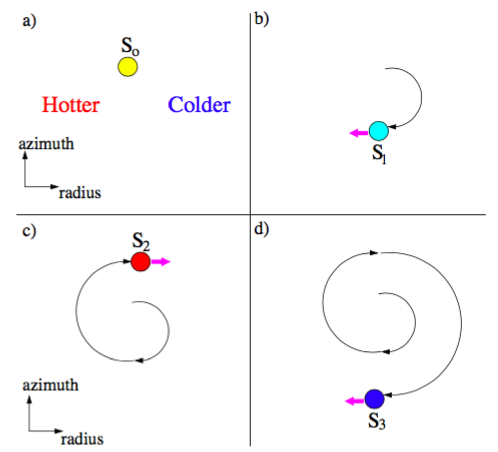}}
\end{center}
\caption[]{A cartoon in four panels indicating the convective overstability
  mechanism. In panel (a) a fluid blob is embedded in a radial entropy
  gradient. In panel (b) it undergoes half an epicycle and returns to
  its original radius with a smaller entropy than when it begun $s_1 <
  s_0$. It hence feels a buoyancy acceleration inwards and the epicycle
  is amplified. The process occurs in reverse once the epicycle is
  complete, shown in panel (c), where now $s_2$ > $s_0$. The oscillations
  hence grow larger and larger. The impression of two-dimensionality is deceptive. 
    To conserve mass, a similar motion must happen in another horizontal layer, 
    in the opposite direction. This requires $k_z\neq 0$. Reproduced from \cite{Latter16}.} 
\label{fig:latter}
\end{figure}

\begin{figure*}
  \begin{center}
    \resizebox{\textwidth}{!}{\includegraphics{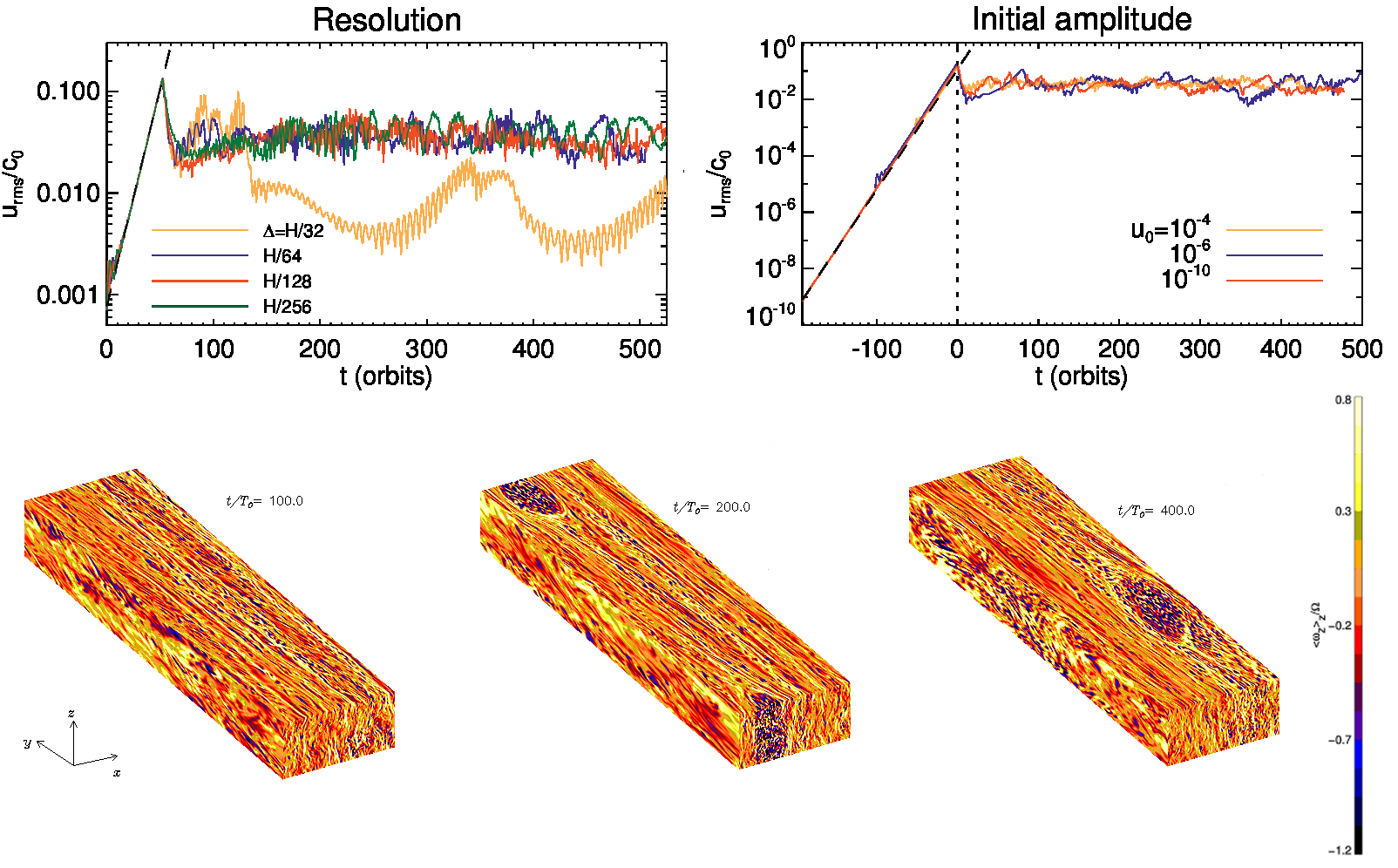}} 
\end{center}
\caption[]{Linear and nonlinear evolution of the convective overstability. Upper panels: Convergence study with resolution (left) and initial amplitude of perturbation (right). Resolution of 64 points per scale height is enough to resolve the overstability, and initial amplitudes as small as $10^{-10}$ times the sound speed ($c_0$) are enough to lead to growth, demonstrating the linear nature of the process. The linear growth rate (black dashed curve) is well reproduced in all cases. Lower panels: With the linear overstability raising the amplitude of the initial fluctuations to nonlinear levels, a large-scale vortex is generated. The panels show the vertical vorticity. The Reynolds stress saturates at $\alpha\approx\ttimes{-3}$. Adapted from \cite{Lyra14}.}
\label{fig:figure3d}
\end{figure*}

\begin{figure}
  \begin{center}
    \resizebox{\columnwidth}{!}{\includegraphics{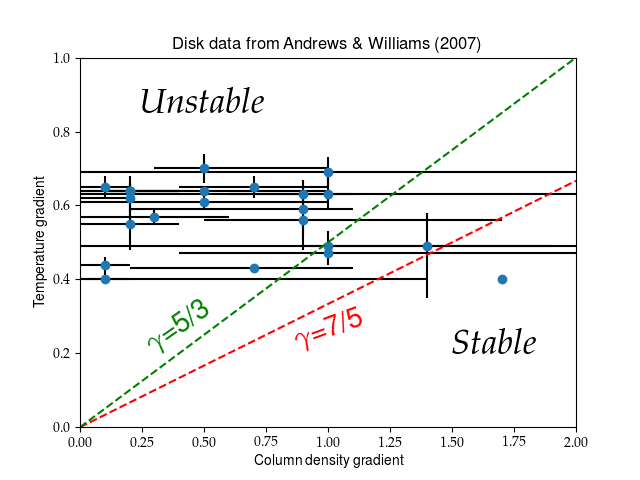}}
\end{center}
\caption[]{The sign of the square of the Brunt-V\"ais\"al\"a frequency
  defines the stability criterion, here shown as a function of the
  density and temperature power-law indices. The plot shows the lines for two
  values of $\gamma$. Above (below) the respective line the system is
  unstable (stable). Adapted from \cite{Lyra14}. The dots represent some 
disk data for 
  which temperature and density gradients have been estimated. The SBI may
  be an important process for the majority of disks in the sample.} 
\label{fig:stability}
\end{figure}

The  {second} hydrodynamical instability discovered is the convective 
overstability, which, as the name suggests, is a convective process. 
It contrasts with classical convection because the direction it 
occurs is the radial, not the vertical one. It manifests itself as 
epicyclic oscillations that are amplified by buoyancy, hence the name. 

The COV shares many similarities with the subcritical baroclinic instability (SBI), 
a process discovered earlier, when \cite{KlahrBodenheimer03}, 
modeling disks with negative entropy gradients, found baroclinic growth of vortices.
No linear growth was found in linear analysis \citep{Klahr04}, hinting at nonlinear origin. The nature 
of the instability was clarified in the works of \cite{Petersen+07a,Petersen+07b}, 
who highlighted the importance of finite thermal inertia. When the 
thermal time is comparable to the eddy turnover time, the vortex is 
able to establish an entropy gradient around itself that compensates 
the large-scale entropy gradient that created it. The mechanism 
was elucidated by \cite{LesurPapaloizou10}: the cycle starts with a fluid parcel 
buoyantly lifted to an outer orbit.  As it travels outwards, it finds itself surrounded by gas of
lower entropy. The resulting buoyancy lifts it even further
outwards. Deflected by the Coriolis force and following the vortex streamline, the fluid now 
travels along the azimuthal direction. In this leg, the fluid thermalizes, cooling down. As it cools, 
it gets denser, and sinks back inwards.  The last azimuthal leg will thermalize the fluid, heating it 
back to the temperature it had in the beginning, closing the cycle. The original radial entropy gradient 
and the azimuthal temperature gradient established by the thermalization 
lead to a non-zero baroclinic term $\grad{p}\times\grad{\rho}$, generating vorticity 
locally and amplifying the vortex. A respectable amount of angular momentum is 
transported by spiral density waves excited by the vortex \citep{HeinemannPapaloizou09,HeinemannPapaloizou12}, 
leading to $\alpha$ values in the vicinity of $10^{-3}$. 

 {Notice that the process requires the radial entropy gradient and the radial pressure gradient to be aligned (to give buoyancy) and thermal relaxation or diffusion (to establish the azimuthal gradient).} If the gas is locally isothermal a gas parcel has always the same temperature as the
surroundings and buoyancy cannot act to amplify the motion.  {If the gas
is adiabatic a gas parcel cannot exchange heat so an outward/inward motion cannot
thermalize and return to its equilibrium position.} The SBI
operates optimally for cooling times similar to the dynamical time. 

 {The convective overstability obeys practically the same mechanism. Indeed, \cite{Klahr04} had 
already noted that the SBI could be regarded as a convective instability in the 
radial direction, but modified by rotation and shear. The difference is the unstable mode. 
While the SBI uses a large scale vortex, in the COV the fluid executes a small epicycle. This unveils 
also a crucial difference between the two mechanisms: while the SBI is nonlinear, the COV is linear. 
The cartoon of \fig{fig:latter} illustrates the growth of a particular mode. 
As a gas parcel moves into areas colder (hotter) gas, it loses (gains) heat, compresses (expands), and because of the extra buoyancy, 
overshoots the original amplitude of oscillation. The cartoon is illustrative and does 
not exactly represent the 3D axisymmetric most unstable mode. Being axisymmetric, 
the motion is 1D, not necessitating the azimuthal direction. The need for the vertical direction, 
shown by the $k_z$ dependency in the dispersion relation comes about because of mass conservation: 
if a gas parcel moves inwards, an equal amount of gas must move outwards. 
In order to conserve mass, the motion shown in the cartoon must be matched by the same 
motion executed in another horizontal layer, in the opposite direction.}

The overstability exists both if the cooling is provided by thermal
relaxation \citep{Lyra14} or by thermal diffusion \citep{Latter16}. In the case that the cooling time $\tau_r$ is 
finite, the Rayleigh and the Solberg-H{\o}iland criteria are but the isothermal and adiabatic limits of the more general  
anelastic dispersion relation \citep{KlahrHubbard14,Lyra14,Latter16}

\beq
\omega^3 + \frac{i}{\gamma \tau_r} \omega^2 - \omega(\kappa^2 + N_R^2) - \frac{i}{\gamma\tau_r}\kappa^2 = 0.
\eeq

Thermal relaxation and thermal diffusion should mimic the limits of cooling in optically thin and in the
optically thick regimes. The difference is that thermal relaxation
happens at all scales at the same time, as expected if photons are
simply allowed to escape. The mode is therefore ballistic, with all
scales of the disk executing the same oscillation  {(in the 
local box only; in a global disk the epicyclic frequency is 
a function of radius and height).} The growth rate is 

\beq
\sigma = -\frac{1}{2} \left[\frac{\mu^2 \beta N_R^2}{\beta^2 + \mu^2\left(\kappa^2 + N_R^2\right)} \right]
\eeq

\noindent where $\beta = 1/\gamma\tau$ is the inverse of the cooling
time and $\mu=k_z/k$, with $k_z$ the vertical wavenumber, and $k^2
= k_R^2 + k_z^2$, where $k_R$ is the radial
wavenumber. 

The most unstable cooling time is 

\beq
\tau_{\rm max} = \frac{1}{\gamma}\left|\frac{k}{k_z}\right|
\frac{1}{\sqrt{\kappa^2 + N_R^2}}
\eeq 

\noindent with associated growth rate 

\beq
\sigma_{\rm max} = -\frac{1}{4}\left|\frac{k_z}{k}\right|
\frac{N_R^2}{\sqrt{\kappa^2 + N_R^2}}
\eeq 

\noindent The fastest growth rates occur for $k_z \gg k_R$, i.e., relatively flat and radially elongated epicycles{\footnote{ {In the local box we could set 
$k_R=0$ and obtain a channel flow. Yet, while insightful, channel flows are an artifact of local boxes, not possible in global disks 
where $\kappa$ is a function of radius. Also, setting $k_R=0$ violates the short wave approximation for obtaining the dispersion relation. It suffices to set $k_z \gg k_R$ to obtain the 
maximum growth rate. Interestingly, this contrasts with the dominant VSI mode, which is the very opposite, $k_R \gg k_z$, a tall slender mode. This should help future simulations prone 
to both instabilities to be able to distinguish between the processes.}}}. If we also assume Keplerian disks
($\kappa=\varOmega\gg |N_R|$), 

\beqn
\tau_{\rm max} &=& \frac{1}{\gamma\varOmega}\\
\sigma_{\rm max} &=& -\frac{N_R^2}{4\varOmega}.
\eeqn

Diffusion on the other hand is characterized by a random
walk of photons, with a characteristic length scale set by its 
Laplacian scaling.  \cite{Latter16} finds the same growth rate in the
diffusion regime, now with a length-dependent cooling time 
$\beta = \chi k_z^2$, where $\chi$ is the thermal
diffusivity. Maximum growth occurs for $\beta=\varOmega$, setting the
wavenumber of maximum growth $k_z=\sqrt{\chi/\varOmega}$, with same
growth rate as in the optically thin cooling law. 

We show in 
\fig{fig:figure3d} (left panel) the evolution of a 3D simulation of
convective overstability.  We set a box of 
size $4H\times 16H\times 2H$, with resolution $256\times256\times128$ in 
$x$, $y$, and $z$, respectively, including the linearized pressure
gradient. When the initial $k_z$ mode saturates (at 50
orbits), a sharp rise in enstrophy
occurs, developing into a large scale
 vortex. The saturated state is identified with the SBI. 

Some words are warranted about the robustness of the COV as
an astrophysical process relevant to disks.  The
condition that $N_R^2 < 0$ requires (for negative radial pressure gradient) 

\beq
q_\rho - \frac{q_T}{(\gamma-1)} < 0
\label{eq:reqN}
\eeq

\noindent where $q_\rho$ is the power law index of the density gradient and $q_T$ the power law index of the 
temperature gradient. For a column density following a power law 

 \beq
\varSigma\propto \rho H \propto r^{-q_\Sigma}
\eeq

\noindent we have for a Keplerian disk 

\beq
q_\Sigma = q_\rho +\frac{q_T}{2} -\frac{3}{2}
\eeq

\noindent The requirement of \eq{eq:reqN} is then 

\beq
2q_\Sigma < q_T  \frac{(\gamma+1)}{(\gamma-1)} -3 
\eeq

\noindent which, for $\gamma=7/5$ means 

\beq
q_\Sigma < 3\left(q_T-\frac{1}{2}\right)
\eeq 

\noindent For $q_T=1/2$ the column density has to be flat or
increasing with distance in order to lead to instability, which is
not reasonable. For $q_T=3/4$ the onset of instability in the midplane 
corresponds to $q_\Sigma=3/4$ (also for $\gamma=7/5$). While this is consistent 
with the range of $q_\Sigma\approx [0.4,1.0]$ found in the observations of \citet{Andrews+09}, 
 {it lies in the middle of the range and thus the COV may not be robust in the midplane.}

 {That does not mean that the COV is not present in general. The VSI does not formally exist in the 
midplane either, yet the unstable layers above and below are enough to drive the whole disk into a 
turbulent state as shown in numerical simulations. Indeed, for the COV, there exists the possibility of unstable 
radial entropy gradients away from the midplane, even when the midplane is stable. The entropy gradient is

\beq
\frac{\partial \ s(R,z)}{\partial \ln R} = \frac{\partial \ln T(R,z)}{\partial \ln R} - \left(\gamma-1\right) \frac{\partial \ln \rho (R,z)}{\partial \ln R}
\eeq

\noindent which, for the isothermal stratification given by \eq{eq:density-stratification}, becomes

\beq
\frac{\partial \ s(r,Z)}{\partial \ln R} = - q_T + \left(\gamma-1\right) \left[ q_\rho - \frac{z^2}{2H^2} \left(3-q_T\right) \right]
\eeq

For the MMSN model \citep{Hayashi81}, with $q_T=0.5$ and $q_\Sigma = 1.5$, then for $\gamma=1.4$ the 
entropy gradient, which is 0.6 at the midplane, flips sign and becomes negative at $z = \sqrt{1.2}H \sim H$. 
Whether this is near the midplane enough 
to render the whole disk column turbulent remains to be shown by numerical simulations.

Notice also that yet another entropy gradient is to be considered for the nonlinear sibling of the COV, the SBI. 
Being essentially a vertically global COV with $k_z=0$, the relevant quantities
to be checked for radial buoyancy are the column density gradient, and the vertically integrated 
pressure and entropy. In this case, the vertical integration of the polytropic equation of state passes from $p \propto \rho^{1+1/n}$
to $P \propto \Sigma^{1+1/\tilde{n}}$, with \citep{Goldreich+86}

\beq
\tilde{n} = n+\sfrac{1}{2}
\eeq

\noindent which is equivalent to an effective adiabatic index

\beq
\tilde{\gamma} = \frac{3\gamma-1}{\gamma+1}.
\eeq

\noindent and the condition of $N^2<0$ now becomes simply $q_T = (\tilde{\gamma} - 1) q_\varSigma$. We overplot in \fig{fig:stability} densities 
and temperature gradients for a sample of disks from \cite{AndrewsWilliams07}. According to this criterion 
we see that most disks in the sample should be susceptible to the SBI.

Finally,} we caution that although the COV exists in the local box, 
its existence has not yet been established in global disks. The different 
between these models is tied to what we just discussed previously. 
In a  global disk the entropy gradient is allowed to adjust to the flow, 
whereas in the local box it is hardcoded and enforced. Global simulations of the COV 
would be useful to resolve the issue. If the thermal diffusion is set by the turbulence itself, with $\chi = \alpha\varOmega H^2$, then the maximally growing wavelength of the instability is $\sqrt{\alpha}H$, or roughly 0.07$H$. For $H/R = 0.05 - 0.1$, between 1500 and 3000 points in the radial direction are needed to resolve the instability in global models.

\subsubsection{Finite Amplitude Initiated Turbulence: the Zombie Vortex Instability}
 {
Prior to the discovery that the MRI is a viable process to drive turbulence in protoplanetary disks,
researchers long struggled (in vain) to identify a purely hydrodynamic {\it{supercritical}} route to turbulence.
During the heyday of the MRI, many adherents of the hydrodynamic route began examining the possibility that 
a {\it {subcritical}} turbulence transition may be possible in protoplanetary disks \citep{Balbus_2003}.  
In this picture it is argued that while a Keplerian disk initiated with
a low-amplitude velocity field should quickly shear away incipient shear driven dynamic activity,
a relatively long-lived turbulent state could take root if the initial flow field were either of sufficiently high-amplitude
(i.e., with speeds approximately that of the local $c_s$) or seeded with perturbations
that result in strong transient growth which might trigger could trigger a secondary transition
in the flow \citep{Chagelishvili+2003,Yecko_2004}.
\footnote{ {{The idea of transient growth
is that certain small amplitude perturbations can result in large amplitude fluctuations in the perturbation energy in a strongly sheared flow.  This notion has been shown to explain the transition
to turbulence in viscous three-dimensional flows which are otherwise linearly stable \citep{Butler_Farrell_1992}.  For Keplerian disks, the
amplification factor in the perturbation energy, given by $G_{{\rm max}}$, was shown
in \cite{Yecko_2004} and \cite{Maretzke+2014} to have a Re$^{2/3}$ dependence.
For a comprehensive review of the ideas of transient growth in strongly sheared flows see \cite{Schmid_Henningson_2001}.}}}}

 {
\cite{Richard_Zahn_1999} examined the earliest
Taylor-Couette experiments -- those of Wendt's 1933 experiments report and repeated by G.I. Taylor
\citep{Taylor_1936} -- that are both stable to the  Taylor-Couette instability and
 support a flow profile exhibiting an increase of 
angular momentum with axial distance.  Interpreting this as reasonably mimicking a Keplerian flow and extrapolating the measured
laboratory torques (and corresponding effective turbulent viscosity) onto flows with the astronomical values of Re appropriate to protoplanetary disks, they argue that such disks should easily support a turbulent state owing to their small viscosities even with the smallest of initial triggers.  Follow-up 
laboratory work to test this proposition has been inconclusive with contradictory results reported \citep{Ji+06,Paoletti+12} and theoretical
analysis of the laboratory setting identify the source of instability as emanating from the axial boundaries of the experiment causing the laboratory measured outward transport of angular momentum \citep{Avila_2012}. 
{Subsequent higher Re laboratory studies \citep{Edlund_Ji_2014,Edlund_Ji_2015} and 
numerical experiment \citep{Lopez_Avila_2017} has shown that the observed turbulence increasingly
recedes and becomes localized to the axial caps of the system and is entirely the result of triggered Ekman boundary layer flow phenomenon.  These authors also show that as Re is increased,
the flow increasingly appears laminar and self-similar within the bulk interior of chamber.  The results of these efforts showcase the importance of doggedly assessing the character
of these systems at increasingly higher values of
Re as well as highlighting the dangers of deriving conclusions regarding the behavior of protoplanetary disks
based on relatively low Re quasi-Keplerian laboratory flows and their numerical experiments \citep{Balbus_2017}.}

\par
Nevertheless, the idea that finite amplitude perturbations may play a role in the life of a protoplanetary disk has lead to the discovery of several processes that 
can lead to dynamically interesting end-states like the Rossby Wave Instability \citep{Lovelace+99} and the so-called Subcritical
Baroclinic Instability \citep{KlahrBodenheimer03,Petersen+07a,LesurPapaloizou10,LyraKlahr11}.  Each of these processes appear to generally lead to the 
emergence of coherent, quasi-steady, and relatively
stable  vortex structures and
while these are interesting in their own right -- e.g., in their ability to attract and accumulate dust grains required
for planet assembly -- they themselves do not appear to play a role in driving 
hydrodynamic turbulence in the sense we envision.
\par
\medskip
The zombie vortex instability (ZVI) is a recently identified third hydrodynamical process
that has the potential to initiate widespread turbulence in protoplanetary disks. Unlike
the VSI and the COV, the ZVI is
a finite amplitude instability.  It differs from the aforementioned finite amplitude examinations in that it 
requires there to be a sufficiently large vorticity field as opposed to a relatively large amplitude velocity perturbation.}
\par
\begin{figure}
  \begin{center}
    \resizebox{.48\textwidth}{!}{\includegraphics{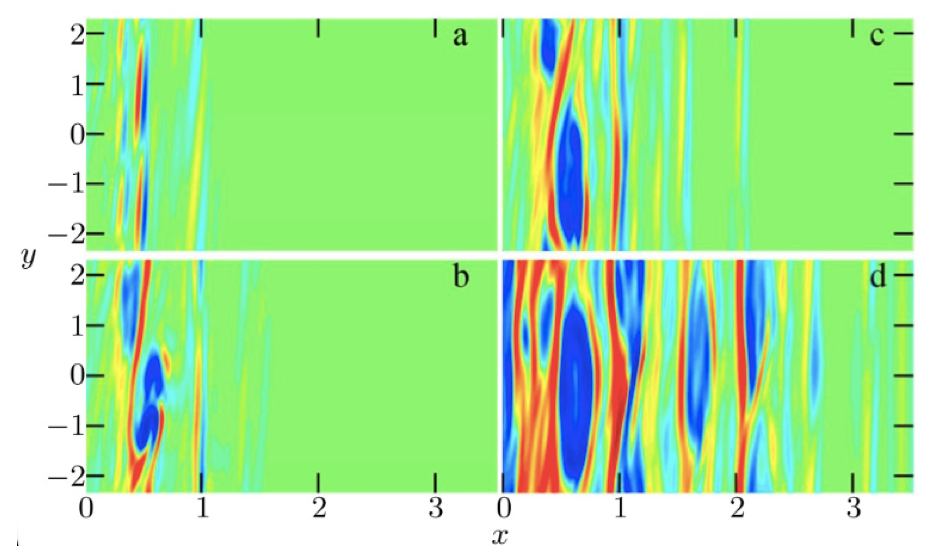}}
\end{center}
\caption[]{Cascade of baroclinic critical layers resulting in
  turbulence. Color coded is the Rossby number (gas
vorticity over Keplerian vorticity) in an $x–y$ plane
off the midplane ($z=0.4\Delta$). The simulation was initialized with
a vortex in the midplane, not visible at this height. The vortex
layers associated with baroclinic critical layers are seen at the
locations $x/\Delta = m$ for non-zero integer $m$. Anticyclonic
vorticity is indicated by blue, while cyclonic vorticity is red; the
darkest red/blue colors correspond to ${\rm Ro} =\pm 0.10$, while
green indicates {\rm Ro = 0}. (a) t = 64 orbits, (b) t = 256 orbits,
(c) t = 576 orbits, and (d) t = 2240 orbits. Reproduced from \cite{Marcus+13}.} 
\label{fig:zombies}
\end{figure}

Its 
mechanism of how a vortex column or sheet can render a flow unstable and induce spreading can be
 rationalized in a simple model, as shown in \fig{fig:zombies}, which is based on the original
findings of \citet{Marcus+13}. In this study, a tubular vortex filament is oriented
in the azimuthal direction in a constantly vertically stratified fluid in a 
local shearing box setting with a locally applied Keplerian shear. 
Azimuthal vibrations
of the vortex filament (Rossby waves) induce a response in the fluid
around it with an associated frequency. At some radial distance
from the filament, this Rossby wave frequency will 
appear Doppler shifted due to the Keplerian shear. This Doppler shifted 
frequency
will find resonance with a buoyancy frequency at some radial distance away in the disk,
amplifying the localized buoyant perturbations.   
The location and very narrow radial extent where this resonance operates is called
a "critical layer".
The amplified perturbation drives the generation of new filaments which
also turn into vortices.  In doing so, the newly generated filaments excite other buoyant resonances further on and, thereby, repeating
the process and setting the whole disk into a turbulent state
\citep{Marcus+13,Umurhan+16ZVI}
 \footnote{ The name of ``zombie'', a fictional undead creature resulting
from the reanimation of a corpse, is given because these vortices happen
in the Ohmic "dead" zones of protoplanetary disks. Because the process 
gives rise to new vortices by the excitation of
critical layers -- that in turn replicate the process -- the effect reminds
one of the spread of an infection. Another analogy would be that of tuning a guitar. Strike a note in a string that another string is tuned to, and
the resonance will make the unstruck string vibrate. An attempt to
rename the process ``Guitar String Instability'' will not be undertaken.}.
\par
There are minimum criteria for filaments to cause this kind of eruption.  The
first of these is that the Rossby number (Ro) of the anomalous vorticity must exceed
a critical threshold.  The Ro is assessed based on the velocity fluctuation, $v$,
around the base Keplerian state $\overline v$.  We are reminded that in a fluid shearing box rotating around the central star with angular frequency $\varOmega$, the Keplerian shear is $\overline v= -(3/2)\varOmega x$, where $x$ is the radial coordinate with respect to $r_0$.
If the radial extent of the filament is $\ell_0$ and we take $v$ to be the deviation
azimuthal velocity characterized by some speed scale $\delta v_0$, then
an estimate for Ro is
\beq
{{\rm {Ro}}} \equiv \left|\pderiv{v}{x}\right|\Big/2\varOmega_0 \approx \frac{\delta v}{2\varOmega \ell_0}. 
\eeq
Typically, the nature of the filament will be identified as ``cyclonic" if
${\rm sgn}\left(\partial v/\partial x\right)> 0$ and ``anti-cyclonic" otherwise (not
including zero).  \cite{Marcus+13} and\citet{Marcus+15} showed that filaments
must have $\Ro>\Ro_c \approx 0.2$ in order for the ZVI to be initiated.  
However, \cite{Umurhan+16ZVI} also showed that the ZVI may be instigated for
lower values of $\Ro_c$ approaching even zero.   What is important to note here is
that minimum requirement for the filament's vorticity must have its $\Ro>\Ro_c$, so 
it implies that {\emph{there is no minimum perturbation velocity needed}} to get the instability to go. Only the filament's vorticity matters since a radially-thin azimuthally elongated filament with velocity scale $\delta v_0$ can satisfy the minimum vorticity requirement so long as $\delta v_0/\ell_0$ remains finite
as $\ell_0 \rightarrow 0$ (see also discussion in \sect{sect:2D_turbulence}).
 
 \begin{figure}
  \begin{center}
    \resizebox{0.45\textwidth}{!}{\includegraphics{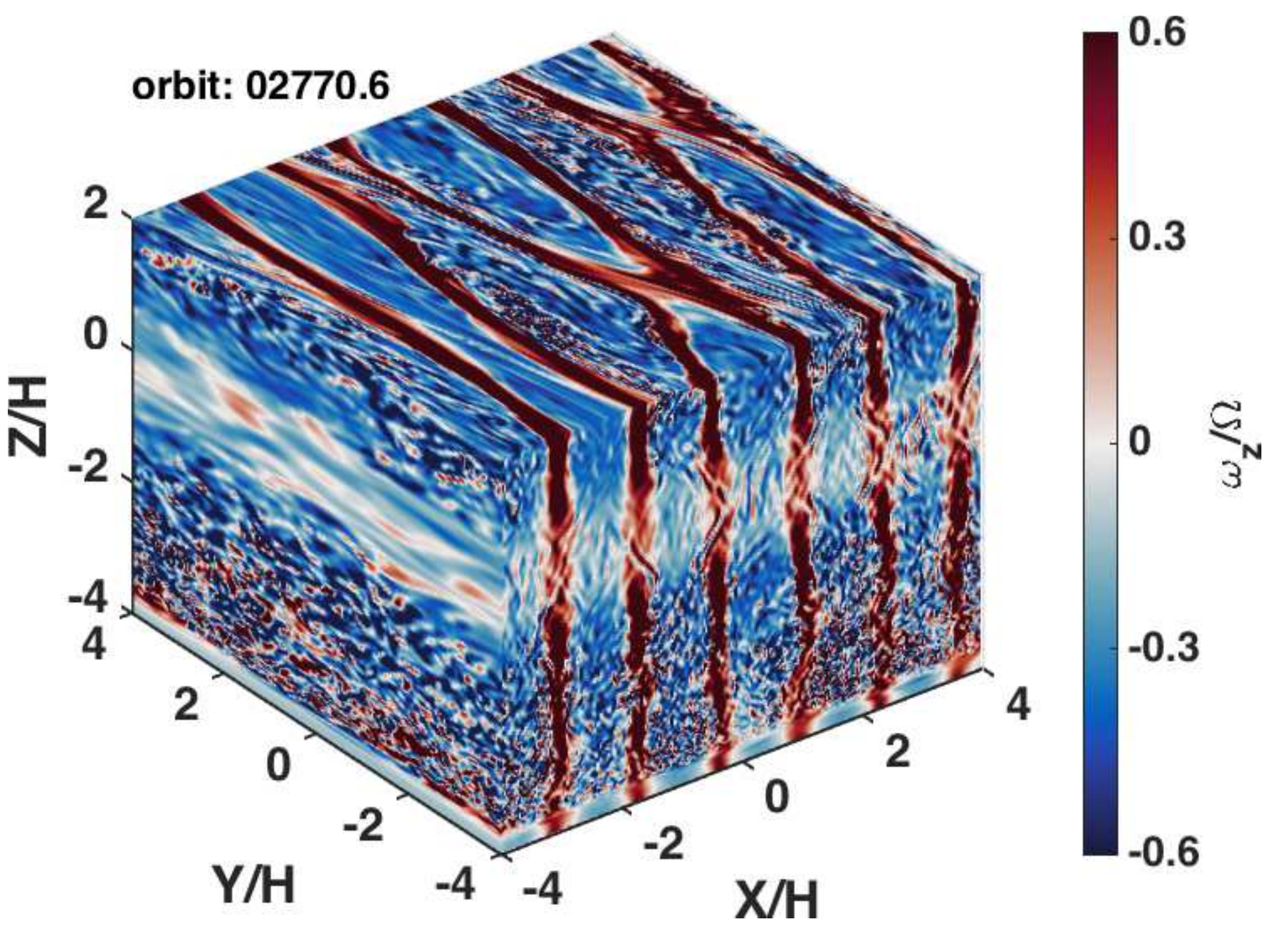}}
\end{center}
\caption[]{Anomalous vertical vorticity profile from fully developed ZVI simulations. 
Notice the banded structure in vertical vorticity, implying weak undulations atop
the basic Keplerian profile.
Reproduced from \citet{Barranco+18}.} 
\label{fig:FullyDevelopedZVI}
\end{figure} 
\par
Because the ZVI relies on excitation of baroclinic critical
layers,
its growth times are relatively long once $\Ro>\Ro_c$, 
with growth rates of $< 0.05-0.1 \varOmega^{-1}$.
The ZVI operates most effectively in the limit of very long cooling times as
demonstrated by
\citet{LesurLatter16}, who examined  the ZVI under a variety
of thermal cooling times (whether driven by optically thin or optically
thick cooling) and showed that the critical cooling time, $\tau_r$, to be longer than 10
orbits in order for the ZVI to be dynamically significant.  This corresponds
to approximately a minimum gas opacity of $>50{\rm cm}^2/{\rm g}$.
This can be rationalized since the ZVI resonance mechanism requires
buoyancy oscillations to be
maintained over long timescales. 
For this reason, the second criterion is that the process requires
the gas to be close to adiabatic, with very long cooling times. 
\par
The spawned 
vorticity will have an anomalous vorticity that is jet-like \citep{Umurhan+16ZVI}, i.e., a side-by-side pair of vortex filaments of alternating signed vorticity. The radial size of this anomalous jet is the size of the critical layer and, furthermore, it is the anticyclonic part of this jet that undergoes secondary transition.
\par
\begin{figure}
  \begin{center}
    \resizebox{0.9\columnwidth}{!}{\includegraphics{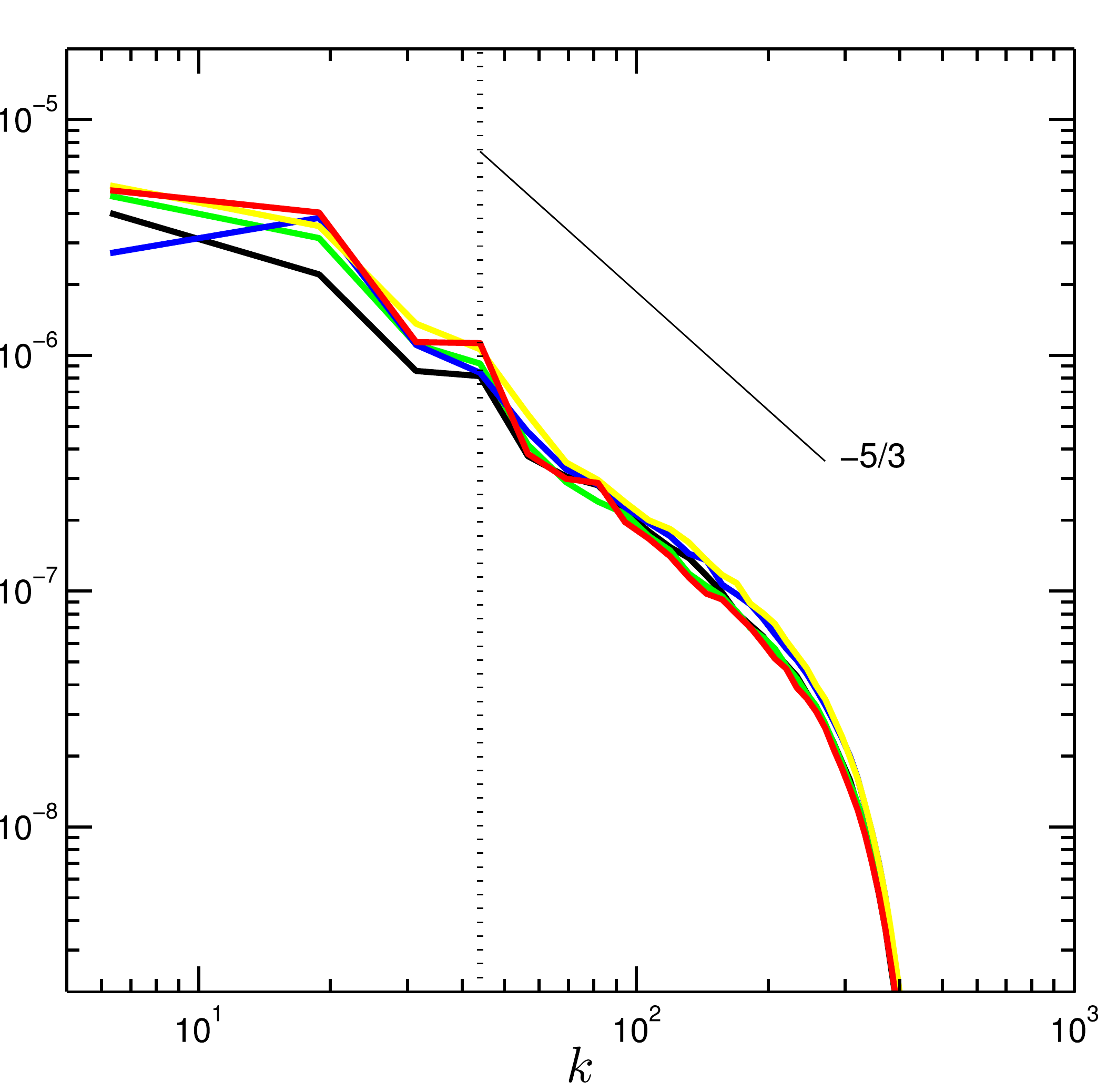}}
\end{center}
\caption[]{ {Energy spectra of the non-Keplerian component of kinetic energy in ZVI turbulence, based on simulations in the shearing box
whose box size $=H$.  The different lines represent different initial conditions: Kolmogorov noise with
initial rms Mach number of 0.01 (black); Kolmogorov noise with initial rms
Mach number of 0.007 (green); Kolmogorov noise with initial rms Mach
number of 0.004 (blue); noise with an energy spectral index of 1.0 and initial
rms Mach number of 0.007 (red); and a run initialized with an isolated anticyclonic vortex (yellow). 
At late-time flows are all attracted to the same energy spectrum, regardless of the initial conditions. 
Vertical lines correspond to scales $\approx 0.1 H$ which marks
the beginning of the $k^{-5/3}$ spectrum. For $80 < k < 300$, the spectra are approximately Kolmogorov. 
Based on reported values of $\alpha \sim 4 \times 10^{-4}$ and $h = 0.05$, 
a nominal value of the Rhines scale of $\lambda_b = \pi H$, based on \eq{eq:disk_Rhines}, although
see text for caveats.}
Reproduced
  from \citet{Marcus+16}.}
\label{fig:ZVI_spectrum}
\end{figure}

\citet{Marcus+15,Marcus+16} showed that the ZVI leads to widespread turbulence
in realistic shearing box models in which the vertical component of gravity shows the proper
vertical dependence $g\sim-\varOmega^2 z$. These studies show that the ZVI appears at locations away from the midplane where the stratification is strong and subsequently spreads down toward the midplane, suggesting that the process can operate even when the stratification is weak, which is consistent with predictions made in analytical work \citep{Umurhan+16ZVI}. 
\citet{Marcus+16} also show that the saturated state has a turbulent spectrum ${\cal E} \sim k^{-5/3}$ suggesting that a downscale direct energy cascade has been resolved in these
simulations (\fig{fig:ZVI_spectrum}). We note that the starting point of the inertial spectrum appears to correspond to the critical layer scaling (see below).  
The emergence of large-scale organized zonal flows (see below) suggests
that a subtle interplay of both processes may be at play.  \par
The injection scales are those of the critical layers.  Unlike in purely hydrodynamic shear flows, where the size of the critical layer is set by the viscosity of the gas and scales
like Re$^{-1/3}$, the size of the critical layers ($\Delta_c$) driving the ZVI is set by the intrinsic linear growth rate $\sigma$  with an approximate
scaling given by $\Delta_c \approx \sqrt{N/\varOmega}\cdot\sqrt{\sigma/\varOmega} k_z^{-1}$ where $k_z$ is the vertical wavenumber of the fastest growing mode expressed in inverse units of $H$ \citep{Umurhan+16ZVI}.  As the growth rate depends upon the vorticity of the triggering filament and the Brunt-V\"ais\"al\"a frequency, it is difficult to assess generalized expected length scale for $\Delta$.  Results of published simulations,
together with these analytical considerations, suggest that $\Delta_c\sim 0.1 H$ or less.
\par

\begin{figure*}
  \begin{center}
    \resizebox{0.95\textwidth}{!}{\includegraphics{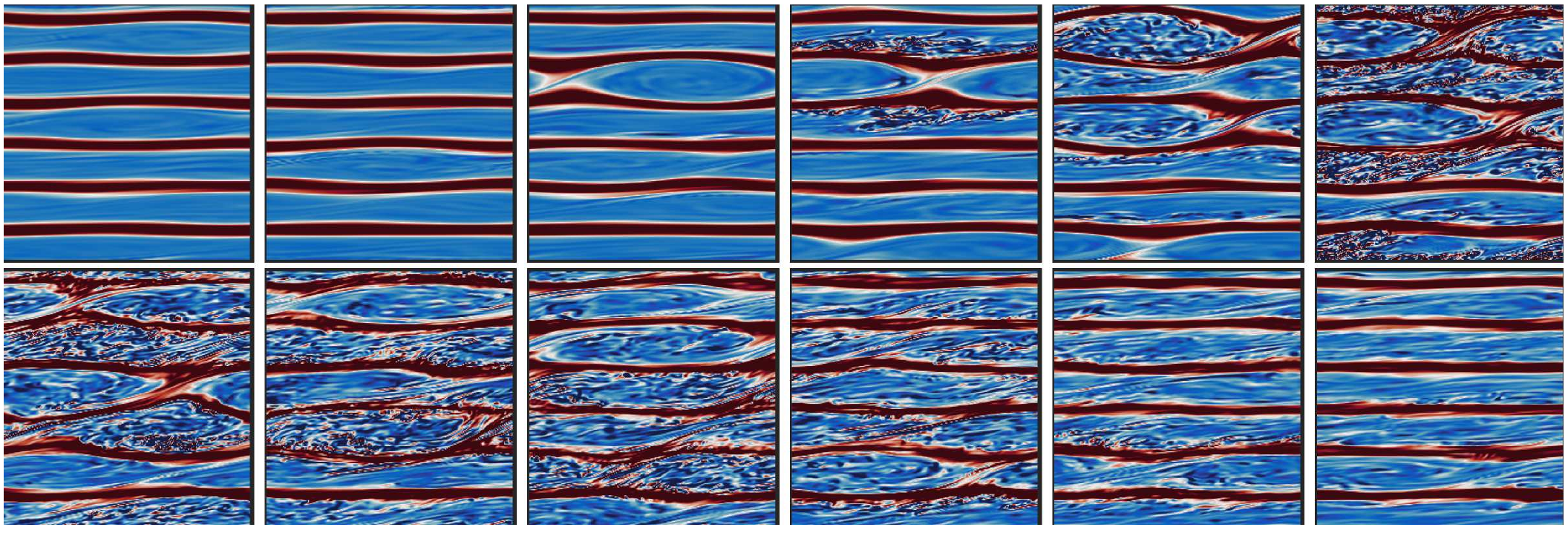}}
\end{center}
\caption[]{ZVI burst cycle shown as radial-azimuthal cuts at 2 scale heights of the 
anomalous vertical vorticity.  Each successive panel corresponds to 3.5 orbit times.  
Banded structure undergoes a breakdown, triggering critical layer excitations along the
way.  Banded structure restored after 50 orbit times.  Complete cycle is approximately 100 orbit times. 
Reproduced from \citet{Barranco+18}.} 
\label{fig:ZVI_Burst_Cycle}
\end{figure*}

\cite{LesurLatter16} also find that
the saturated state of the ZVI is similar to that of the COV and VSI,
with anticyclonic vortices and cyclonic sheets in the largest
scales. These authors conclude that the small scales appear 
characterized by more isotropic turbulence.
\citet{Barranco+18} examine similar behavior but they also show something
quite remarkable.  They show that long-time simulations of the ZVI leads to the emergence of large scale, axisymmetric, radially periodic zonal flow.  Each successive zonal flow sheet exhibits periodic cyclonic and anti-cyclonic behavior with a radial wavelength of about 1 scale height. Moreover, this periodic
flow exhibits {\emph{temporal intermittency}} in which the anti-cyclonic portion of the zonal flow periodically breaks-down after 50-100 orbit times, but eventually the flow reorganizes into the zonal flow again -- presumably driven by the small scale ZVI itself (\fig{fig:ZVI_Burst_Cycle}). 
 {Might this be an indication of an inverse cascade of energy and its transfer onto large scale zonal flows as might be expected by the Rhines mechanism?  This remains to be fully explored.}
\par
Is this to be expected for protoplanetary disks?  
As we stated in the introduction of this section, the ZVI is not a linear instability in the usual sense, but
rather a finite-amplitude process. It requires
a seed vortex filament (or a spatially distributed spectrum)
to launch the process with the vorticity of the filament being an order 1 fraction of the Keplerian flow itself. Generating such an initial
finite amplitude state in order to spawn the ZVI cascade/spread
was raised by \cite{Umurhan+16ZVI} and \cite{LesurLatter16} 
as a concern that limits the applicability of the ZVI in actual
disks.  \citet{Marcus+16} address this matter stating an initial flow
state structured with an energy spectrum $\sim k^{-a}$, with $1<a<3$, will always
support a small scale exhibiting anomalous vorticity satisfying the minimum requirement.  For example,
following our discussion in \sect{sect:3D_Kolmogorov},
a Kolmogorov spectrum ($a=5/3$) has an anomalous velocity $\delta v \sim \epsilon^{1/3}k^{-1/3}$. Its corresponding vorticity scale would be $\omega \sim \delta v/\ell = \delta v \cdot k \sim \epsilon^{1/3} k^{2/3}$.  Thus, even for a weak initial Kolmogorov energy spectrum
${\cal E}_0(k)$,
-- whose total integrated energy is controlled by $\epsilon^{1/3}$ -- there always exists a large enough value of $k_0$
 in which $\omega_0$ satisfies the minimum criterion -- provided, of course, $k_0<k_{diss}
 = 2\pi/\ell_{diss}$.
 \footnote{We note that a spectrum of the sort envisioned contains disturbances of every shape for any given value of $k$.  It is that part of the flow field that gives rise to filaments with little azimuthal structure that would trigger the instability.}
\par
Whether or not this process can be self-sustaining and lead to
a turbulent state under a wide umbrella of conditions appropriate to
the Ohmic zone remains to be determined. It would seem that its
self-sustainability depends centrally on what way the original
disturbances are structured and how they arrive. 
Although \cite{Marcus+16} find that
the critical finite amplitude in the velocity field to be very small (${\rm Ma} \approx
10^{-6}$), sustaining a coherent perturbation for long enough times
may still be of concern. However keeping in mind the foregoing discussion,
the minimum velocity scale quoted in \cite{Marcus+16} is a value inferred based on the limited
resolution of their simulations and is probably indicative of a maximum lower
limit because higher resolved simulations would permit the introduction 
of radially narrower filaments that both satisfy the requisite anomalous vorticity
criterion while exhibiting even
smaller values of Ma.
Also, \cite{LesurLatter16} and Marcus and co-workers find the instability
numerically when hyperdiffusion is used - hyperdiffusion is an instance
of Large Eddy Simulation (LES, see further below).
When Laplacian viscosity
is used, \cite{LesurLatter16} find that the Reynolds numbers must significantly exceed $10^7$ in order
for the process to lead to sustained disorganized activity.  
While the Re for disks easily satisfies this criterion, it still remains to be understood how the resonance mechanism can 
sustainably extract energy from the critical layer and what might be the overall structure of the
disk after ZVI turbulence has acted on it.
\par
Where might the ZVI be active?  There are several locations in the disk
that, if seeded with suitable initial vorticity field, may erupt
into ZVI turbulence (see further in \sect{sect:synthesis}). Another
idea we conjecture is that ZVI may emerge perhaps near the boundary
of magnetically active regions or near disk zones with pressure bumps, i.e., locations that 
can support radially
localized Rossby waves.  
For example, simulations of protoplanetary disks 
indicate that strong zonal flow features appear at the boundary
separating MHD active zones from purely hydrodynamic zones \citep{VarniereTagger06,LyraMacLow12,Lyra+15}. 
Such zonal flows support Rossby waves.  Since
the ZVI mechanism requires there
to be a Rossby wave to resonate with a buoyancy oscillation somewhere in the hydrodynamic
regions of the fluid, 
the zonal flow characterizing this transition region might act like a large amplitude loudspeaker
by providing the energy -- in the form of the parent Rossby wave -- to set off
the eruption that could lead to ZVI turbulence spreading deep into the Ohmic zone of the disk.  

\section{Synthesis : A butcher diagram for hydro instabilities}
\label{sect:synthesis}

\begin{table*}
\caption[]{Hydrodynamical instabilities summary characteristics.}
\label{table:thermal}
\begin{center}

\begin{tabular}{lccccccc}\hline
Instability                         & Violation of                        & Mechanism Type                      & Linear growth           & Length scale      & Opacity & Thermal & $\alpha$ \\
                                         & Rayleigh criterion              &                                                 &  rate                          & of linear growth &$\kappa$ (\sfrac{cm$^2$}{g})&time $(\varOmega\tau)$ &            \\\hline
Vertical Shear   & $d\varOmega/dz \neq 0$ & Angular momentum exchange & $\sqrt{m}|q| h \varOmega/4$& $\pi|q|hH$ & $<$ 1 & $ \ll 1$& $10^{-4}-10^{-3}$\\
                        &                                          & between adjacent elements.      & && & & \\
Convective       & $N_R^2<0$                          & Buoyant amplification              & $|N^2|/4\varOmega$& $\sqrt{\chi/\varOmega}$& $1-50$& $\sim 1$& $10^{-4}-10^{-3}$\\
                        &                                          & of epicyclic oscillations.           & & & & & \\
Zombie Vortex & $N_z^2>0$                  & Resonance between Rossby   & -- & -- & $>$ 50&  $ \gg 1$ & $10^{-4}-10^{-3}$\\
                        &                                         & and buoyancy frequency.          & & & &   & \\\hline
\end{tabular}

\end{center}
\end{table*}

The three hydrodynamical instabilities operate optimally in different cooling
times, thus existing in very different regimes of opacity. In the
adiabatic case ($\varOmega\tau \gg 1$) the ZVI dominates. In the
isothermal case ($\varOmega\tau \ll 1$), the VSI dominates. In between
these extrema, the COV operates ($\varOmega\tau \sim 1$). 
\par
 {
For what material conditions do these cooling constraints correspond?  
Reviewing the expression for the optically thick limit for thermal relaxation,
i.e., the $\ell_{rad} \ll \ell$
expression of Eq. (\ref{eq:relaxationtimes}), it is evident that it depends upon
the effective material opacities, density, temperature and length scale of perturbation, $\ell$.
\cite{Nelson_etal_2013} examined cooling times to assess these conditions assuming values of $\ell \sim 0.1 H$ (being
typical of the fastest growing mode of the VSI) 
together with
values of the expected midplane densities and radial temperature profiles based on the canonical 
flared minimum mass solar nebula disk models of \cite{ChiangGoldreich97}.  Such disk models
predict midplane densities and radial temperature profiles that are
\beqn
& & \rho_{\rm mid} \approx 2.7 \times 10^{-9} F (R/{\rm AU})^{-39/14} \ {\rm g/cm}^3, \\
& & T = 120 (R/AU)^{-3/7} \ {\rm K}, 
\eeqn
respectively, where $F$ is the relative mass of the disk in units of minimum solar mass models.  The corresponding
cooling times are
\beq
\varOmega\tau/2\pi = 168 F^2 \left(\frac{\kappa_R}{{\rm cm}^2/{\rm g}}\right) \left(\frac{\ell}{R}\right)^2 
\left(\frac{R}{20 {\rm AU}}\right)^{-53/14}.
\eeq
The onset of the VSI for regions at around 10 AU around a 1 M$_\odot$ star with $F=1$ requires opacities of
$\kappa < 1 {\rm cm}^2/{\rm g}$.\footnote{Note that the opacity is the effective opacity of a fluid mixture
of gas and dust.  Typically minimum mass solar nebula models assume a 1 percent content of dust grains.}
Since the ZVI also has growth on similar scales the corresponding requirements suggest
$\kappa > 50 {\rm cm}^2/{\rm g}$
for the ZVI to be minimally operative.
Even though the COV operates at larger scales, a reasonable estimate of the
opacity is $1 < \kappa < 50 {\rm
  cm}^2/{\rm g}$.}

\begin{figure*}
  \begin{center}
    \resizebox{.9\textwidth}{!}{\includegraphics{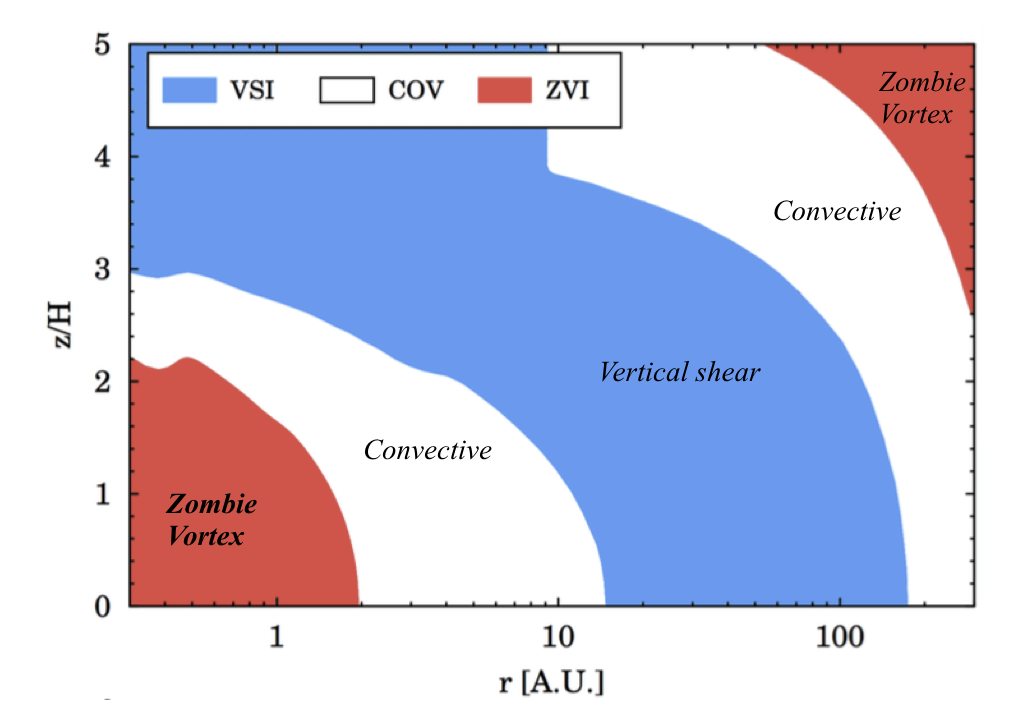}}
\end{center}
\caption[]{A ``butcher diagram'' for hydrodynamical
  instabilities. Disk opacities define cooling times that, in turn,
  determine which hydrodynamical instability process is dominant. The ZVI
  dominates for long cooling times, happening at the optically thick
  inner disk. The COV requires moderate cooling times, happening in
  the 10\,AU region. The VSI happens where the disk is more
  isothermal. As one moves outwards, the disk becomes more optically
  thin and thus the cooling time increases again, leading to a new
  zone of COV and eventually another adiabatic region for the
  ZVI. The regions where ZVI dominate may be too ionized and thus
  prone to MRI (in the inner disk) or magnetocentrifugal winds. Adapted from \cite{Malygin+17}. 
  The different instabilities have different saturation mechanisms: the VSI saturates via RWI, the COV via SBI, 
    and the ZVI via itself and RWI.} 
\label{fig:butcher}
\end{figure*}

The cooling time requirement means that the ZVI will be more important
in the very inner disk (inwards of 1\,AU) , {\bf where the gas} is more adiabatic;
the COV in the region 1-10\,AU; and the VSI in the outer disk beyond
10\,AU which tends to be more isothermal. As we go further outwards,
the very outer disk beyond 100\,AU starts to become too optically thin
and the cooling time goes up again, being unstable to COV. Eventually,
if the disk is too optically thin and the cooling time long enough,
the disk will be unstable to ZVI. That invites a mapping of
instabilities that we call a ``butcher diagram''
\footnote{Inspired by diagrams delineating various cuts of beef.}
, in
\fig{fig:butcher}. 

\begin{figure*}
  \begin{center}
    \resizebox{.33\textwidth}{!}{\includegraphics{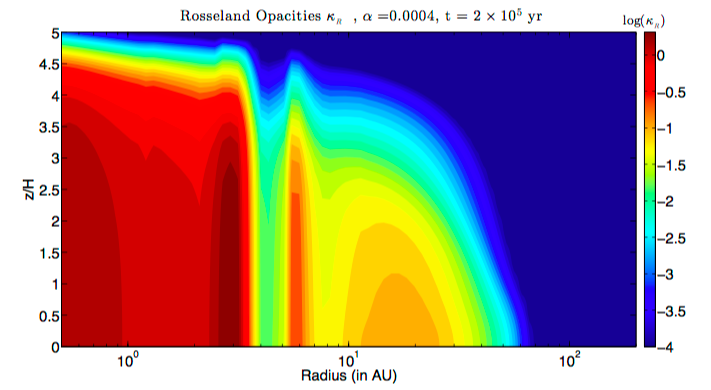}}
    \resizebox{.33\textwidth}{!}{\includegraphics{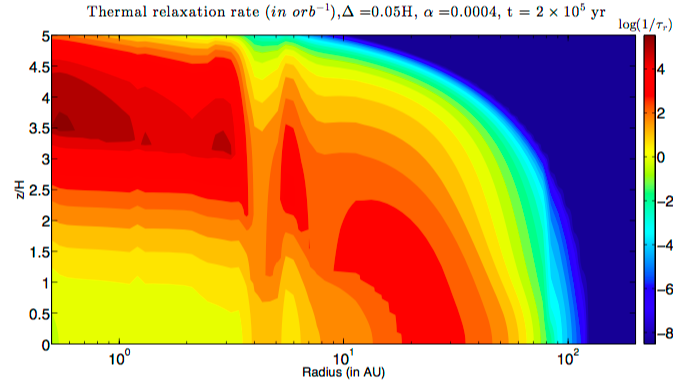}}
    \resizebox{.33\textwidth}{!}{\includegraphics{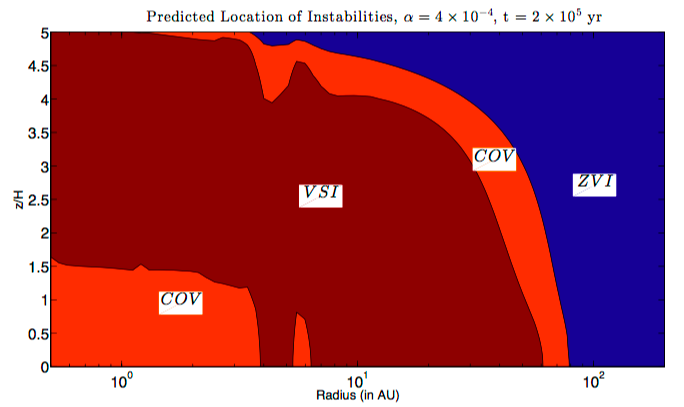}}
\end{center}
\caption[]{Thermal properties of global alpha-disk models.  Left: Rosseland mean opacities after $2\times 10^5$ yr. Opacities quoted in units of cm$^2$/g. Middle: Derived thermal relaxation times after $2\times 10^5$ yr. Cooling time $\tau$ in unit of local orbit times around a 1 $M_\odot$ star. Right: A butcher diagram indicating locations where various instabilities are operating based on output of \cite{Estrada+16}: $2\times 10^5$ years, $\alpha =4\times 10^{-4}$ and dust porosities of zero.  Adapted from \citet{Umurhan+2017}.} 
\label{fig:butcher2}
\end{figure*}

The actual mapping of the instabilities depends on translating gas densities, dust densities, 
and temperatures, to opacities and thence to optical thicknesses and cooling times. As such, it will 
depend heavily on the underlying disk model assumed. The specific map shown in \fig{fig:butcher} is taken from 
\cite{Malygin+17}, and uses the axisymmetric disk model of \cite{Hartmann+98}, also used in \cite{Dzyurkevich+13}, 

\beq
\varSigma = \varSigma_c \left(\frac{r}{r_c}\right)^{-q_\Sigma} {\rm exp}\left[-\left(\frac{r}{r_c}\right)^{2-q_\Sigma} \right]
\eeq

\noindent with 

\beq
  \varSigma_c = (2-q_\Sigma) \frac{M_{\rm disk}}{2\pi r^2_c}
\eeq

The stellar mass is $M_\star = 1 M_\odot$, the reference column density is $\varSigma_c = 1700$ g\,cm$^{-2}$ 
and the power laws is taken as the median of \cite{Andrews+09}, $q_\Sigma=0.9$. The cutoff radius is $r_c=40$\,AU. 
This profile yields a good match to both CO and continuum observations \citep{Hughes+08}, and matches the minimum 
mass solar nebula up to 100\,AU. The temperature chosen is that of a passive disk \citep{ChiangGoldreich97}

\beq
T = T_0 \left(\frac{1\,{\rm AU}}{R}\right)^{1/2}
\eeq

\noindent with $T_0 = 280$\,K. Vertical temperature gradients are ignored, and a lower limit of 10\,K 
set beyond 250\,AU. Using the dust opacities from \cite{Semenov+03} and a well-mixed dust at 
the dust-to-gas ratio of $10^{-2}$, one obtains the mapping in \fig{fig:butcher} for the appropriate cooling times 
described in the text. 
 {Ignoring vertical temperature gradients is valid for passive-disks \citep{ChiangGoldreich97}.  However, disks
in which there is some amount of viscous heating will show potentially strong deviations from
nearly uniform and this is expected to be the case if the turbulence is relatively strong $\alpha > 10^{-2}$ like, for example, the global disk evolution models considered in \cite{Bitsch+2015}.  For the kinds of hydrodynamic instabilities discussed here, wherein  $\alpha < 10^{-3}$,
viscous heating is expected to be a secondary effect.}

Using the coagulation/fragmentation and dust evolution model of \cite{Estrada+16}, and assuming equal Rosseland mean and Planck mean opacities, the temperature profiles are largely locally isothermal (with midplane to lid-temperature variations of less than 1\,K). The radial temperature dependence is given by $T \sim R^{-3/7}$ for disk radii $>3$\,AU, typical of flared disks and consistent with theoretical predictions \citep{ChiangGoldreich97}.  \citet{Umurhan+2017} displays the effective dust opacities in a disk experiencing some amount of alpha-disk turbulence according to the disk evolution model of \citet{Estrada+16}.  The model compiled distributes dust particles in vertical height according to the turbulent intensity ($\alpha \sim 4\times 10^{-4}$) and the dust size which together yield a turbulent Stokes number.  \citet{Estrada+16} 
determine the size of particles as a function of disk radius and epoch.
The left panel of \fig{fig:butcher2} shows a profile of the internal opacities at a model after $t = 2 \times 10^5$ years of dust evolution. The thermal relaxation times can be very roughly represented as either dominated by radiative diffusion (optically thick) or radiative cooling (optically thin) regimes and are given by \eq{eq:relaxationtimes}. For optically thick conditions, the middle panel of \fig{fig:butcher2} shows $\tau$ for fluid structures
with scales $\ell = 0.05H$, typical of the VSI. The right panel of \fig{fig:butcher2} displays a butcher diagram of the instabilities considering these thermal relaxation times. The VSI dominates the turbulence in the range 3-90 AU while the COV occurs generally in the inner disk (<3 AU). The ZVI is prominent in outer solar nebula ($>$100 AU). This particular combination of gas and dust model never becomes optically thick enough for the ZVI in the inner disk. 

We stress that at the time of this review no model combining all the three (or even two) of the instabilities has been attempted. Therefore, the 
boundaries between the instabilities in the butcher diagram are ill-defined. As long as the entropy gradient is negative, the COV has the 
least stringent thermal constrains, effectively existing for all finite cooling times. The VSI, on the other hand, seems to be the most robust in respect to the violation 
of the Solberg-H{\o}iland criterion, as stellar irradiation should always enforce some amount of vertical shear. The ZVI seems to be less robust, as it requires a 
``prime mover'' in the form of a sufficiently large finite-amplitude disturbances
to be emplaced inside the dead zone. However, we conjecture that this initial finite-amplitude perturbation should be present in the RWI-unstable transition between the MRI zone and the Ohmic zone in the inner disk, where the thermal conditions are also adiabatic.  {We note, although, that the ZVI, being present in more adiabatic conditions, will probably be self-regulated at small $\alpha$ values. At the high densities of the optically thick depths of the disk, modest amounts of accretional stresses, perhaps as little as $\alpha\sim10^{-5}$ are sufficient to generate enough heat to bring the 
stratification to a profile that shuts down the ZVI. \table{table:thermal} summarizes the parameters of the instabilities. }

\section{Near future goals for numerical simulations of disk turbulence}
\label{sect:future} 
The near and long term goals of numerical modeling of turbulence in the protoplanetary disk
is to establish the conditions on which dust particles grow into planetesimals on a 0.5Ma to 2.3Ma time frame after the appearance of Calcium Aluminium Inclusions (CAI's).  
This means properly characterizing the turbulent velocity field on scales where the 
dust particles accumulate.  From our standpoint it is incumbent on 
the fluid dynamicists and theorists to establish, with some confidence, 
the quality of the energy spectrum at
the relevant scales and the first step toward this is identifying and quantifying the 
shape of the turbulent energy spectrum in the inertial regime.  If the resulting turbulent
state is both steady and the turbulence in the inertial range is indeed self-similar, 
then upon its resolution the inertial
spectrum could be extrapolated down to the smaller scales where dust collection is thought
to be actively taking place. \par
A case that illustrates the path forward is the state of our understanding of the VSI. 
Simulations conducted thus far have been only moderately resolved.  For example, in the original
simulations conducted by \citet{Nelson_etal_2013} where the radial domain was 
a radial annulus from $R=R_0$ to $R=2R_0$, the conditions of the disk were such that
the fastest growing mode had a wavelength of $\ell_0=0.008 R_0$.  Note that on these scales, the pressure scale height is $\sim 0.05 R_0$.  These simulations resolved this fastest growing mode with about 15 grid points.  With $\ell_0$ being nominally the injection scale, in order to begin seeing the emergence of an inertial regime requires at least 10-100 times the resolution achieved in those simulations.  For the kinds of global scale simulations currently feasible, such high resolution studies is still quite prohibitive.\par
 One way around this is to consider very narrow global simulations or appeal to the
 often maligned shearing box setting \citep{Goldreich_Lynden-Bell_1965,UmurhanRegev04,Latter_Pap_2017}, which represents a model facsimile of a box section of a disk orbiting the central star at
 the reference radius $R=R_0$.  The spatial scales of the box are usually in units of the local scale height so, in principle, higher resolution simulations are possible.  For example, simulations of the ZVI are performed in this model setup and, as the results of \citet{Marcus+16} shows, they were able to resolve a Kolmogorov like spectrum.  In the context of the VSI, for example, adjustments to the formulation of the shearing box equations that are suitable for the mechanism may be needed, like the composite local-global model suggested by \citet{McNally_Pessah_2015}.
 If a shearing box model is used then the global scale variations that give rise to each of the three
 instabilities must be carefully represented within the confines of these models. For example, the global scale variation of the temperature field must be faithfully represented on the small
 scales in just the right way so that the instability is triggered.  Careful thought must also
 be given to the choice of boundary conditions adopted, making sure that they do not contaminate the results obtained. 
 \par

 Because the molecular viscosity of H$_2$ is so small, it is practically impossible at this
 time to conduct direct numerical simulations (DNS) that resolve the flow down to the dissipation scales.
 Since energy cascades toward small scales in fully turbulent supercritical 3D flows
 \footnote{...and if 2D flow, then enstrophy cascades toward small scales.},
 unless the natural dissipation scales are resolved in flow modeling, then power can build up at the grid scales of a simulation leading to unrealistic and, often times, numerically unstable situations.
 Large Eddy Simulation (LES, \citet{Davidson_2004}) is one way to confront this problem.
 An LES simulation will introduce a high order diffusion operator, like $\nu_{4}\nabla^4$ or $\nu_{8}\nabla^8$
 or even $\nu_{16}\nabla^{16}$, into the equations of momentum conservation.  The overall strength of the dissipation is controlled by the "hyperdiffusion" coefficient $\nu_n$, where $n$ is the order of the Laplacian operator.  With a suitably chosen value of the coefficient 
$\nu_n$ a hyperdiffusion operator permits the normal cascade of energy through the inertial regime of a turbulent flow but will increasingly dissipate power as the grid scales are approached in a simulation.  This is intended to mimic the continued
transference of power to the unresolved dissipation scales by artificially
 dissipating the energy flux at the resolved scale of the simulation.  This is usually 
a delicate procedure requiring careful tuning of $\nu_n$.  LES, or any other equivalent small-scale energy dissipation procedure, is an essential tool for numerical
modeling of protoplanetary disk turbulence
that will be necessary for the foreseeable future.

\section{Conclusion}
\label{sect:conclusions}

The past decade of disk research have seen the emergency of disk
instability mechanisms that provide hydrodynamical turbulence in the
Ohmic dead zone. They violate the Rayleigh criterion in different ways. The
vertical shear instability (VSI) by having $d\varOmega/dz \neq 0$, the
convective overstability (COV) by the presence of entropy gradients with
$N_R^2<0$, and the zombie vortex instability (ZVI) with $N_z^2 > 0$. 
Because of these requirements, they exist in very different regimes of opacity: 
the ZVI in adiabatic regions, the VSI in isothermal ones, and the COV in between. 
These can be mapped into different locations, depending on the particular disk model 
(\fig{fig:butcher} and \fig{fig:butcher2}). The instabilities have different turbulent responses, with $\alpha$ values
around $10^{-3}$. Their properties are summarized in \table{table:thermal}.

Of the three processes, the most robust seems to be the VSI, as the unstable entropy gradients
necessary for the COV are  {not necessarily} realized in all disks, and the first
perturbation necessary for the ZVI is difficult to maintain over long
times. The three instabilities saturate into three-dimensional vortices. The VSI because
of RWI and Kelvin-Helmholtz instability of the thin tall sheets. The COV
because it builds up the finite difference perturbations that lead to
subcritical baroclinic instability. The ZVI also via
RWI and Kelvin-Helmholtz instability. All vortices decay due to elliptical
instability in the cores \citep{Pierrehumbert86,Bayly86,LesurPapaloizou09}, 
that leads to a direct enstrophy cascade.  { In saturation, vortex strength 
should be defined by a balance between the saturation mechanism, 
that feeds the vortices (RWI for VSI/ZVI, SBI for COV), and the rate of 
decay via elliptic instability \citep{Lyra13}. Weaker vortices (of larger aspect ratio) linger for 
longer because the growth rates of elliptic instability decrease with vortex aspect ratio.}

In this review, we have specifically used the term {\it Ohmic zone} instead of the 
traditional ``dead'' zone.  We could have used ``hydrodynamical'' zone since Ohmic effects, 
in the sense of magnetic Reynolds numbers close to 1, are not at play. The Ohmic zone is 
deep into the resistive regime and any magnetic effect is irrelevant. Yet, we keep the name 
Ohmic for juxtaposition to the other non-ideal MHD effects: the Hall effect and ambipolar 
diffusion, that may be at play in very low density regions of the disk \citep{Wardle99,Wardle07,Lesur+14}. 
The Hall effect leads to the Hall shear instability if the magnetic field and the angular momentum vectors 
are aligned. Ambipolar diffusion leads to magnetocentrifugal winds, evidence for which may have been 
recently discovered \citep{Simon+16}. The very high up region of the 
butcher diagram (\fig{fig:butcher})  {are probably prone to ambipolar diffusion, 
providing a ``lid'' for all instabilities; if the ionization fraction is high enough 
the MRI can be reactivated. The outer ZVI region may thus be completely quenched, by ambipolar diffusion or the MRI itself. 
The outer COV region is low density enough for ambipolar diffusion to 
operate. The VSI region may also operate optimally within the region where the Hall 
effect is the dominant source of resistivity. It remains to be shown how the instabilities 
here summarized behave in connection with these non-ideal terms. }

The importance of characterizing the shape and quality of turbulence and its influence
upon the growth of planetesimals, especially at the smallest scales, is one of the main messages of this review.  While we have not delved into how the turbulence mitigates the growth of planetesimals, we have emphasized that the dynamical state of the protoplanetary disk is the fundamental canvas on which the particle growth narrative is etched.
With this in mind, we conclude with another quote from the ancients, this time of Hesiod, who
writes of the birth of the Earth (``Gaia"),
\begin{quote}
{\it In the beginning there was Chaos; but then there came to be
Gaia, the broad-breasted, the ever-secure seat of all
immortals who dwell in the peaks of snowy Olympos.}\par
\medskip
Hesiod {\emph{Theogony}} 116-118\footnote
{Translation by Professor Osman S. Umurhan who notes that ``$\chi\alpha\omicron\sigma$" or ``Chaos" represents not disorder in the modern sense, but a chasm or
abyss, whose properties appear to be a dark, gaping space. Later in Hesiod's account
this very chasm appears to have the appropriate material ``to catch fire" from Zeus'
thunderbolt.}
\end{quote}
This vision strikes as modern as well: if one substitutes ``chaos" with  {\emph{turbulence}} and ``Gaia" with planetesimals, then we have a picture describing elements
of today's scientific paradigm of planet formation.\\

\acknowledgments
{\bf{\emph{Acknowledgements.}}} 
 {We thank Joe Barranco, Konstantin Batygin, Luca Biancofiore, Til Birnstiel,  Omer Blaes, Axel Brandenburg, Simon Casassus, Thayne Currie, Jeff Cuzzi, Kees Dullemond, Natalia Dzyurkevich, Paul Estrada, Mario Flock, Sebastien Fromang, Oliver Gressel, Nili Harnik, Eyal Heifetz, Tobias Heinemann, Thomas Henning, Anders Johansen, Hubert Klahr, Willy Kley, Min-Kai Lin, Mordecai Mac Low, Philip Marcus, Fr\'ed\'eric Masset, Colin McNally, Helo\"ise M\'eh\'eut, Farzana Meru, Alessandro Morbidelli, Krzyzstof Mizerski, Richard Nelson, Chris Ormel, Sebasti\'an P\'erez, Oded Regev, Luca Ricci, Giora Shaviv, Jake Simon, Neal Turner, Nienke van der Marel, Peggy Varni\`ere, and Andrew Youdin, for their support and valuable scientific discussions over the years. We thank Willy Kley and Henrik Latter for permission to use their figures, and for comments on the draft. We thank the referee, Hubert Klahr, for substantial suggestions that vastly improved the quality of this review. We also appreciate comments from Hantao Ji and Andrew Youdin correcting several omissions from an earlier version of the work. We express gratitude to Professor Osman Umurhan for help with classical period translations. We are especially indebted to James Y.-K. Cho for illuminating subtleties characterizing the nature of atmospheric turbulence. 
\par
W.L. acknowledges support of Space Telescope Science Institute through grant HST-AR-14572 and the NASA Exoplanet Research Program through grant 16-XRP16\_2-0065. O.M.U. acknowledges support from the NASA Astrophysics Theory Program through grant NNX17AK59G.}

\bibliographystyle{apj}
\bibliography{ms}

\begin{thebibliography}{172}
\expandafter\ifx\csname natexlab\endcsname\relax\def\natexlab#1{#1}\fi

\bibitem[{{Abramowicz} {et~al.}(1984){Abramowicz}, {Livio}, {Piran}, \&
  {Wiita}}]{Abramowicz+84}
{Abramowicz}, M.~A., {Livio}, M., {Piran}, T., \& {Wiita}, P.~J. 1984, \apj,
  279, 367

\bibitem[{{Andrews} \& {Williams}(2007)}]{AndrewsWilliams07}
{Andrews}, S.~M., \& {Williams}, J.~P. 2007, \apj, 659, 705

\bibitem[{{Andrews} {et~al.}(2009){Andrews}, {Wilner}, {Hughes}, {Qi}, \&
  {Dullemond}}]{Andrews+09}
{Andrews}, S.~M., {Wilner}, D.~J., {Hughes}, A.~M., {Qi}, C., \& {Dullemond},
  C.~P. 2009, \apj, 700, 1502

\bibitem[{{Ansdell} {et~al.}(2018){Ansdell}, {Williams}, {Trapman}, {van
  Terwisga}, {Facchini}, {Manara}, {van der Marel}, {Miotello}, {Tazzari},
  {Hogerheijde}, {Guidi}, {Testi}, \& {van Dishoeck}}]{Ansdell+18}
{Ansdell}, M. {et~al.} 2018, \apj, 859, 21

\bibitem[{{Arlt} \& {Urpin}(2004{\natexlab{a}})}]{Arlt_Urpin_2004}
{Arlt}, R., \& {Urpin}, V. 2004{\natexlab{a}}, \aap, 426, 755

\bibitem[{{Arlt} \& {Urpin}(2004{\natexlab{b}})}]{ArltUrpin04}
---. 2004{\natexlab{b}}, \aap, 426, 755

\bibitem[{{Aumann}(1985)}]{Aumann85}
{Aumann}, H.~H. 1985, \pasp, 97, 885

\bibitem[{{Avila}(2012)}]{Avila_2012}
{Avila}, M. 2012, Physical Review Letters, 108, 124501

\bibitem[{{Bai}(2015)}]{Bai15}
{Bai}, X.-N. 2015, \apj, 798, 84

\bibitem[{{Bai} \& {Stone}(2011)}]{BaiStone11}
{Bai}, X.-N., \& {Stone}, J.~M. 2011, \apj, 736, 144

\bibitem[{{Bai} \& {Stone}(2017)}]{BaiStone17}
---. 2017, \apj, 836, 46

\bibitem[{{Balbus}(2003)}]{Balbus_2003}
{Balbus}, S.~A. 2003, \araa, 41, 555

\bibitem[{{Balbus}(2017)}]{Balbus_2017}
---. 2017, Journal of Fluid Mechanics, 824, 1

\bibitem[{{Balbus} \& {Hawley}(1991)}]{BalbusHawley91}
{Balbus}, S.~A., \& {Hawley}, J.~F. 1991, \apj, 376, 214

\bibitem[{{Balbus} \& {Terquem}(2001)}]{BalbusTerquem01}
{Balbus}, S.~A., \& {Terquem}, C. 2001, \apj, 552, 235

\bibitem[{{Barker} \& {Latter}(2015{\natexlab{a}})}]{BarkerLatter15}
{Barker}, A.~J., \& {Latter}, H.~N. 2015{\natexlab{a}}, \mnras, 450, 21

\bibitem[{{Barker} \& {Latter}(2015{\natexlab{b}})}]{Barker_Latter_2015}
---. 2015{\natexlab{b}}, \mnras, 450, 21

\bibitem[{{Barranco} {et~al.}(2018){Barranco}, {Pei}, \&
  {Marcus}}]{Barranco+18}
{Barranco}, J., {Pei}, S., \& {Marcus}, P. 2018, ArXiv e-prints

\bibitem[{{Batygin}(2018)}]{Batygin18}
{Batygin}, K. 2018, \aj, 155, 178

\bibitem[{{Bayly}(1986)}]{Bayly86}
{Bayly}, B.~J. 1986, Physical Review Letters, 57, 2160

\bibitem[{{Beckwith} {et~al.}(1990){Beckwith}, {Sargent}, {Chini}, \&
  {Guesten}}]{Beckwith+90}
{Beckwith}, S.~V.~W., {Sargent}, A.~I., {Chini}, R.~S., \& {Guesten}, R. 1990,
  \aj, 99, 924

\bibitem[{{Benz}(2000)}]{Benz00}
{Benz}, W. 2000, \ssr, 92, 279

\bibitem[{{Biancofiore} \& {Umurhan}(2018)}]{BiancofioreUmurhan18}
{Biancofiore}, L., \& {Umurhan}, O.~M. 2018, ArXiv e-prints

\bibitem[{{Bitsch} {et~al.}(2015){Bitsch}, {Johansen}, {Lambrechts}, \&
  {Morbidelli}}]{Bitsch+2015}
{Bitsch}, B., {Johansen}, A., {Lambrechts}, M., \& {Morbidelli}, A. 2015, \aap,
  575, A28

\bibitem[{{Blaes} \& {Balbus}(1994)}]{BlaesBalbus94}
{Blaes}, O.~M., \& {Balbus}, S.~A. 1994, \apj, 421, 163

\bibitem[{{Boffetta} \& {Ecke}(2012)}]{Boffetta_Ecke_2012}
{Boffetta}, G., \& {Ecke}, R.~E. 2012, Annual Review of Fluid Mechanics, 44,
  427

\bibitem[{{Boffetta} \& {Musacchio}(2010)}]{Boffetta_Musacchio_2010}
{Boffetta}, G., \& {Musacchio}, S. 2010, \pre, 82, 016307

\bibitem[{{Brauer} {et~al.}(2007){Brauer}, {Dullemond}, {Johansen}, {Henning},
  {Klahr}, \& {Natta}}]{Brauer+07}
{Brauer}, F., {Dullemond}, C.~P., {Johansen}, A., {Henning}, T., {Klahr}, H.,
  \& {Natta}, A. 2007, \aap, 469, 1169

\bibitem[{{Bryan} {et~al.}(2018){Bryan}, {Benneke}, {Knutson}, {Batygin}, \&
  {Bowler}}]{Bryan+18}
{Bryan}, M.~L., {Benneke}, B., {Knutson}, H.~A., {Batygin}, K., \& {Bowler},
  B.~P. 2018, Nature Astronomy, 2, 138

\bibitem[{{Butler} \& {Farrell}(1992)}]{Butler_Farrell_1992}
{Butler}, K.~M., \& {Farrell}, B.~F. 1992, Physics of Fluids A, 4, 1637

\bibitem[{{Cabot}(1996)}]{Cabot+96}
{Cabot}, W. 1996, \apj, 465, 874

\bibitem[{{Cameron}(1978)}]{Cameron78}
{Cameron}, A.~G.~W. 1978, Moon and Planets, 18, 5

\bibitem[{{Chagelishvili} {et~al.}(2003){Chagelishvili}, {Zahn}, {Tevzadze}, \&
  {Lominadze}}]{Chagelishvili+2003}
{Chagelishvili}, G.~D., {Zahn}, J.-P., {Tevzadze}, A.~G., \& {Lominadze}, J.~G.
  2003, \aap, 402, 401

\bibitem[{{Chiang} \& {Goldreich}(1997)}]{ChiangGoldreich97}
{Chiang}, E.~I., \& {Goldreich}, P. 1997, \apj, 490, 368

\bibitem[{{Davidson}(2004)}]{Davidson_2004}
{Davidson}, P.~A. 2004, {Turbulence : an introduction for scientists and
  engineers} (Oxford)

\bibitem[{{de Val-Borro} {et~al.}(2007){de Val-Borro}, {Artymowicz},
  {D'Angelo}, \& {Peplinski}}]{deValBorro+07}
{de Val-Borro}, M., {Artymowicz}, P., {D'Angelo}, G., \& {Peplinski}, A. 2007,
  \aap, 471, 1043

\bibitem[{{Diels} \& {Kranz}(1961)}]{Diels_Kranz_1961}
{Diels}, H., \& {Kranz}, W. 1961, {Die Fragmente der Vorsokratiker, vols. I and
  II.} (Berlin Weidmann)

\bibitem[{{Dubrulle} {et~al.}(1995){Dubrulle}, {Morfill}, \&
  {Sterzik}}]{Dubrulle+95}
{Dubrulle}, B., {Morfill}, G., \& {Sterzik}, M. 1995, Icarus, 114, 237

\bibitem[{{Dzyurkevich} {et~al.}(2013){Dzyurkevich}, {Turner}, {Henning}, \&
  {Kley}}]{Dzyurkevich+13}
{Dzyurkevich}, N., {Turner}, N.~J., {Henning}, T., \& {Kley}, W. 2013, \apj,
  765, 114

\bibitem[{{Edlund} \& {Ji}(2014)}]{Edlund_Ji_2014}
{Edlund}, E.~M., \& {Ji}, H. 2014, \pre, 89, 021004

\bibitem[{{Edlund} \& {Ji}(2015)}]{Edlund_Ji_2015}
---. 2015, \pre, 92, 043005

\bibitem[{{Elsasser} \& {Staude}(1978)}]{ElsasserStaude78}
{Elsasser}, H., \& {Staude}, H.~J. 1978, \aap, 70, L3

\bibitem[{{Estrada} {et~al.}(2016){Estrada}, {Cuzzi}, \& {Morgan}}]{Estrada+16}
{Estrada}, P.~R., {Cuzzi}, J.~N., \& {Morgan}, D.~A. 2016, \apj, 818, 200

\bibitem[{{Flock} {et~al.}(2017){Flock}, {Nelson}, {Turner}, {Bertrang},
  {Carrasco-Gonz{\'a}lez}, {Henning}, {Lyra}, \& {Teague}}]{Flock_etal_2017}
{Flock}, M., {Nelson}, R.~P., {Turner}, N.~J., {Bertrang}, G.~H.-M.,
  {Carrasco-Gonz{\'a}lez}, C., {Henning}, T., {Lyra}, W., \& {Teague}, R. 2017,
  \apj, 850, 131

\bibitem[{{Fricke}(1968)}]{Fricke_1968}
{Fricke}, K. 1968, \zap, 68, 317

\bibitem[{{Frisch}(1995)}]{Frisch_1995}
{Frisch}, U. 1995, {Turbulence} (Cambridge University Press)

\bibitem[{{Gammie}(1996)}]{Gammie96}
{Gammie}, C.~F. 1996, \apj, 457, 355

\bibitem[{{Gamow} \& {Hynek}(1945)}]{Gamow_Hynek_1945}
{Gamow}, G., \& {Hynek}, J.~A. 1945, \apj, 101, 249

\bibitem[{{Garaud} \& {Lin}(2004)}]{GaraudLin04}
{Garaud}, P., \& {Lin}, D.~N.~C. 2004, \apj, 608, 1050

\bibitem[{{Goldreich} {et~al.}(1986){Goldreich}, {Goodman}, \&
  {Narayan}}]{Goldreich+86}
{Goldreich}, P., {Goodman}, J., \& {Narayan}, R. 1986, \mnras, 221, 339

\bibitem[{{Goldreich} \& {Lynden-Bell}(1965)}]{Goldreich_Lynden-Bell_1965}
{Goldreich}, P., \& {Lynden-Bell}, D. 1965, \mnras, 130, 125

\bibitem[{{Goldreich} \& {Schubert}(1967)}]{Goldreich_Schubert_1967}
{Goldreich}, P., \& {Schubert}, G. 1967, \apj, 150, 571

\bibitem[{{Goldreich} \& {Ward}(1973)}]{GoldreichWard73}
{Goldreich}, P., \& {Ward}, W.~R. 1973, \apj, 183, 1051

\bibitem[{{Goodman} \& {Xu}(1994)}]{Goodman_Xu_1994}
{Goodman}, J., \& {Xu}, G. 1994, \apj, 432, 213

\bibitem[{{Gressel} {et~al.}(2015){Gressel}, {Turner}, {Nelson}, \&
  {McNally}}]{Gressel+15}
{Gressel}, O., {Turner}, N.~J., {Nelson}, R.~P., \& {McNally}, C.~P. 2015,
  \apj, 801, 84

\bibitem[{{Haghighipour} \& {Boss}(2003)}]{HaghighipourBoss03}
{Haghighipour}, N., \& {Boss}, A.~P. 2003, \apj, 583, 996

\bibitem[{{Hartmann} {et~al.}(1998){Hartmann}, {Calvet}, {Gullbring}, \&
  {D'Alessio}}]{Hartmann+98}
{Hartmann}, L., {Calvet}, N., {Gullbring}, E., \& {D'Alessio}, P. 1998, \apj,
  495, 385

\bibitem[{{Hayashi}(1981)}]{Hayashi81}
{Hayashi}, C. 1981, Progress of Theoretical Physics Supplement, 70, 35

\bibitem[{{Heinemann} \& {Papaloizou}(2009)}]{HeinemannPapaloizou09}
{Heinemann}, T., \& {Papaloizou}, J.~C.~B. 2009, \mnras, 397, 52

\bibitem[{{Heinemann} \& {Papaloizou}(2012)}]{HeinemannPapaloizou12}
---. 2012, \mnras, 419, 1085

\bibitem[{{Hughes} {et~al.}(2008){Hughes}, {Wilner}, {Qi}, \&
  {Hogerheijde}}]{Hughes+08}
{Hughes}, A.~M., {Wilner}, D.~J., {Qi}, C., \& {Hogerheijde}, M.~R. 2008, \apj,
  678, 1119

\bibitem[{{Ji} {et~al.}(2006){Ji}, {Burin}, {Schartman}, \& {Goodman}}]{Ji+06}
{Ji}, H., {Burin}, M., {Schartman}, E., \& {Goodman}, J. 2006, \nat, 444, 343

\bibitem[{{Johansen} {et~al.}(2006){Johansen}, {Henning}, \&
  {Klahr}}]{Johansen+06}
{Johansen}, A., {Henning}, T., \& {Klahr}, H. 2006, \apj, 643, 1219

\bibitem[{{Johansen} {et~al.}(2007){Johansen}, {Oishi}, {Mac Low}, {Klahr},
  {Henning}, \& {Youdin}}]{Johansen+07}
{Johansen}, A., {Oishi}, J.~S., {Mac Low}, M.-M., {Klahr}, H., {Henning}, T.,
  \& {Youdin}, A. 2007, \nat, 448, 1022

\bibitem[{{Johansen} \& {Youdin}(2007)}]{JohansenYoudin07}
{Johansen}, A., \& {Youdin}, A. 2007, \apj, 662, 627

\bibitem[{{Kippenhahn} {et~al.}(2012){Kippenhahn}, {Weigert}, \&
  {Weiss}}]{Kippenhahn_etal_2012}
{Kippenhahn}, R., {Weigert}, A., \& {Weiss}, A. 2012, {Stellar Structure and
  Evolution} (Springer-Verlag)

\bibitem[{{Klahr}(2004)}]{Klahr04}
{Klahr}, H. 2004, \apj, 606, 1070

\bibitem[{{Klahr} \& {Hubbard}(2014)}]{KlahrHubbard14}
{Klahr}, H., \& {Hubbard}, A. 2014, \apj, 788, 21

\bibitem[{{Klahr} \& {Bodenheimer}(2003)}]{KlahrBodenheimer03}
{Klahr}, H.~H., \& {Bodenheimer}, P. 2003, \apj, 582, 869

\bibitem[{{Koller} {et~al.}(2003){Koller}, {Li}, \& {Lin}}]{Koller+03}
{Koller}, J., {Li}, H., \& {Lin}, D. 2003, \apj, 596, L91

\bibitem[{{Krall} \& {Trivelpiece}(1973)}]{KrallTrivelpiece73}
{Krall}, N.~A., \& {Trivelpiece}, A.~W. 1973, {Principles of plasma physics}

\bibitem[{{Kunz}(2008)}]{Kunz08}
{Kunz}, M.~W. 2008, \mnras, 385, 1494

\bibitem[{{Kunz} \& {Lesur}(2013)}]{KunzLesur13}
{Kunz}, M.~W., \& {Lesur}, G. 2013, \mnras, 434, 2295

\bibitem[{{Latter}(2016)}]{Latter16}
{Latter}, H.~N. 2016, \mnras, 455, 2608

\bibitem[{{Latter} \& {Papaloizou}(2017)}]{Latter_Pap_2017}
{Latter}, H.~N., \& {Papaloizou}, J. 2017, \mnras, 472, 1432

\bibitem[{{Latter} \& {Papaloizou}(2018)}]{LatterPapaloizou18}
---. 2018, \mnras, 474, 3110

\bibitem[{{Lesur} {et~al.}(2014){Lesur}, {Kunz}, \& {Fromang}}]{Lesur+14}
{Lesur}, G., {Kunz}, M.~W., \& {Fromang}, S. 2014, \aap, 566, A56

\bibitem[{{Lesur} \& {Papaloizou}(2009)}]{LesurPapaloizou09}
{Lesur}, G., \& {Papaloizou}, J.~C.~B. 2009, \aap, 498, 1

\bibitem[{{Lesur} \& {Papaloizou}(2010)}]{LesurPapaloizou10}
---. 2010, \aap, 513, A60

\bibitem[{{Lesur} \& {Latter}(2016)}]{LesurLatter16}
{Lesur}, G.~R.~J., \& {Latter}, H. 2016, \mnras, 462, 4549

\bibitem[{{Li} {et~al.}(2001){Li}, {Colgate}, {Wendroff}, \& {Liska}}]{Li+10}
{Li}, H., {Colgate}, S.~A., {Wendroff}, B., \& {Liska}, R. 2001, \apj, 551, 874

\bibitem[{{Li} {et~al.}(2000){Li}, {Finn}, {Lovelace}, \& {Colgate}}]{Li+00}
{Li}, H., {Finn}, J.~M., {Lovelace}, R.~V.~E., \& {Colgate}, S.~A. 2000, \apj,
  533, 1023

\bibitem[{{Lin} \& {Papaloizou}(1980)}]{LinPapaloizou80}
{Lin}, D.~N.~C., \& {Papaloizou}, J. 1980, \mnras, 191, 37

\bibitem[{{Lin}(2014)}]{Lin14}
{Lin}, M.-K. 2014, \mnras, 437, 575

\bibitem[{{Lin} \& {Youdin}(2015)}]{Lin_Youdin_2015}
{Lin}, M.-K., \& {Youdin}, A.~N. 2015, \apj, 811, 17

\bibitem[{{Lopez} \& {Avila}(2017)}]{Lopez_Avila_2017}
{Lopez}, J.~M., \& {Avila}, M. 2017, Journal of Fluid Mechanics, 817, 21

\bibitem[{{Lovelace} {et~al.}(1999){Lovelace}, {Li}, {Colgate}, \&
  {Nelson}}]{Lovelace+99}
{Lovelace}, R., {Li}, H., {Colgate}, S., \& {Nelson}, A. 1999, \apj, 513, 805

\bibitem[{{Lynden-Bell} \& {Pringle}(1974)}]{Lynden-BellPringle74}
{Lynden-Bell}, D., \& {Pringle}, J.~E. 1974, \mnras, 168, 603

\bibitem[{{Lyra}(2013)}]{Lyra13}
{Lyra}, W. 2013, in European Physical Journal Web of Conferences, Vol.~46,
  European Physical Journal Web of Conferences, 04003

\bibitem[{{Lyra}(2014)}]{Lyra14}
{Lyra}, W. 2014, \apj, 789, 77

\bibitem[{{Lyra} {et~al.}(2009){Lyra}, {Johansen}, {Klahr}, \&
  {Piskunov}}]{Lyra+09b}
{Lyra}, W., {Johansen}, A., {Klahr}, H., \& {Piskunov}, N. 2009, \aap, 493,
  1125

\bibitem[{{Lyra} \& {Klahr}(2011)}]{LyraKlahr11}
{Lyra}, W., \& {Klahr}, H. 2011, \aap, 527, A138

\bibitem[{{Lyra} \& {Mac Low}(2012)}]{LyraMacLow12}
{Lyra}, W., \& {Mac Low}, M.-M. 2012, \apj, 756, 62

\bibitem[{{Lyra} {et~al.}(2015){Lyra}, {Turner}, \& {McNally}}]{Lyra+15}
{Lyra}, W., {Turner}, N.~J., \& {McNally}, C.~P. 2015, \aap, 574, A10

\bibitem[{{Lyttleton}(1972)}]{Lyttleton72}
{Lyttleton}, R.~A. 1972, \mnras, 158, 463

\bibitem[{{Mac Low} {et~al.}(1995){Mac Low}, {Norman}, {Konigl}, \&
  {Wardle}}]{MacLow+95}
{Mac Low}, M.-M., {Norman}, M.~L., {Konigl}, A., \& {Wardle}, M. 1995, \apj,
  442, 726

\bibitem[{{Malygin} {et~al.}(2017){Malygin}, {Klahr}, {Semenov}, {Henning}, \&
  {Dullemond}}]{Malygin+17}
{Malygin}, M.~G., {Klahr}, H., {Semenov}, D., {Henning}, T., \& {Dullemond},
  C.~P. 2017, \aap, 605, A30

\bibitem[{{Manger} \& {Klahr}(2018)}]{MangerKlahr18}
{Manger}, N., \& {Klahr}, H. 2018, \mnras

\bibitem[{{Marcus} {et~al.}(2016){Marcus}, {Pei}, {Jiang}, \&
  {Barranco}}]{Marcus+16}
{Marcus}, P.~S., {Pei}, S., {Jiang}, C.-H., \& {Barranco}, J.~A. 2016, \apj,
  833, 148

\bibitem[{{Marcus} {et~al.}(2015){Marcus}, {Pei}, {Jiang}, {Barranco},
  {Hassanzadeh}, \& {Lecoanet}}]{Marcus+15}
{Marcus}, P.~S., {Pei}, S., {Jiang}, C.-H., {Barranco}, J.~A., {Hassanzadeh},
  P., \& {Lecoanet}, D. 2015, \apj, 808, 87

\bibitem[{{Marcus} {et~al.}(2013){Marcus}, {Pei}, {Jiang}, \&
  {Hassanzadeh}}]{Marcus+13}
{Marcus}, P.~S., {Pei}, S., {Jiang}, C.-H., \& {Hassanzadeh}, P. 2013, Physical
  Review Letters, 111, 084501

\bibitem[{{Maretzke} {et~al.}(2014){Maretzke}, {Hof}, \&
  {Avila}}]{Maretzke+2014}
{Maretzke}, S., {Hof}, B., \& {Avila}, M. 2014, Journal of Fluid Mechanics,
  742, 254

\bibitem[{{Marino} {et~al.}(2014){Marino}, {Mininni}, {Rosenberg}, \&
  {Pouquet}}]{Marino_etal_2014}
{Marino}, R., {Mininni}, P.~D., {Rosenberg}, D.~L., \& {Pouquet}, A. 2014,
  \pre, 90, 023018

\bibitem[{{McCaughrean} \& {O'dell}(1996)}]{McCaughreanO'Dell96}
{McCaughrean}, M.~J., \& {O'dell}, C.~R. 1996, \aj, 111, 1977

\bibitem[{{McNally} \& {Pessah}(2015)}]{McNally_Pessah_2015}
{McNally}, C.~P., \& {Pessah}, M.~E. 2015, \apj, 811, 121

\bibitem[{{Meheut} {et~al.}(2010){Meheut}, {Casse}, {Varniere}, \&
  {Tagger}}]{Meheut+10}
{Meheut}, H., {Casse}, F., {Varniere}, P., \& {Tagger}, M. 2010, \aap, 516, A31

\bibitem[{{Mestel}(1965{\natexlab{a}})}]{Mestel65a}
{Mestel}, L. 1965{\natexlab{a}}, \qjras, 6, 161

\bibitem[{{Mestel}(1965{\natexlab{b}})}]{Mestel65b}
---. 1965{\natexlab{b}}, \qjras, 6, 265

\bibitem[{{Nastrom} \& {Gage}(1985)}]{Nastrom_Gage_1985}
{Nastrom}, G.~D., \& {Gage}, K.~S. 1985, Journal of Atmospheric Sciences, 42,
  950

\bibitem[{{Nelson} {et~al.}(2013{\natexlab{a}}){Nelson}, {Gressel}, \&
  {Umurhan}}]{Nelson+13}
{Nelson}, R.~P., {Gressel}, O., \& {Umurhan}, O.~M. 2013{\natexlab{a}}, \mnras,
  435, 2610

\bibitem[{{Nelson} {et~al.}(2013{\natexlab{b}}){Nelson}, {Gressel}, \&
  {Umurhan}}]{Nelson_etal_2013}
---. 2013{\natexlab{b}}, \mnras, 435, 2610

\bibitem[{{O'dell} \& {Wen}(1994)}]{O'dellWen94}
{O'dell}, C.~R., \& {Wen}, Z. 1994, \apj, 436, 194

\bibitem[{{Ogilvie}(2016)}]{Ogilvie16}
{Ogilvie}, G.~I. 2016, Journal of Plasma Physics, 82, 205820301

\bibitem[{{Paardekooper}(2006)}]{Paardekooper06}
{Paardekooper}, S.-J. 2006, PhD thesis, Leiden Observatory, Leiden University,
  P.O.~Box 9513, 2300 RA Leiden, The Netherlands

\bibitem[{{Pandey} \& {Wardle}(2006)}]{PandeyWardle06}
{Pandey}, B.~P., \& {Wardle}, M. 2006, \mnras, 371, 1014

\bibitem[{{Pandey} \& {Wardle}(2008)}]{PandeyWardle08}
---. 2008, \mnras, 385, 2269

\bibitem[{{Paoletti} {et~al.}(2012){Paoletti}, {van Gils}, {Dubrulle}, {Sun},
  {Lohse}, \& {Lathrop}}]{Paoletti+12}
{Paoletti}, M.~S., {van Gils}, D.~P.~M., {Dubrulle}, B., {Sun}, C., {Lohse},
  D., \& {Lathrop}, D.~P. 2012, \aap, 547, A64

\bibitem[{{Pedlosky}(1982)}]{Pedlosky82}
{Pedlosky}, J. 1982, {Geophysical fluid dynamics}

\bibitem[{{Petersen} {et~al.}(2007{\natexlab{a}}){Petersen}, {Julien}, \&
  {Stewart}}]{Petersen+07a}
{Petersen}, M.~R., {Julien}, K., \& {Stewart}, G.~R. 2007{\natexlab{a}}, \apj,
  658, 1236

\bibitem[{{Petersen} {et~al.}(2007{\natexlab{b}}){Petersen}, {Stewart}, \&
  {Julien}}]{Petersen+07b}
{Petersen}, M.~R., {Stewart}, G.~R., \& {Julien}, K. 2007{\natexlab{b}}, \apj,
  658, 1252

\bibitem[{{Pierrehumbert}(1986)}]{Pierrehumbert86}
{Pierrehumbert}, R.~T. 1986, Physical Review Letters, 57, 2157

\bibitem[{{Regev} {et~al.}(2016){Regev}, {Umurhan}, \&
  {Yecko}}]{Regev_Umurhan_Yecko_2016}
{Regev}, O., {Umurhan}, O.~M., \& {Yecko}, P.~A. 2016, {Modern Fluid Dynamics
  for Physics and Astrophysics} (Springer)

\bibitem[{{Ricci} {et~al.}(2008){Ricci}, {Robberto}, \& {Soderblom}}]{Ricci+08}
{Ricci}, L., {Robberto}, M., \& {Soderblom}, D.~R. 2008, \aj, 136, 2136

\bibitem[{{Richard} \& {Zahn}(1999)}]{Richard_Zahn_1999}
{Richard}, D., \& {Zahn}, J.-P. 1999, \aap, 347, 734

\bibitem[{{Richard} {et~al.}(2016){Richard}, {Nelson}, \&
  {Umurhan}}]{Richard_etal_2016}
{Richard}, S., {Nelson}, R.~P., \& {Umurhan}, O.~M. 2016, \mnras, 456, 3571

\bibitem[{{Rucinski}(1985)}]{Rucinski85}
{Rucinski}, S.~M. 1985, \aj, 90, 2321

\bibitem[{{Safronov}(1972)}]{Safronov72}
{Safronov}, V.~S. 1972, {Evolution of the protoplanetary cloud and formation of
  the earth and planets.}

\bibitem[{{Salmeron} \& {Wardle}(2005)}]{SalmeronWardle05}
{Salmeron}, R., \& {Wardle}, M. 2005, \mnras, 361, 45

\bibitem[{{Sargent} \& {Beckwith}(1987)}]{SargentBeckwith87}
{Sargent}, A.~I., \& {Beckwith}, S. 1987, \apj, 323, 294

\bibitem[{{Schmid} \& Henningson(2001)}]{Schmid_Henningson_2001}
{Schmid}, P.~J., \& Henningson, D.~S. 2001, {Stability and Transition in Shear
  Flows} (Springer), 558

\bibitem[{{Sekiya}(1998)}]{Sekiya98}
{Sekiya}, M. 1998, \icarus, 133, 298

\bibitem[{{Semenov} {et~al.}(2003){Semenov}, {Henning}, {Helling}, {Ilgner}, \&
  {Sedlmayr}}]{Semenov+03}
{Semenov}, D., {Henning}, T., {Helling}, C., {Ilgner}, M., \& {Sedlmayr}, E.
  2003, \aap, 410, 611

\bibitem[{{Shakura} \& {Sunyaev}(1973)}]{ShakuraSunyaev73}
{Shakura}, N.~I., \& {Sunyaev}, R.~A. 1973, \aap, 24, 337

\bibitem[{{Shakura} {et~al.}(1978){Shakura}, {Sunyaev}, \&
  {Zilitinkevich}}]{Shakura+78}
{Shakura}, N.~I., {Sunyaev}, R.~A., \& {Zilitinkevich}, S.~S. 1978, \aap, 62,
  179

\bibitem[{{Sheehan} {et~al.}(1999){Sheehan}, {Davis}, {Cuzzi}, \&
  {Estberg}}]{Sheehan+99}
{Sheehan}, D.~P., {Davis}, S.~S., {Cuzzi}, J.~N., \& {Estberg}, G.~N. 1999,
  \icarus, 142, 238

\bibitem[{{Simon} {et~al.}(2018){Simon}, {Bai}, {Flaherty}, \&
  {Hughes}}]{Simon_etal_2018}
{Simon}, J.~B., {Bai}, X.-N., {Flaherty}, K.~M., \& {Hughes}, A.~M. 2018, \apj,
  865, 10

\bibitem[{{Simon} {et~al.}(2016){Simon}, {Pascucci}, {Edwards}, {Feng},
  {Gorti}, {Hollenbach}, {Rigliaco}, \& {Keane}}]{Simon+16}
{Simon}, M.~N., {Pascucci}, I., {Edwards}, S., {Feng}, W., {Gorti}, U.,
  {Hollenbach}, D., {Rigliaco}, E., \& {Keane}, J.~T. 2016, \apj, 831, 169

\bibitem[{{Spiegel}(1957)}]{Spiegel_1957}
{Spiegel}, E.~A. 1957, \apj, 126, 202

\bibitem[{{Stoll} \& {Kley}(2014)}]{Stoll_Kley_2014}
{Stoll}, M.~H.~R., \& {Kley}, W. 2014, \aap, 572, A77

\bibitem[{{Stoll} \& {Kley}(2016)}]{Stoll_Kley_2016}
---. 2016, \aap, 594, A57

\bibitem[{{Stoll} {et~al.}(2017{\natexlab{a}}){Stoll}, {Kley}, \&
  {Picogna}}]{Stoll_Kley_Picogna_2017}
{Stoll}, M.~H.~R., {Kley}, W., \& {Picogna}, G. 2017{\natexlab{a}}, \aap, 599,
  L6

\bibitem[{{Stoll} {et~al.}(2017{\natexlab{b}}){Stoll}, {Picogna}, \&
  {Kley}}]{Stoll_Picogna_Kley_2017}
{Stoll}, M.~H.~R., {Picogna}, G., \& {Kley}, W. 2017{\natexlab{b}}, \aap, 604,
  A28

\bibitem[{{Strom} {et~al.}(1989){Strom}, {Strom}, {Edwards}, {Cabrit}, \&
  {Skrutskie}}]{Strom+89}
{Strom}, K.~M., {Strom}, S.~E., {Edwards}, S., {Cabrit}, S., \& {Skrutskie},
  M.~F. 1989, \aj, 97, 1451

\bibitem[{{Sun} {et~al.}(2017){Sun}, {Rotunno}, \& {Zhang}}]{Sun_etal_2017}
{Sun}, Y.~Q., {Rotunno}, R., \& {Zhang}, F. 2017, Journal of Atmospheric
  Sciences, 74, 185

\bibitem[{{Takata} \& {Stevenson}(1996)}]{TakataStevenson96}
{Takata}, T., \& {Stevenson}, D.~J. 1996, \icarus, 123, 404

\bibitem[{{Tassoul}(1978)}]{Tassoul_1978}
{Tassoul}, J.-L. 1978, {Theory of rotating stars} (Princeton)

\bibitem[{{Taylor}(1936)}]{Taylor_1936}
{Taylor}, G.~I. 1936, Proceedings of the Royal Society of London Series A, 157,
  546

\bibitem[{{Tulloch} \& {Smith}(2006)}]{Tulloch_Smith_2006}
{Tulloch}, R., \& {Smith}, K.~S. 2006, Proceedings of the National Academy of
  Science, 103, 14690

\bibitem[{{Turner} \& {Drake}(2009)}]{TurnerDrake09}
{Turner}, N.~J., \& {Drake}, J.~F. 2009, \apj, 703, 2152

\bibitem[{{Umurhan}(2010)}]{Umurhan10}
{Umurhan}, O.~M. 2010, \aap, 521, A25

\bibitem[{{Umurhan} {et~al.}(2017){Umurhan}, {Estrada}, \&
  {Cuzzi}}]{Umurhan+2017}
{Umurhan}, O.~M., {Estrada}, P.~R., \& {Cuzzi}, J.~N. 2017, in Lunar and
  Planetary Inst.~Technical Report, Vol.~48, Lunar and Planetary Science
  Conference, 2616

\bibitem[{{Umurhan} {et~al.}(2013){Umurhan}, {Nelson}, \&
  {Gressel}}]{Umurhan+13}
{Umurhan}, O.~M., {Nelson}, R.~P., \& {Gressel}, O. 2013, in European Physical
  Journal Web of Conferences, Vol.~46, European Physical Journal Web of
  Conferences, 03003

\bibitem[{{Umurhan} {et~al.}(2016{\natexlab{a}}){Umurhan}, {Nelson}, \&
  {Gressel}}]{Umurhan+16VSI}
{Umurhan}, O.~M., {Nelson}, R.~P., \& {Gressel}, O. 2016{\natexlab{a}}, \aap,
  586, A33

\bibitem[{{Umurhan} \& {Regev}(2004)}]{UmurhanRegev04}
{Umurhan}, O.~M., \& {Regev}, O. 2004, \aap, 427, 855

\bibitem[{{Umurhan} {et~al.}(2016{\natexlab{b}}){Umurhan}, {Shariff}, \&
  {Cuzzi}}]{Umurhan+16ZVI}
{Umurhan}, O.~M., {Shariff}, K., \& {Cuzzi}, J.~N. 2016{\natexlab{b}}, \apj,
  830, 95

\bibitem[{{Urpin}(2003)}]{Urpin_2003}
{Urpin}, V. 2003, \aap, 404, 397

\bibitem[{{Urpin} \& {Brandenburg}(1998)}]{Urpin_Brandenburg_1998}
{Urpin}, V., \& {Brandenburg}, A. 1998, \mnras, 294, 399

\bibitem[{{Vallis}(2006)}]{Vallis06}
{Vallis}, G.~K. 2006, {Atmospheric and Oceanic Fluid Dynamics} (Cambridge
  University Press), 770

\bibitem[{{Varni{\`e}re} \& {Tagger}(2006)}]{VarniereTagger06}
{Varni{\`e}re}, P., \& {Tagger}, M. 2006, \aap, 446, L13

\bibitem[{{Wardle}(1999)}]{Wardle99}
{Wardle}, M. 1999, \mnras, 307, 849

\bibitem[{{Wardle}(2007)}]{Wardle07}
---. 2007, \apss, 311, 35

\bibitem[{{Wardle} \& {K\"onigl}(1993)}]{WardleKonigl93}
{Wardle}, M., \& {K\"onigl}, A. 1993, \apj, 410, 218

\bibitem[{{Wardle} \& {Salmeron}(2012)}]{WardleSalmeron12}
{Wardle}, M., \& {Salmeron}, R. 2012, \mnras, 422, 2737

\bibitem[{{Weidenschilling}(1977)}]{Weidenschilling77a}
{Weidenschilling}, S.~J. 1977, \mnras, 180, 57

\bibitem[{{Weidenschilling}(1980)}]{Weidenschilling80}
---. 1980, \icarus, 44, 172

\bibitem[{{Weidenschilling} \& {Cuzzi}(1993)}]{WeidenschillingCuzzi93}
{Weidenschilling}, S.~J., \& {Cuzzi}, J.~N. 1993, in Protostars and Planets
  III, ed. E.~H. {Levy} \& J.~I. {Lunine}, 1031--1060

\bibitem[{{Yecko}(2004)}]{Yecko_2004}
{Yecko}, P.~A. 2004, \aap, 425, 385

\bibitem[{{Yellin-Bergovoy} {et~al.}(2017){Yellin-Bergovoy}, {Heifetz}, \&
  {Umurhan}}]{Yellin-Bergovoy+17}
{Yellin-Bergovoy}, R., {Heifetz}, E., \& {Umurhan}, O.~M. 2017, ArXiv e-prints

\bibitem[{{Youdin} \& {Johansen}(2007)}]{YoudinJohansen07}
{Youdin}, A., \& {Johansen}, A. 2007, \apj, 662, 613

\bibitem[{{Youdin} \& {Goodman}(2005)}]{YoudinGoodman05}
{Youdin}, A.~N., \& {Goodman}, J. 2005, \apj, 620, 459

\bibitem[{{Youdin} \& {Shu}(2002)}]{YoudinShu02}
{Youdin}, A.~N., \& {Shu}, F.~H. 2002, \apj, 580, 494

\bibitem[{{Young} \& {Read}(2017)}]{Young_Read_2017}
{Young}, R.~M.~B., \& {Read}, P.~L. 2017, Nature Physics, 13, 1135

\end{thebibliography}

\end{document}